\documentclass[twoside,12pt]{article}
\usepackage{epsfig}

\catcode`@=11
\def\@citex[#1]#2{\if@filesw\immediate\write\@auxout{\string\citation{#2}}\fi
  \def\@citea{}\@cite{\@for\@citeb:=#2\do
    {\@citea\def\@citea{,\penalty\@m}\@ifundefined
      {b@\@citeb}{{\bf ?}\@warning
       {Citation `\@citeb' on page \thepage \space undefined}}%
\hbox{\csname b@\@citeb\endcsname}}}{#1}}

\def\citer{\@ifnextchar
[{\@tempswatrue\@citexr}{\@tempswafalse\@citexr[]}}
\def\@citexr[#1]#2{\if@filesw\immediate\write\@auxout{\string\citation{#2}}\fi
  \def\@citea{}\@cite{\@for\@citeb:=#2\do
    {\@citea\def\@citea{--\penalty\@m}\@ifundefined
       {b@\@citeb}{{\bf ?}\@warning
       {Citation `\@citeb' on page \thepage \space undefined}}%
\hbox{\csname b@\@citeb\endcsname}}}{#1}}
\catcode`@=12

\def\Journal#1#2#3#4{{#1} {#2} (#4) #3 }
\def\AIP{{\em AIP Conf. Proc.}} 
\def\PT{{\em Phys. Today}}
\def\eConf{{\em eConf}}
\def\ZETF{{\em Zh. Eksp. Teor. Fiz.}}
\def\JHEP{{\em JHEP}}
\def\JCAP{{\em JCAP}}
\def\JPB{{\em J. Phys.} B}
\def\JPG{{\em J. Phys.} G}

\def\NC{{\em Nuovo Cimento}}
\def\LNC{{\em Lett. Nuovo Cimento}}

\def\NPA{{\em Nucl. Phys.} A}

\def\PPS{{\em Proc. Phys. Soc.} A}
\def\PRO{{\em Prog. Theor. Phys.}}
\def\NPB{{\em Nucl. Phys.} B}
\def\PLA{{\em Phys. Lett.} A}
\def\PLB{{\em Phys. Lett.} B}

\def\PL{{\em Phys. Lett.}}
\def\PRL{\em Phys. Rev. Lett.}
\def\PREV{\em Phys. Rev.}
\def\PREP{\em Phys. Rep.}
\def\PRA{{\em Phys. Rev.} A}
\def\PRD{{\em Phys. Rev.} D}
\def\PRC{{\em Phys. Rev.} C}

\def\SCI{{\em Science}}
\def\PPNP{\em Prog. Part. Nucl. Phys.}
\def\NPPS{{\em Nucl. Phys. Proc. Suppl.} B}
\def\ZPC{{\em Z. Phys.} C}

\def\EPJ{{\em Eur. Phys. J.} C}
\def\EPA{{\em Eur. Phys. J.} A}
\def\EP{{\em Eur. Phys. J.}}
\def\ANNP{\em Ann. Phys. (N.Y.)}
\def\ARNPS{\em Ann. Rev. Nucl. Part. Sci.}
\def\RMP{{\em Rev. Mod. Phys.}}
\def\MPL{{\em Mod. Phys. Lett.} A}

\def\INT{{\em Int. J. Mod. Phys.} A}

\def\AJ{\em Astrophys. J.}
\def\APP{\em Astropart. Phys.}
\def\JETP{\em J. Exp. Theor. Phys.} 
\def\SJNP{\em Sov. J. Nucl. Phys.} 
\def\NATURE{\em Nature}

\newcommand{\lsim}{\buildrel < \over {_\sim}}
\newcommand{\gsim}{\buildrel > \over {_\sim}}
\newcommand{\psibar}{\overline{\psi}}
\newcommand{\ms}{$\overline{\rm MS}$}
\newcommand{\be}{\begin{equation}}
\newcommand{\ee}{\end{equation}}
\newcommand{\bea}{\begin{eqnarray}}
\newcommand{\eea}{\end{eqnarray}}
\newcommand{\ba}{\begin{array}}
\newcommand{\ea}{\end{array}}
\newcommand{\ie}{{\em i.e.,} }

\newcommand{\etal}{{\it et al.}}
\newcommand{\hpinn}{h_{\pi}^1}
\def\bra#1{{\langle#1\vert}}
\def\ket#1{{\vert#1\rangle}}
\newcommand{\sstw}{\sin^2\theta_W}

\begin{document}

\title {
\hfill {\normalsize FT--2004--02} \\  \vspace{-10pt}
\hfill {\normalsize  
} \\ \vspace{10pt}
Low Energy Tests of the Weak Interaction}
\author{J.\ Erler$^1$ and M.J.\ Ramsey-Musolf $^{2,3}$ \\
$^1$Instituto de F\'\i sica, Universidad Nacional Aut\'onoma de M\'exico, 
M\'exico \\
$^2$Kellogg Radiation Laboratory, California Institute of Technology, USA \\
$^3$Department of Physics, University of Connecticut, USA}
\maketitle
\begin{abstract} 
The study of low energy weak interactions of light quarks and leptons continues
to provide important insights into both the Standard Model as well as 
the physics that may lie beyond it. We review the status and future prospects 
for low energy electroweak physics. Recent important experimental and 
theoretical developments are discussed and open theoretical issues are 
highlighted. Particular attention is paid to neutrino physics, searches for 
permanent electric dipole moments, neutral current tests of the running of 
the weak mixing angle, weak decays, and muon physics. We argue that the broad 
range of such studies provides an important complement to high energy collider 
searches for physics beyond the Standard Model. The use of low energy weak 
interactions to probe novel aspects of hadron structure is also discussed.
\end{abstract}
\eject
\tableofcontents

\newpage

\section{Introduction}
\subsection{\it Overview}
The study of low energy weak interactions of light quarks and leptons has 
played an important role in elucidating the structure of the electroweak (EW)
interaction. From the early observations of nuclear $\beta$ decay, through 
the discovery of the neutrino, and up to the observation of neutral current 
(NC) interactions in atomic parity violation (APV) and parity violating (PV) 
electron scattering, these studies have contributed decisively to the shape of 
the Standard Model (SM)~\cite{Weinberg:tq,Salam:rm} as we know it. As of today,
the SM has been tested and confirmed at the 0.1\% level in processes for which
perturbation theory is applicable, and the absence of any substantial 
systematic disagreements with the SM --- other than the non-vanishing neutrino
mass --- indicates that whatever physics lies beyond it must be of 
the decoupling type. Indeed, the search for such \lq\lq new physics" is now 
the driving force in particle physics as well as in various subfields of both 
nuclear and atomic physics. 

This search is motivated by both theoretical and experimental considerations. 
Theoretically, the SM presents a number of unsatisfying features, despite its 
simplicity and phenomenological successes: the hierarchy problem (instability 
of the EW scale), the unexplained origins of mass and violation of discrete 
symmetries ({\em e.g.}, parity), and the lack of unification with gravity 
to name a few examples. Similarly, experimental observations of neutrino 
oscillations, along with cosmological phenomena of dark matter and energy and 
the matter-antimatter asymmetry, have posed puzzles for particle physics that 
cannot be solved within the SM. The quest for answers to these questions is 
clearly one that must be pursued through experiments at high energy colliders,
where direct signatures of new particles would be found. At the same time, 
however, highly precise measurements carried out at lower energies will 
continue to provide important clues as to the shape of the larger framework in 
which the SM must be embedded. 

Historically, precision EW measurements --- at both high and low energies --- 
have provided important insights into various aspects of the SM. For example, 
the stringent limits obtained on the permanent electric dipole moments of 
the neutron and neutral atoms imply that the so-called 
$\theta$-term~\cite{Wilczek:pj} in the $SU(3)_C$ sector of the SM has 
an unnaturally small coefficient, leading one to suspect the existence of some
new symmetry to explain it. From a somewhat different perspective, the study of
EW radiative corrections to precision observables and their dependence on 
the top quark mass, $m_t$, led to a predicted range before the top quark was
discovered at the Tevatron. The agreement between the measured value of $m_t$ 
and the implications of precision measurements provided a stunning confirmation
of the SM at the level of radiative corrections. Finally, the study of weak 
decays of baryons and mesons taught us that flavor mixing among quarks is 
rather minimal, in striking contrast to what has emerged for neutrinos from 
recent neutrino oscillation studies. 

In the future, one would like to obtain analogous insights about the structure
of what will become the new Standard Model, and in this respect, precision EW
measurements performed at a variety of energy scales will continue to be 
needed. In what follows, we review the status of studies invoving low energy 
weak interactions of the lightest quarks and leptons and the role they are 
likely to play in the next decade. After summarizing the status of the SM and 
briefly reviewing the most widely considered scenarios for physics beyond it, 
we address various classes of low energy studies that may shed new light on 
what this physics could be. Our breakdown of these studies includes: NC 
phenomena, such as neutrino and charged lepton scattering; weak decays;
phenomena forbidden or suppressed by symmetries such as CP and lepton flavor; 
and the properties of neutrinos. We also include the anomalous magnetic moment 
of the muon, given its recent high visibility as well as its potential 
sensitivity to new physics (NP) beyond the SM. Among these topics, we 
especially highlight neutrino physics, which has brought about a revolution in
our understanding of the EW interaction in the past few years, and the searches
for permanent electric dipole moments, where advances in experimental 
techniques raise the possibility of another revolution in the next decade. 
Given the limitations of space for this review, there are also topics we have 
chosen to omit, such as the weak decays and flavor oscillations of heavy 
quarks. This choice implies no bias on our part regarding the importance of 
the latter topics, but rather reflects the particular emphasis of this article
on the lightest quarks and leptons (for recent reviews of heavy flavor physics,
see Refs.~\citer{Nir:2001ge,Isidori:2004rd}). We have endeavored to include 
the most recent experimental results, although it is possible, in the course of
compiling this review, we have overlooked developments that occurred after 
a particular section was written.

\begin{figure}[t]
\begin{center}
\begin{minipage}[t]{15.5 cm}
\rotatebox{270}{\epsfig{file=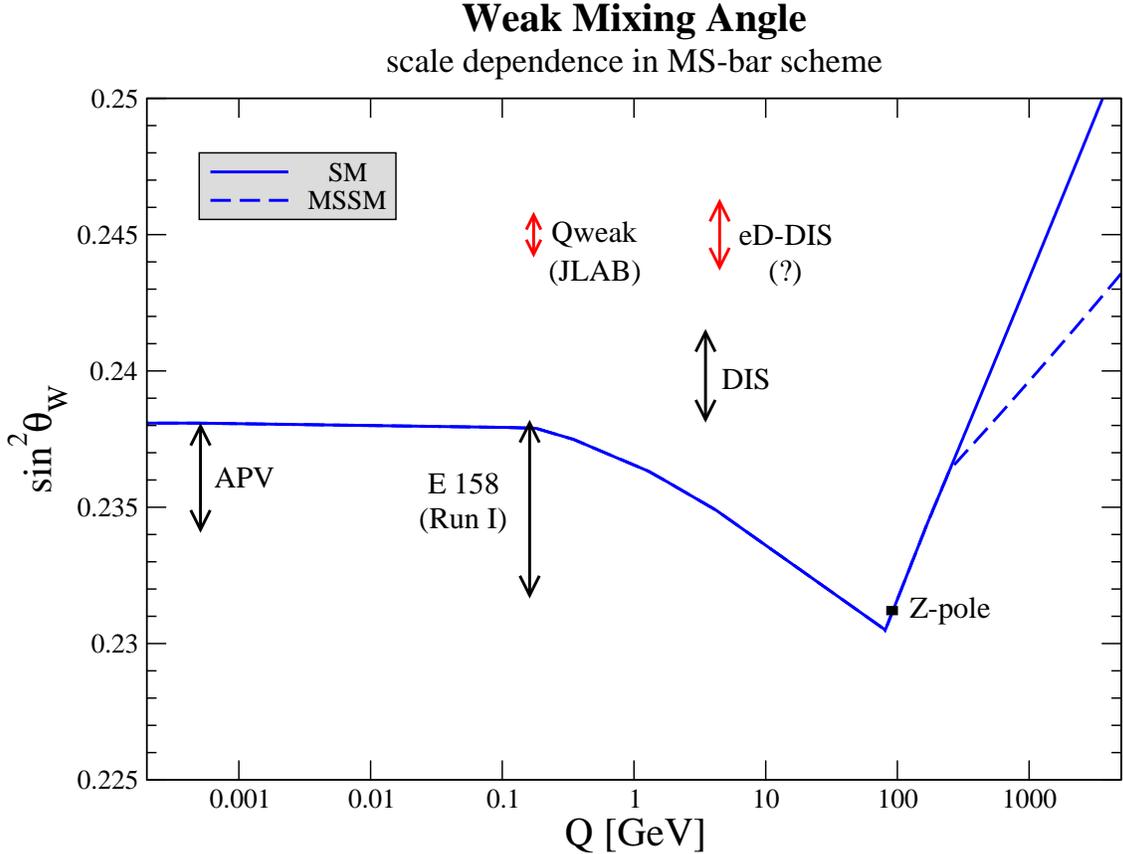,scale=0.6}}
\end{minipage}
\begin{minipage}[t]{16.5 cm}
\caption{Calculated running of the weak mixing angle in the SM, defined in
the $\overline{\rm MS}$ renormalization scheme (the dashed line indicates
the reduced slope typical for the Minimal Supersymmetric Standard Model). Shown
are the results from the $Z$ pole, deep inelastic $\nu$ scattering,
PV M\o ller scattering, and APV (Cs and Tl). Qweak and $e$-D DIS refer to a 
future PV elastic $e^-$-$p$ and a possible deep inelastic $e^-$-D measurement, 
respectively, and have arbitrarily chosen vertical locations.
\label{s2w}}
\end{minipage}
\end{center}
\end{figure}

Finally, we emphasize that within the SM itself, there remain important 
elements of both the weak and strong interaction that remain to be tested or 
more deeply understood. From the standpoint of the EW interaction, one
of the most fundamental quantities that has yet to be fully explored is 
the weak mixing angle, $\theta_W$ (defined in subsection~\ref{SM}). The scale 
dependence of gauge couplings above the weak scale can be used to predict 
the value of $\sin^2\theta_W$ at the weak scale in various grand unified 
theories (GUTs), a feature that has been used --- in conjunction with 
measurements of the weak mixing angle at the $Z$ pole --- to test or rule out 
various scenarios. On the other hand, the SM makes a definite prediction for 
the running of $\sin^2\theta_W$ below the $Z$ 
pole~\cite{Marciano:pd,Czarnecki:2000ic}. Unlike the gauge couplings of QED and
QCD, whose running below the weak scale has been stringently tested in 
a variety of ways, the low scale running of $\sin^2\theta_W$ has been tested
precisely in only a handful of experiments. This situation is summarized in 
Figure~\ref{s2w}. To date, only four types of experiments that probe 
$\sin^2\theta_W$ below the $Z$ pole with high precision have been completed or 
approved: cesium~APV~\cite{Wood:zq,Bennett:1999pd}, deep-inelastic
neutrino-nucleus scattering~\cite{Zeller:2001hh}, PV M\o ller
scattering~\cite{Anthony:2003ub}, and PV elastic electron-proton 
scattering~\cite{Armstrong:2001}. Unlike the situation with the QED and QCD
couplings, the experimental picture of the low scale running of the weak mixing
angle is not entirely settled. There exists a need for additional precision 
EW measurements at low energies that would further test 
this fundamental prediction of the SM. 

Perhaps a broader arena of SM physics that still presents a variety of 
challenges to particle and nuclear physicists is that of non-perturbative 
strong interactions. While the predictions of QCD for perturbative phenomena, 
heavy quark systems, and chiral dynamics have been tested and confirmed at 
an impressive level, deep questions remain as to the origins of confinement, 
the nature of chiral symmetry breaking, and the attendant implications for 
the quark and gluon substructure of matter. The recent flurry of activity
involving baryons with exotic quantum numbers, such as the $\Theta^+$ (see,
{\em e.g.}, Ref.~\cite{Close:2004da} and references therein), only deepens 
the sense of mystery that surrounds non-perturbative QCD. While a review of 
these issues clearly lies outside the purview of this article, we do note that
--- in the past decade --- considerable attention has been paid to the use of 
the low energy EW interaction as a tool to study certain aspects of 
non-perturbative QCD. Consequently, we also include a short synopsis of 
EW probes of the strong interaction at the end of the article.

In the next Subsection, we briefly review the basic structure of the SM, with 
the main purpose to fix our notation and conventions.

\subsection{\it The Standard Model}
\label{SM}
The Standard Model of the strong and EW interactions is based on gauge 
interactions with gauge group,
\be
   SU(3)_C \times SU(2)_L \times U(1)_Y,
\label{su321}
\ee
where the factors refer to color, isospin, and hypercharge, respectively. The 
corresponding gauge couplings are denoted by $g_s \equiv \sqrt{4\pi\alpha_s}$, 
$g$, and $g'$. The latter are often traded for the weak mixing angle, 
$\theta_W$, and the electric unit charge, $e$, which are given by,
\be
\ba{l}
   \sin^2\theta_W = 1 - \cos^2\theta_W = \frac{g'^2}{g^2 + g'^2}, \\[4pt]
   e \equiv \sqrt{4\pi\alpha} = g \sin\theta_W = g' \cos\theta_W.
\ea
\ee

Three generations of fermions are known to transform under the SM gauge 
group~(\ref{su321}) in the representations and with hypercharges as shown in 
Table~\ref{tab:smfermions} for the first generation.
\begin{table}
\begin{center}
\begin{minipage}[t]{16.5 cm}
\caption{Left-handed fermion representations of the first SM generation. 
The corresponding right-handed fermions transform in the complex conjugate 
representations.}
\label{tab:smfermions}
\vspace*{4pt}
\end{minipage}
\begin{tabular}{|cccr|cccr|}
\hline
&&&&&&&\\[-8pt]
${1\over 2}(1 - \gamma_5)\Psi$ & $SU(3)_C$ & $SU(2)_L$ & $U(1)_Y$ &
${1\over 2}(1 - \gamma_5)\Psi$ & $SU(3)_C$ & $SU(2)_L$ & $U(1)_Y$ \\[4pt]
\hline
&&&&&&&\\[-8pt]
$E = \left( \ba{c} \nu_e \\ e^- \ea \right)$ & {\bf 1} & {\bf 2} & $-1/2$ &
$\ba{c} \bar{\nu}_e  \\ e^+ \ea$ & $\ba{c} {\bf 1} \\ {\bf 1} \ea$ & 
$\ba{c} {\bf 1} \\ {\bf 1} \ea$ & $\ba{c} 0 \\ + 1 \ea$ \\[12pt]
$Q = \left( \ba{c} u \\ d \ea \right)$ & {\bf 3} & {\bf 2} & +1/6 &
$\ba{c} \bar{u} \\ \bar{d} \ea$ & $\ba{c} \bar{\bf 3} \\ \bar{\bf 3} \ea$ &
$\ba{c} {\bf 1} \\ {\bf 1} \ea$ & $\ba{c} -2/3 \\ +1/3 \ea$ \\[-8pt]
&&&&&&&\\
\hline
\end{tabular}
\end{center}
\end{table}     
Table~\ref{tab:smfermions} includes a left-handed anti-neutrino, $\bar{\nu}$,
(and correspondingly a right-handed neutrino) which transforms trivially under 
the SM gauge group, but is needed if one wishes to construct a Dirac neutrino 
mass term. Since $\bar{\nu}$ does not participate in the SM interactions it is 
often referred to as {\em sterile\/}. Notice that,
\be
   \sum\limits_{\rm quarks} Y = \sum\limits_{\rm leptons} Y = 0,
\label{mixed}
\ee
independently for quarks (lower part of Table~\ref{tab:smfermions}) and leptons
(upper part) of each generation which ensures cancellation of mixed 
$U(1)_Y$--$SU(3)_C$ and $U(1)_Y$--gravitational anomalies, respectively. 
Similarly, the pure $U(1)_Y$ and mixed $U(1)_Y$--$SU(2)_L$ gauge anomalies 
cancel by virtue of,
\be
   \sum_{\rm all} Y^3 = \sum_{\rm doublets} Y T_3^2 = 0,
\label{other}
\ee
where $T_3 = \pm 1/2$ is the third component of isospin.  Gauge invariance 
forces the gauge vector bosons to transform in the adjoint representation, \ie 
the gluons, $G_\mu^\alpha$, as $({\bf 8},{\bf 1},0)$, the isospin fields, 
$W_\mu^i$, as $({\bf 1},{\bf 3},0)$, and the Abelian hypercharge field, 
$B_\mu$, as $({\bf 1},{\bf 1},0)$. 

In addition, the SM contains a scalar Higgs field, $\phi$, transforming in 
the $({\bf 1},{\bf 2}, +1/2)$ representation, and with potential,
\be
  V(\phi) = m_\phi^2\phi^\dagger\phi + {\lambda^2\over 2} (\phi^\dagger\phi)^2.
\label{Higgs}
\ee
Assuming $m_\phi^2 < 0$, $\phi$ develops a vacuum expectation value (VEV) 
related to the Fermi constant, $G_F$,
\be
   |<0|\phi|0>| = {\sqrt{ - m_\phi^2}\over \lambda} \equiv {v\over\sqrt{2}}
                = \sqrt{{1\over 2\sqrt{2} G_F}} = 174.104 \pm 0.001\mbox{ GeV},
\label{VHiggs}
\ee
and $V(\phi)$ spontaneously breaks part of the gauge symmetry,
\be
   SU(2)_L \times U(1)_Y \rightarrow U(1)_{\rm EM}.
\label{smbreaking}
\ee
The one remaining physical Higgs degree of freedom, $H = (0,\phi^0/\sqrt{2})$, 
acquires a mass given by $M_H = \lambda v$. The others become the longitudinal
components of the EW gauge bosons,
\be
   W^\pm_\mu = {1\over \sqrt{2}} (W^1_\mu \mp i W^2_\mu), \hspace{50pt}
   Z_\mu^0   = \cos\theta_W W^3_\mu - \sin\theta_W B_\mu,
\label{gaugebosons}
\ee
which are the mediators of the weak charged and neutral currents with masses,
$M_W = T_3 (\phi) g v = g v/2 = M_Z \cos\theta_W$. The photon, 
\be 
   A_\mu = \cos\theta_W B_\mu + \sin\theta_W W^3_\mu, 
\ee
remains massless and couples to the electromagnetic charge as given by 
the Gell-Mann-Nishijima relation, $Q = T_3 + Y$.

Quarks and charged leptons receive masses through Yukawa interactions, 
exemplified here for the first generation,
\be
   {\cal L}_Y = \lambda_e \bar{E}_L \phi e^-_R + \lambda_d \bar{Q}_L \phi d_R +
   \lambda_u \bar{Q}_L (i \tau_2 \phi^*) u_R + {\rm h.c.},
\label{yukawa}
\ee
where $\tau_2$ is a Pauli matrix.  For example, the electron mass is given by,
\be
   m_e = \lambda_e {v\over \sqrt{2}}.
\label{fermionmass}
\ee
After symmetry breaking, factors of $1/\sqrt{2}$ also appear in the effective 
$Hf\bar{f}$-Yukawa couplings, and could be entirely removed by a redefinition 
of the $\lambda_f$. Likewise, one could remove the extra factor of three 
multiplying the $-i \lambda^2 H^4$ self-coupling. However, with the conventions
in Eqs.~(\ref{Higgs}) and~(\ref{yukawa}), $\lambda$ can be directly compared to
the gauge couplings (for example, in the context of supersymmetry discussed in
Section~\ref{beyond}) and to $\lambda_f$ (for example, in the context of 
superstring theories and other types of NP addressing gauge-Yukawa coupling 
unification). With these definitions and conventions the fermion Lagrangian 
takes the form,
$$
   {\cal L}_F = \sum \psibar_f \left( i \not\hbox{\hglue -2pt}\partial - m_f
              - {\lambda_f\over\sqrt{2}} \right) \psi_f
              - \frac{g}{2\sqrt 2} \sum \psibar_f \gamma^\mu (1 - \gamma_5) 
                (T^+\; W_\mu^+ + T^-\; W_\mu^- ) \psi_f
$$
\be
              - e \sum Q_f \psibar_f \gamma^\mu \psi_f A_\mu 
              - \frac{\sqrt{g^2 + g'^2}}{2} \sum \psibar \gamma^\mu
                (v_f - a_f \gamma_5) \psi Z_\mu^0,
\label{Lagrangian} 
\ee
where $T^\pm$ are the raising and lowering operators of isospin, and where
the vector and axial vector $Z$ couplings are given by,
\be
   v_f \equiv T_3^f - 2 Q_f \sin^2 \theta_W, \hspace{50pt}
   a_f \equiv T_3^f.
\label{zcouplings}
\ee
The fact that $\sin^2\theta_W \approx 1/4$ has two favorable experimental 
consequences for observables that are proportional to $v_\ell$, because for 
charged leptons ($\ell = e,\mu,\tau$) $v_\ell \ll 1$. They possess an enhanced 
sensitivity to $\sin^2 \theta_W$ and, in particular at low energies, possible 
contributions from NP are relatively enhanced. 

In the three-generation SM, the couplings $\lambda_f$ become matrix valued.
By means of bi-unitary transformations, $U^f_L \lambda_f (U^f_R)^\dagger$, one
passes to the mass eigenstates, producing flavor changing charged currents in 
the left-handed quark sector through the unitary Cabibbo-Kobayashi-Maskawa 
matrix~\cite{Cabibbo:yz,Kobayashi:fv}, 
\be
   V_{\rm CKM} = U_L^u (U_L^d)^\dagger.
\label{vckm}
\ee
The magnitudes of four of the elements of $V_{\rm CKM}$ are sufficient 
to determine the whole matrix up to a single sign ambiguity\footnote{This
ambiguity can be resolved by studying CP violation in the kaon system.}. Thus,
over-constraining $V_{\rm CKM}$ can provide a valuable test of the SM. 
This will be addressed in Section~\ref{CKM}.

\subsection{\it Status of the Standard Model}
\label{SMstatus}
With the notable exception of the Higgs sector, the basic structure of the SM
has been well established since the top quark~\cite{Abe:1995hr}, $t$, and 
the $\tau$-neutrino~\cite{Kodama:2000mp}, $\nu_\tau$, were observed 
directly in the 1990s. Moreover, the $e^+ e^-$ colliders LEP~1 and SLC provided
very high precision experiments at the $Z$ pole testing many of the relations 
presented above at the level of quantum corrections.  LEP~2 (up to center of 
mass energies of 209~GeV) and the Tevatron at Run I (in $p\bar{p}$ collisions 
at 1.8~TeV) explored the highest energies to date. It should be stressed, 
however, that there are many aspects of the SM that are difficult to study at 
the high energy frontier, but which can be suitably addressed at much lower 
energies. This is done by exploiting aspects of the weak interaction, such as
parity violation, that cleanly separate it from the strong and electromagnetic 
ones. The purpose of this article is to review these low energy tests; here we
give a short account of the experimental situation at high energies.

The first part of Table~\ref{tab:zpole} shows the $Z$ lineshape and leptonic 
\begin{table}
\begin{center}
\begin{minipage}[t]{16.5 cm}
\caption[]{Results from precision measurements at the $Z$ pole, and the $W$ 
boson and top quark masses. Shown are the measurement values compared to the SM
prediction obtained from a global analysis of high and low energy experiments
using the FORTRAN package GAPP~\cite{Erler:1999ug}. The uncertainties in the
SM predictions reflect the uncertainties in the SM parameters, which are 
determined self-consistently, {\em i.e.}, we do not use external 
constraints.  This yields, {\em e.g.}, $\alpha_s (M_Z) = 0.121 \pm 0.002$.
The deviations from the predictions (in terms of the pull) are also shown.  
The experimental uncertainties of many of the observables are mutually 
correlated. The largest (anti)-correlations occur between $R_e$ and $A_{FB}(e)$
($-37\%$) and between $\sigma_{\rm had}$ and $\Gamma_Z$ 
($-30\%$)~\cite{Abbaneo:2002}. All experimental correlations are taken into 
account in the fits. The theoretical uncertainty from unknown higher orders in 
the prediction for the QCD correction enters commonly $\Gamma_Z$, 
$\sigma_{\rm had}$, and the leptonic ratios $R_\ell$. We estimate 
this uncertainty to affect $\alpha_s (M_Z)$ at the level of 
$5\times 10^{-4}$~\cite{Erler:1999ug}, which we currently neglect against 
the much larger experimental errors.}
\label{tab:zpole}
\vspace*{4pt}
\end{minipage}
\begin{tabular}{|ll|c|c|c|r|}
\hline
&&&&&\\[-8pt]
Quantity & & Group(s) & Value & Standard Model & Pull\\[4pt]
\hline
&&&&&\\[-8pt]
$M_Z$ & [GeV] & LEP &$ 91.1876 \pm 0.0021 $&$ 91.1874 \pm 0.0021 $&$ 0.1$\\
$\Gamma_Z$ & [GeV] & LEP &$ 2.4952 \pm 0.0023 $&$  2.4972 \pm 0.0011 $&$-0.9$\\
$\Gamma({\rm inv})$ & [MeV] & [derived quantity]  &$ 499.0\pm 1.5   $& $501.74\pm 0.15$  &---\\
$\sigma_{\rm had}$ & [nb] & LEP & $41.541\pm 0.037$ & $41.470\pm 0.010$&$1.9$\\
$R_e$    & & LEP & $20.804 \pm 0.050$ & $20.753 \pm 0.012$ & $ 1.0$\\
$R_\mu$  & & LEP & $20.785 \pm 0.033$ & $20.754 \pm 0.012$ & $ 1.0$\\
$R_\tau$ & & LEP & $20.764 \pm 0.045$ & $20.799 \pm 0.012$ & $-0.8$\\
$A_{FB} (e)$ & & LEP &   $0.0145 \pm 0.0025$ & $0.01639 \pm 0.00026$ & $-0.8$\\
$A_{FB} (\mu)$ & & LEP & $0.0169 \pm 0.0013$ &       $"$      & $ 0.4$\\
$A_{FB} (\tau)$ & & LEP& $0.0188 \pm 0.0017$ & '' & $ 1.4$\\[4pt]
\hline
&&&&&\\[-8pt]
$R_b$ & & LEP + SLD & $0.21644 \pm 0.00065$ & $0.21572 \pm 0.00015$ & $ 1.1$\\
$R_c$ & & LEP + SLD & $ 0.1718 \pm 0.0031 $ & $0.17231 \pm 0.00006$ & $-0.2$\\
$R_{s,d}/R_{(d+u+s)}$ \hspace*{-20pt} & & OPAL & $0.371\pm 0.023$   &
$0.35918\pm 0.00004$ & $0.5$\\
$A_{FB} (b)$ & & LEP & $0.0995 \pm 0.0017$ & $0.1036 \pm 0.0008$ & $-2.4$\\
$A_{FB} (c)$ & & LEP & $0.0713 \pm 0.0036$ & $0.0741 \pm 0.0007$ & $-0.8$\\
$A_{FB} (s)$ & & DELPHI + OPAL & $0.0976\pm 0.0114$&$0.1038\pm 0.0008$&$-0.5$\\
$A_b$ & & SLD & $0.922 \pm 0.020$ & $0.93477 \pm 0.00012$ & $-0.6$\\
$A_c$ & & SLD & $0.670 \pm 0.026$ & $0.6681  \pm 0.0005 $ & $ 0.1$\\
$A_s$ & & SLD & $0.895 \pm 0.091$ & $0.93571 \pm 0.00010$ & $-0.4$\\[4pt]
\hline
&&&&&\\[-8pt]
$A_{LR}$(hadrons) \hspace*{-20pt} & & SLD & $0.15138\pm 0.00216$ & 
$0.1478\pm 0.0012$&$ 1.6$\\
$A_{LR}$(leptons) \hspace*{-20pt} & & SLD & $0.1544 \pm 0.0060 $ & '' &$ 1.1$\\
$A_\mu$            & & SLD & $0.142  \pm 0.015  $ &          ''       &$-0.4$\\
$A_\tau$           & & SLD & $0.136  \pm 0.015  $ &          ''       &$-0.8$\\
$A_e (Q_{LR})$     & & SLD & $0.162  \pm 0.043  $ &          ''       &$ 0.3$\\
$A_\tau ({\cal P}_\tau)$ & & LEP & $0.1439 \pm 0.0043$ &     ''       &$-0.9$\\
$A_e ({\cal P}_\tau)$    & & LEP & $0.1498 \pm 0.0049$ &     ''       &$ 0.4$\\
$Q_{FB}$ & & LEP & $0.0403 \pm 0.0026$ & $0.0424 \pm 0.0003 $ & $-0.8$\\[4pt]
\hline
&&&&&\\[-8pt]
$m_t$ & [GeV] & Tevatron &  $174.3 \pm 5.1$   &  $174.4 \pm 4.4$   & $0.0$\\
$M_W$ & [GeV] &    LEP   & $80.447 \pm 0.042$ & $80.391 \pm 0.019$ & $1.3$\\
$M_W$ & [GeV] & Tevatron + UA2 & $80.454 \pm 0.059$ &     ''      &$ 1.1$\\[-8pt]
&&&&&\\
\hline
\end{tabular}
\end{center}
\end{table}     
forward-backward (FB) cross section asymmetry measurements from LEP~1.  Besides
$M_Z$, they include the total $Z$ decay width, $\Gamma_Z$, the hadronic cross 
section on top of the $Z$ resonance, $\sigma_{\rm had}$, and for each lepton 
flavor the ratio of hadronic to leptonic partial $Z$ widths, $R_\ell$, and 
the FB-asymmetry,
\be
   A_{FB}(\ell) = {3\over 4} A_e A_\ell,
\label{afbell}
\ee
defined in terms of the asymmetry parameters,
\be
   A_f = {2 v_f a_f \over v^2_f + a^2_f} = 
         \frac{1 - 8 T_3^f Q_f \sin^2\theta^{\rm eff}_f}
 {1 - 8 T_3^f Q_f \sin^2\theta^{\rm eff}_f + 8 Q_f^2 \sin^4\theta^{\rm eff}_f}.
\label{af}
\ee
Here, $\sin^2\theta^{\rm eff}_f = \kappa_f \sin^2\theta_W$ is an effective 
mixing angle where (flavor dependent) radiative corrections are absorbed into 
form factors, $\kappa_f$. The $\kappa_f$ are renormalization-scheme dependent, 
while the $\sin^2\theta^{\rm eff}_f$ are not.  The results in the first part of
the Table are obtained simultaneously from a fit to cross section data, and
therefore mutually correlated.  The invisible $Z$ partial width, 
$\Gamma({\rm inv})$, is derived from $\Gamma_Z$, $\sigma_{\rm had}$, and 
the $R_\ell$, and not an independent input.  It is smaller by about two 
standard deviations than the prediction for the SM with three {\em active\/} 
(participating in the weak interaction) neutrinos. This deviation can be traced
to $\sigma_{\rm had}$ which deviates by a similar amount.  Conversely, one can 
allow the number of active neutrinos, $N_\nu$, as an additional free parameter 
in the fit.  This yields $N_\nu = 2.986 \pm 0.007$ again showing a 2~$\sigma$ 
deviation. The experimental results discussed in this paragraph are 
final~\cite{Abbaneo:2002}.

The second part of Table~\ref{tab:zpole} shows the results from $Z$ decays into
heavy flavors~\cite{Abbaneo:2002} ($b$ and $c$ quarks), as well as into hadrons
with non-vanishing strangeness~\citer{Ackerstaff:1997rc,Abe:2000uc}, but 
the latter with much poorer precision. For each of these three quark flavors 
($q = b,c,s$) the partial $Z$ width normalized to the hadronic partial width, 
$R_q$, has been obtained, as well as the FB-asymmetry, $A_{FB}(q)$, in analogy
to Eq.~(\ref{afbell}), and the combined left-right (LR) forward-backward 
asymmetry,
\be
   A^{FB}_{LR}(q) = {3\over 4} A_q.
\label{afblr}
\ee
The heavy flavor results are obtained from a multi-parameter fit including
a variety of phenomenological parameters and experimental information from both
LEP and the SLC~\cite{Abbaneo:2002}.  The results for $q = b$ and $c$ are 
therefore mutually correlated.  $A_{FB}(b)$ is proportional to both $A_e$ and 
$A_b$, but since $v_e \ll 1$ (see Section~\ref{SM}) it is primarily sensitive
to $A_e$.  Similar statements are true of other FB-asymmetries, as well, but
$A_{FB}(b)$ stands out because $b$ quark tagging while challenging is easier
than tagging of other quarks, and quarks have both bigger cross sections and
bigger asymmetries implying smaller statistical uncertainties as compared to
leptons.  Therefore, $A_{FB}(b)$ provides one of the best determinations
of the weak mixing angle.  It shows a 2.4~$\sigma$ deviation, and (through 
one-loop radiative corrections) by itself favors larger values for $M_H$. It is
tempting to suggest effects of NP in the factor $A_b$ appearing in $A_{FB}(b)$ 
to reconcile this deviation and the disagreement with $A_{LR}$ discussed in
the following paragraph. However, one would need a $(19 \pm 7)$\% radiative 
correction to $\kappa_b$ while typical EW radiative corrections are at the sub
percent level. NP entering at tree level is generally not 
resonating\footnote{A counter example is an extra neutral gauge boson mixing
with the ordinary $Z$.} and therefore suppressed relative to the $Z$ 
contribution.  In any case, $R_b$ is in reasonable agreement with the SM and 
one needs to find NP modifying $A_b$, but not $R_b$, which in general requires
some tuning of parameters.  Most of the experimental results discussed in this 
paragraph are still preliminary but close to final~\cite{Abbaneo:2002}.

The third part of Table~\ref{tab:zpole} shows further measurements proportional
to $v_\ell$.  The LR cross section asymmetry, $A_{LR} = A_e$, from the SLD 
Collaboration for hadronic~\cite{Abe:2000dq} and leptonic final 
states~\cite{Abe:2000hk} show a combined deviation of 1.9~$\sigma$ from the SM 
prediction. In contrast to $A_{FB}(b)$, it favors small values for $M_H$, which
are excluded by the direct searches at LEP~2~\cite{Holzner:2002ft},
\be
    M_H \geq 114.4 \mbox{ GeV } (95\%~\mbox{CL}).
\label{directmh}
\ee
Through LR-FB asymmetries for $\mu$ and $\tau$ final states SLD determines 
$A_\mu$ and $A_\tau$~\cite{Abe:2000hk} (cf.\ Eq.~(\ref{afblr})) while $A_e$ 
from polarized Bhabba scattering is included in $A_{LR}({\rm leptons})$.  
The $\tau$ polarization asymmetry, ${\cal P}_\tau$, is a measurement of 
$A_\tau$, and its angular dependence, ${\cal P}_\tau^{\rm FB}$, yields another 
determination of $A_e$~\cite{Abbaneo:2002}. Finally, the hadronic charge 
asymmetry at LEP~\cite{Abbaneo:2002} is the weighted sum\footnote{The LEP 
groups quote results on $Q_{FB}$ as measurements of $\sin^2\theta^{\rm eff}_e$
setting the $R_q$ and $A_q$ to their SM values.},
\be
   Q_{FB} = (\sum\limits_{q = d,s,b} - \sum\limits_{q = u,c}) R_q A_{FB} (q),
\ee
while the LR charge asymmetry from SLD~\cite{Abe:1996ef} offers a further (less
precise) value of $A_e$.  The experimental results discussed in this paragraph 
are final~\cite{Abbaneo:2002}.
 
\begin{figure}[t]
\begin{center}
\begin{minipage}[t]{11 cm}
\epsfig{file=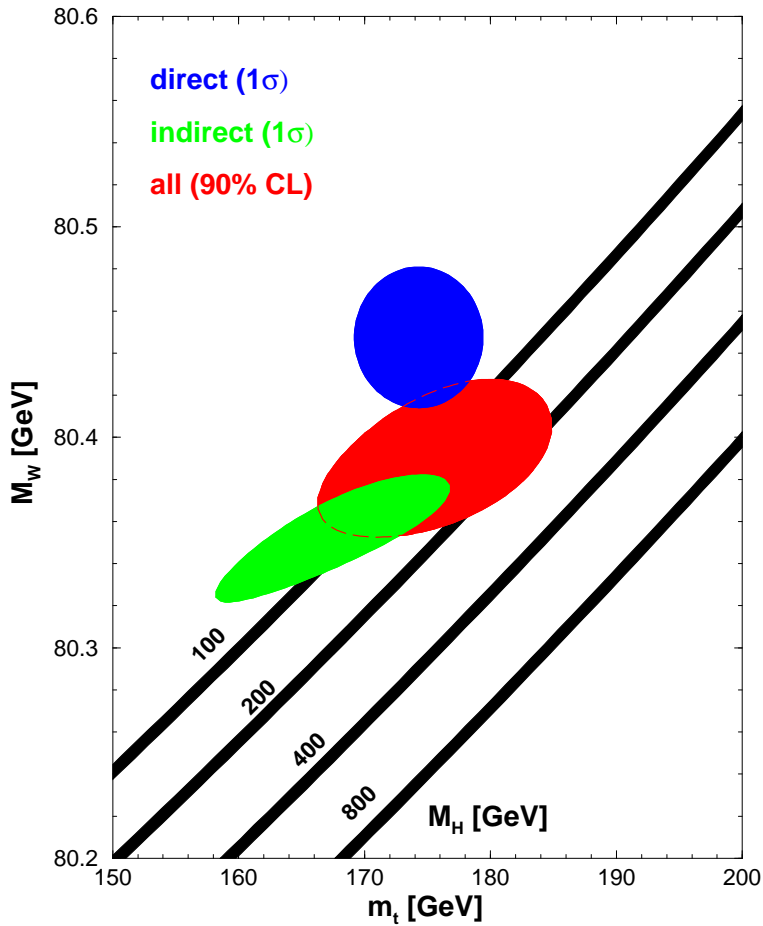,scale=0.6}
\end{minipage}
\begin{minipage}[t]{16.5 cm}
\caption{One-standard-deviation (39.35\%~CL) regions in the $M_W$-$m_t$ plane
for the direct and indirect data.  The combined 90\%~CL contour 
($\Delta\chi^2 = 4.605$) is also shown.  The widths of the $M_H$ bands 
represent the theoretical uncertainty in the SM prediction 
($\alpha_s (M_Z) = 0.120$).
\label{mwmt}}
\end{minipage}
\end{center}
\end{figure}

The last part of Table~\ref{tab:zpole} shows the direct $m_t$ measurement from 
the Tevatron~\cite{Abbott:1998dc,Abe:1998bf}, as well as $M_W$ from
LEP~2~\cite{Abbaneo:2002} and $p\bar{p}$ 
collisions~\citer{Alitti:1991dk,Affolder:2000bp}. The combined $M_W$ is 
1.7~$\sigma$ higher than the SM expectation. Just as $A_{LR}$ it favors smaller
values of $M_H$. We compare these mass measurements with all other (indirect) 
data, and the SM prediction for various values of $M_H$ in Figure~\ref{mwmt}. 
The definition of $m_t$, shown in 
Table~\ref{tab:zpole} and Figure~\ref{mwmt} is the pole mass, which is 
(approximately) the {\rm kinematic\/} mass studied at 
the Tevatron~\cite{Smith:1996xz}. For all other quark masses, as well as $m_t$ 
appearing in radiative corrections, the \ms-definition is used.  CDF and D\O\ 
finalized their results on $m_t$ and $M_W$ from Run~I. $M_W$ from LEP~2 is 
preliminary and final state interaction effects are still under study. Not 
shown in Table~\ref{tab:zpole} are the bottom and charm quark masses, $m_b$ and
$m_c$, which enter the SM predictions of numerous observables.  In particular,
$m_c$ is needed to compute $\alpha(M_Z)$ when the renormalization group (RG)
based calculation of Ref.~\cite{Erler:1998sy} is used. Rather than including 
external constraints, $m_b$ and $m_c$ are constrained using a set of inclusive
QCD sum rules~\cite{Erler:2002bu} and are recalculated in each call within 
the fits as functions of $\alpha_s$.

The deviations described above may be due to unaccounted for experimental or
theoretical effects, physics beyond the SM, or simply fluctuations.
New experimental information is needed to clear the situation. Despite of 
these open problems it must be stressed that the overall agreement 
between the data and the SM is reasonable. The $\chi^2$ per effective degree of
freedom of the global fit is 49.1/40 where the probability for a larger 
$\chi^2$ is 15\%. One can thus conclude that the SM has been successfully 
tested as the correct theory up to energy scales of ${\cal O} (M_Z)$ both 
at the tree level and at the level of loop corrections. At low 
energies, NP can therefore only enter as a small perturbation of the SM 
contribution.

\begin{figure}[t]
\begin{center}
\begin{minipage}[t]{11 cm}
\epsfig{file=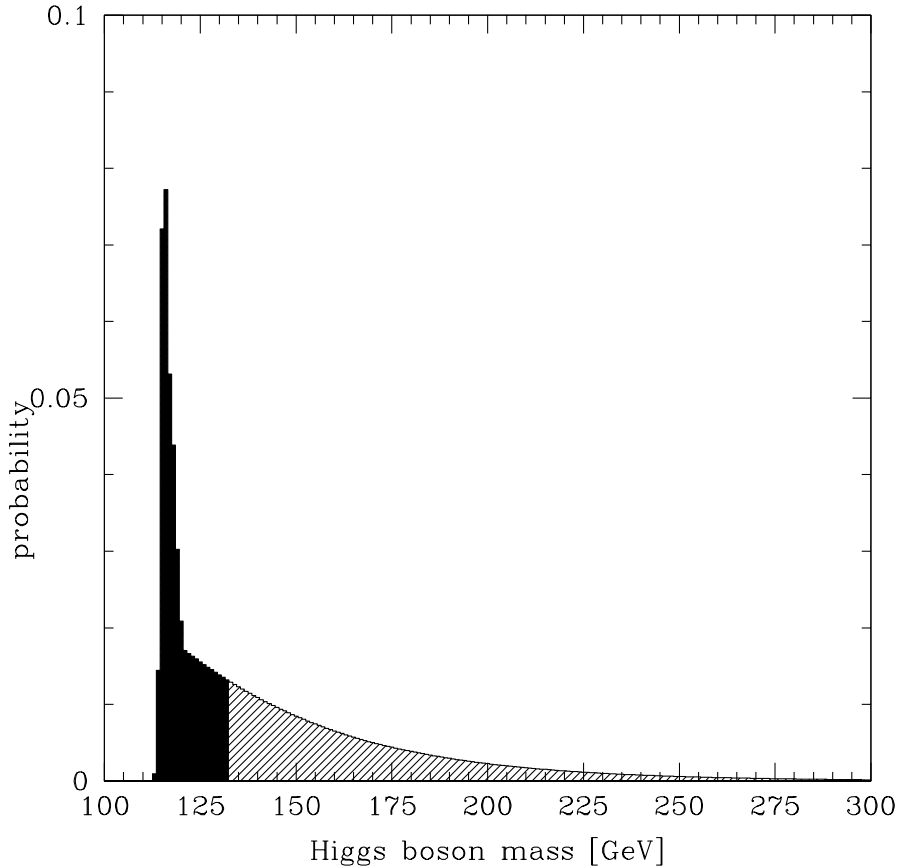,scale=0.6}
\end{minipage}
\begin{minipage}[t]{16.5 cm}
\caption{Probability density for $M_H$ obtained by combining precision data
with direct search results at LEP~2.  The peak is due to the candidate Higgs
events observed at LEP (updated from Ref.~\cite{Erler:2000cr}). The full and
shaded areas each contain 50\% probability.
\label{higgspdf}}
\end{minipage}
\end{center}
\end{figure}

The global fit to all precision data currently favors values for the Higgs
boson mass,
\be
   M_H = 85^{+49}_{-32}\mbox{ GeV},
\ee
where the central value is slightly below the lower LEP~2 limit~(\ref{directmh}).  
If one includes the Higgs search information from LEP~2,
one obtains the probability density shown in Figure~\ref{higgspdf}.  

{\em Note added:} After the completion of this Section, a reanalysis of $m_t$ 
from the lepton + jets channel by the D\O\ Collaboration~\cite{Abazov:2004cs}
and the first results from the Tevatron Run II became available.
The new average, $m_t = 177.9 \pm 4.4$~GeV~\cite{Erler:2004nh}, and some new
results on $M_W$ are driving the extracted $M_H$ to values very close
to the direct limit~(\ref{directmh}).

\subsection{\it Physics Beyond the Standard Model}
\label{beyond}
Despite the phenomenological success of the SM, it is almost certainly only
the low energy approximation of a more fundamental theory.  One reason is its
arbitrariness, since the gauge group~(\ref{su321}) and the fermion (and Higgs)
representations in Table~\ref{tab:smfermions} are {\em ad hoc}, as are 
the values of the gauge and Yukawa couplings, along with the parameters 
entering $V_{\rm CKM}$ and $V(\phi)$. A second reason is that gravity is not 
included in quantized form and it is unlikely that classical General Relativity
can coexist (at a fundamental level) with quantum theories of the other 
interactions. The third class of arguments can be phrased as 
{\em fine-tuning\/} or {\em naturalness\/} problems, of which there are several
within the SM.  

By far the most serious one is associated with the cosmological constant, 
$\Lambda_{\rm C}$, which on general grounds should be of order the reduced 
Planck scale, $\kappa_P = 2.4 \times 10^{18}$~GeV, (the only fundamental mass 
scale) raised to the forth power. In addition, one expects contributions 
associated with the EW and QCD phase transition scales, while on the other 
hand, cosmological observations~\cite{Primack:2002th} suggest,
$\Lambda_{\rm C} = (2.3 \pm 0.2 \mbox{ meV})^4$.  No convincing mechanism is
known to explain this small observed value.

Another naturalness problem is related to the topological CP-odd $\theta$-term 
of QCD~\cite{Wilczek:pj}, which is power counting renormalizable and indeed
logarithmically divergent.  In the SM this divergence is induced by the complex
phase appearing in $V_{\rm CKM}$.  Residually, $\theta \lsim 10^{-10}$, 
to account for the limit on the electric dipole moment of $^{199}$Hg (see 
Section~\ref{CPandT}). There is an analogous term for $SU(2)_L$, and also for
the gravitational interaction~\cite{'tHooft:bx}.

Third, radiative corrections to the $m_\phi^2$ term in Eq.~(\ref{Higgs}) are
quadratically divergent, which strongly suggests one of three logical 
possibilities: (i) $m_\phi^2$ is of order $\kappa_P$ which is excluded both on
formal and phenomenological grounds; (ii) various independent Planck scale 
contributions to $m_\phi^2$ conspire to give a residual result not much larger
than the weak scale, implying fine tuning at the level of 1 part in $10^{16}$;
(iii) physics beyond the SM exists with manifestations at scales not much 
larger than the weak scale, so as to serve as a regulator for 
the quadratically divergent loop integrals.  Examples for the latter include 
dynamical symmetry breaking with a composite rather than a fundamental Higgs, 
supersymmetry (SUSY), and large extra dimensions.  

For all these reasons there is generally a strong believe that there is NP
beyond the SM. Moreover, the stability of the Higgs potential against large
radiative corrections as discussed in the previous paragraph suggests that at 
least some manifestations of the NP should take place at scales not much larger
than $v$. 

In this review, we will frequently refer to SUSY~\cite{Wess:1973kz} as 
an illustrative example for physics beyond the SM. It solves the problem of 
quadratically divergent loop corrections~\cite{Dimopoulos:1981zb,Sakai:1981gr} 
through the introduction of superpartners of the SM particles, which precisely 
cancel the SM contributions. In fact, in the case of exact (unbroken) SUSY
the superpotential which gives rise to Yukawa 
interactions\footnote{The superpotential versions of both ${\cal L}_Y$ and 
${\cal L}_Y^{\rm SUSY}$ discussed later also contribute to the scalar 
potential, but this will not affect the discussion in this review.} is 
unmodified by radiative corrections by virtue of a non-renormalization 
theorem~\cite{Grisaru:1979wc,Fischler:1981zk}. In simple SUSY\footnote{Theories
of extended four-dimensional SUSY do not allow chiral fermions and are 
therefore excluded as effective theories in four dimensions~\cite{Witten:nf}.}
the gauge interactions are logarithmically divergent as in the SM. Thus, gauge
couplings still evolve according to the RG, but unlike in the SM, in the MSSM 
they approximately converge to a common value at a scale
$M_U \approx 2 \times 10^{16}$~GeV~\citer{Dimopoulos:1981yj,Amaldi:1987fu}.

As for the SUSY Higgs sector, the $\lambda_u$-term in Eq.~(\ref{yukawa}) can 
not be supersymmetrized (complex conjugation is not a supersymmetric 
operation), and therefore a second Higgs doublet, $H_u$, with opposite 
hypercharge to the standard Higgs, $H_d \equiv \phi$, needs to be introduced 
for up-type quark mass terms.  This is also necessary to cancel the gauge 
anomaly contribution of the Higgs-fermions (Higgsinos). As a consequence,
there are now five physical Higgs degrees of freedom: the two CP-even, 
$H_{1,2}$, with $M_{H_1} \leq M_{H_2}$ by convention, the CP-odd, $A$, and 
the charged pair, $H^\pm$. 

In the minimal supersymmetric SM (MSSM), the $\lambda^2$-term in 
Eq.~(\ref{Higgs}) arises from the supersymmetric gauge kinetic terms, leading 
to the prediction,
\be
   \lambda = T_3 (\phi) \sqrt{g^2 + {g^\prime}^2}.
\ee
The mass term in Eq.~(\ref{Higgs}) is replaced by a bilinear term, 
$\mu H_u H_d$, in the superpotential.  Again, $\mu \ll \kappa_P$ is assumed
rather than predicted, but now this hierarchy is {\em technically natural\/},
{\em i.e.}, stable under radiative corrections. Since $\lambda$ is no longer
a free parameter, the Higgs boson masses are calculable from other inputs, and
one obtains the tree level upper bound $M_{H_1} \leq M_Z$. Radiative 
corrections~\citer{Carena:1995wu,Brignole:2001jy} move this bound 
to~\cite{Degrassi:2002fi},
\be
   M_{H_1} \lsim 134 \pm 5 \pm 3^{+2}_{-0} \mbox{ GeV},
\ee
where the first error is induced by $m_t$, the second is the theory uncertainty
in the prediction, and the last one corresponds to shifting the SUSY breaking 
scale (see next paragraph) from 1 to 2~TeV.

The non-observation of the superpartners implies that SUSY is broken.  
The mechanism of SUSY breaking is one of the outstanding theoretical problems, 
and is currently treated phenomenologically by allowing the most general 
set~\cite{Girardello:1981wz} of soft SUSY breaking parameters: bilinear scalar 
masses, $m_{ij}^2$, gauge fermion (gaugino) masses, $M_i$, and trilinear scalar
couplings, $A_{ijk}$. The quantity $\mu$ and various of the soft SUSY breaking
parameters are assumed to trigger EW symmetry breaking, but within the MSSM it
is not understood why $\mu$ is of the order the soft masses~\cite{Kim:1983dt} 
(the so-called $\mu$-problem). The VEV, $v$, now receives contributions from 
both Higgs doublets, $v^2 = v_1^2+v_2^2$, with $v_1/v_2 \equiv \tan\beta$. 

With the gauge and particle choice of the SM, baryon number, $B$, and lepton
number, $L$, are conserved, ensuring a stable proton lifetime. In the MSSM,
however, additional Yukawa couplings to the ones in Eq.~(\ref{yukawa}) are 
allowed by the symmetries of the model,
\be
   {\cal L}_Y^{\rm SUSY} = 
      \lambda \bar{E}_L E_L e^-_R + \lambda^\prime \bar{Q}_L E_L d_R + 
      \lambda^{\prime\prime} \bar{u}_R \bar{d}_R \bar{d}_R +
      \mu^\prime \bar{E}_L H_u,
\label{yukawasusy}
\ee
where any one factor in the trilinear couplings is bosonic, while the other two
are fermionic. The first and the third term in Eq.~(\ref{yukawasusy}) are
antisymmetric in the $SU(2)$ and $SU(3)$ indices, respectively, so that they 
need to be antisymmetrized with respect to family space, as well.  The third 
term violates $B$, the others violate $L$. One can eliminate all terms in 
Eq.~(\ref{yukawasusy}) by a so-called $R$-symmetry\footnote{To this end, it is
sufficient to employ the discrete $R$-parity subgroup, 
$R_P = (-1)^{3 B + L + 2 S}$, where $S$ refers to
spin.}~\cite{Salam:1974xa,Fayet:1974pd}, which is consistent with SUSY although
fermions transform differently under it than their scalar partners. If $R_P$ is
exact, the lightest supersymmetric particle would be stable and a cold dark 
matter candidate. If $R_P$ is broken, $B$ or $L$ or both would be violated. For
more details on SUSY and its phenomenology, see 
the reviews~\cite{Fayet:1976cr,Sohnius:qm} 
and~\cite{Nilles:1983ge,Haber:1984rc}, respectively. 

Another possible type of NP involves additional gauge interactions, in 
particular extra Abelian group factors. An extra $U(1)^\prime$ by itself does 
not solve any of the problems associated with the SM. But additional 
$U(1)^\prime$ symmetries are predicted in many extensions of the SM, including 
technicolor~\cite{Farhi:1980xs}, Grand Unified 
Theories~\cite{Langacker:1980js}, SUSY, Kaluza-Klein theories~\cite{Duff:hr},
or string theories~\cite{Green:sp,Hewett:1988xc}. Moreover, they can solve 
the $\mu$-problem of the MSSM~\cite{Cvetic:1995rj}, and forbid the terms in 
Eq.~(\ref{yukawasusy}). They generally do not spoil the successful gauge 
coupling unification, and $U(1)^\prime$ symmetries can be found addressing all
these issues simultaneously, while at the same time being free of 
anomalies~\cite{Erler:2000wu}. The precision $Z$ pole observables discussed in
Section~\ref{SMstatus} mainly constrain the mixing of the corresponding 
$Z^\prime$ with the ordinary $Z$. Low energy observables, especially from APV
as discussed in Section~\ref{APV}, provide important mass and coupling 
constraints on potential $Z^\prime$s. We now turn to a detailed discussion of 
these observables.

\section{Weak Neutral Current Phenomena}
\subsection{\it Neutrino Scattering}
\label{nun}
Assuming an arbitrary gauge theory, one can write the four-Fermi interaction 
Lagrangian for the weak processes discussed in this and in the next Section in 
model independent form. For NC neutrino scattering one has,
\be
  {\cal L}_{\rm NC}^{\nu f} = - \frac{G_F}{\sqrt 2} 
  \bar{\nu} \gamma^\mu (1 - \gamma_5) \nu \bar{f} \gamma_\mu 
  \left[ \epsilon_L(f) (1 - \gamma_5) + \epsilon_R(f) (1 + \gamma_5) \right] f,
\label{nuf}
\ee
where the SM tree level prediction for the effective couplings can be obtained 
from Eqs.~(\ref{Lagrangian}) and (\ref{zcouplings}),
\be
   \epsilon_L (f) = T_3^f - Q_f \sin^2\theta_W, \hspace{50pt}
   \epsilon_R (f) =       - Q_f \sin^2\theta_W. 
\label{epsilons}
\ee
EW radiative corrections~\cite{Marciano:pd,Bellucci:1981bs,Bardin:1980fe} 
modify Eq.~(\ref{epsilons}) through the NC $\rho$-parameter, which appears when
the $Z$ propagator is expressed in terms of $G_F$, through the RG evolution of
the weak mixing angle, and through further propagator, vertex, and box 
contributions.

Historically, deep inelastic scattering (DIS) of neutrinos off nuclei, 
$\nu_\mu N \rightarrow \nu_\mu X$, with $X$ an arbitrary hadronic final state,
provided the discovery of the weak NC~\cite{Hasert:ff}. For measurements of 
$\sin^2\theta_W$, it is advantageous to choose $N$ approximately isoscalar. 
Many theoretical uncertainties cancel in the ratio of neutral-to-charged 
current cross sections~\cite{LlewellynSmith:ie},
\be
   R_\nu       = {\sigma_{\nu N}^{\rm NC}\over \sigma_{\nu N}^{\rm CC}} 
               = {\sigma(\nu_\mu N \rightarrow \nu_\mu X)\over 
                  \sigma(\nu_\mu N \rightarrow \mu^- X)}
               = g_L^2 + r g_R^2, \hspace{50pt}
   R_{\bar\nu} = {\sigma_{\bar\nu N}^{\rm NC}\over \sigma_{\bar\nu N}^{\rm CC}}
               = {\sigma(\bar\nu_\mu N \rightarrow \bar\nu_\mu X)\over 
                  \sigma(\bar\nu_\mu N \rightarrow \mu^+ X)}
               = g_L^2 + {g_R^2\over \bar{r}},
\ee
where at tree level and ignoring sea quarks,
\be
   g_L^2 = \epsilon_L(u)^2 + \epsilon_L(d)^2 = 
           {1\over 2} - \sin^2\theta_W + {5\over9} \sin^4\theta_W,\hspace{50pt}
   g_R^2 = \epsilon_R(u)^2 + \epsilon_R(d)^2 = {5\over9} \sin^4\theta_W.
\ee
$r$ and $\bar{r}$ are related to the charged current (CC) ratio, 
\be
   r \sim {1\over \bar{r}} \sim 
   {\sigma_{\bar\nu N}^{\rm CC}\over \sigma_{\nu N}^{\rm CC}} \sim {1\over 2},
\ee
which depends on the experimental details and is measured directly. $R_\nu$ is 
more sensitive to the weak mixing angle than $R_{\bar\nu}$, and both are 
sensitive to charm threshold effects which introduce the dominant theoretical 
uncertainty. This uncertainty (and some others) largely cancels in 
the ratio~\cite{Paschos:1972kj},
\be
   R^- = \frac{\sigma_{\nu N}^{\rm NC} - \sigma_{\bar\nu N}^{\rm NC}}
              {\sigma_{\nu N}^{\rm CC} - \sigma_{\bar\nu N}^{\rm CC}}
       = \frac{R_\nu - r R_{\bar\nu}}{1 - r} 
       = g_L^2 - g_R^2 
       = {1\over 2} - \sin^2\theta_W, 
\ee
which was used by the NuTeV Collaboration~\cite{Zeller:2001hh} to measure 
the weak mixing angle precisely off the $Z$ pole. NuTeV was the first
experiment of this type with a clean $\bar\nu_\mu$-beam at its disposal which
is necessary to measure $R^-$. In the presence of NP, however, which will in 
general affect $\nu_\mu$ and $\bar\nu_\mu$ cross sections differently, one 
should rather monitor $R_\nu$ and $R_{\bar\nu}$ independently, or equivalently,
$g_L^2$ and $g_R^2$. The uncertainty from the charm threshold is modeled using
the slow rescaling prescription~\cite{Barnett:1976kh,Georgi:1976ve} 
parametrized using an effective charm mass, $m_c^{\rm eff}$. In practice, NuTeV
fits to $g_L^2$, $g_R^2$, and $m_c^{\rm eff}$ with an external constraint. As 
is shown in Table~\ref{tab:dis1}, there is a 2.9~$\sigma$ deviation in $g_L^2$,
reflecting 
\begin{table}
\begin{center}
\begin{minipage}[t]{16.5 cm}
\caption[]{Results from deep inelastic neutrino-nucleus scattering. Shown are 
the measurement values, the SM prediction and the pull as in 
Table~\ref{tab:zpole}. The second errors (where shown) are theoretical. 
The experimental results are updates of the original CDHS~\cite{Blondel:1989ev}
and CHARM~\cite{Allaby:1987vr} publications for which we used the analysis of 
Ref.~\cite{Perrier:qg}. The coefficients entering the linear 
combinations~(\ref{lc}) of the various measurements are given in 
Table~\ref{tab:dis2}.}
\label{tab:dis1}
\vspace*{4pt}
\end{minipage}
\begin{tabular}{|l|c|c|c|r|}
\hline
&&&&\\[-8pt]
Quantity & Group & Value & Standard Model & Pull \\[4pt]
\hline
&&&&\\[-8pt]
$g_L^2$      & NuTeV  & $0.30005\pm 0.00137$ & $0.30397\pm 0.00023$ &  $-2.9$\\
$g_R^2$      & NuTeV  & $0.03076\pm 0.00110$ & $0.03005\pm 0.00004$ &  $ 0.6$\\
$R^\nu$      &CCFR    &$0.5820\pm 0.0027\pm 0.0031$&$0.5833\pm 0.0004$&$-0.3$\\
$R^\nu$      &CDHS    &$0.3096\pm 0.0033\pm 0.0028$&$0.3092\pm 0.0002$&$ 0.1$\\
$R^\nu$      &CHARM   &$0.3021\pm 0.0031\pm 0.0026$&                  &$-1.7$\\
$R^{\bar\nu}$&CDHS       &$0.384\pm 0.016\pm 0.007$&$0.3862\pm 0.0002$&$-0.1$\\
$R^{\bar\nu}$&CHARM      &$0.403\pm 0.014\pm 0.007$&                  &$ 1.0$\\
$R^{\bar\nu}$&CDHS (1979)&$0.365\pm 0.015\pm 0.007$&$0.3817\pm 0.0002$&$-1.0$\\[-8pt]
&&&&\\
\hline
\end{tabular}
\end{center}
\end{table}     
the quoted 3~$\sigma$ deviation~\cite{Zeller:2001hh} in the weak mixing angle. 

There are various attempts to explain the apparent deviation of the NuTeV 
result from the SM prediction. It cannot be excluded that nuclear shadowing
effects are large enough to induce shifts comparable to the NuTeV 
discrepancy~\cite{Miller:2002xh}. However, both $R_\nu$ and $R_{\bar\nu}$ are 
seen lower than the SM prediction, while nuclear shadowing --- when modeled 
using vector meson dominance --- predicts a larger reduction in the CC than in
the NC~\cite{Zeller:2002et}. Moreover, the deviation in $R_\nu$ is larger than
the one in $R_{\bar\nu}$ while from nuclear shadowing one would expect 
the opposite.  Shadowing and other nuclear effects are also discussed in
Ref.~\cite{Kovalenko:2002xe}. An asymmetry in the first moments of the strange
sea distributions, $S - \bar{S}$, could reduce the discrepancy by as much as 
50\%~\cite{Davidson:2001ji}, although NuTeV finds a very small asymmetry from 
a next-to-leading order (NLO) analysis of dimuon production cross 
sections~\cite{Goncharov:2001qe}. Isospin symmetry 
violation~\cite{Zeller:2002du}, particularly violation of charge symmetry under
which protons and neutrons and simultaneously $u$ and $d$ quarks are 
interchanged, can possibly be responsible for the NuTeV effect. However, firm
conclusions cannot be drawn before a global analysis of parton distribution
functions (PDFs) allowing departures from isospin symmetry and an asymmetric
strange sea has been performed~\cite{Bodek:1999bb}. A similar remark applies to
QCD corrections which could have significant impact on the individual ratios
$R_\nu$ and $R_{\bar\nu}$ (but not $R^-$)~\cite{Davidson:2001ji}: NuTeV 
currently uses leading order (LO) expressions and LO PDFs; conclusions about 
the size of QCD corrections must await a NLO analysis including NLO PDFs. For
recent discussions of QCD corrections and parton structure relevant to NuTeV,
see Refs.~\cite{Dobrescu:2003ta,Kretzer:2003wy}.

If none of the SM effects discussed in the previous paragraph are ultimately 
able to explain the NuTeV results, they might be due to NP (or a fluctuation).
Most NP scenarios have difficulty explaining the NuTeV deviation. For example,
inclusion of supersymmetric radiative corrections as well as $R_P$ violating 
(RPV) interactions generally tend to increase the value of $R^\nu$ and is 
unlikely to help account for the NuTeV results~\cite{Kurylov:2003by}. 
A $Z^\prime$ boson would affect the $\epsilon_{L,R} (f)$ couplings, but it is 
difficult to explain the entire discrepancy that way, unless one chooses 
a family non-universal $Z^\prime$~\cite{Langacker:2000ju} with carefully tuned
couplings, which could also shed light on some of the other 
anomalies~\cite{Erler:1999nx} discussed in Section~\ref{SMstatus}. Similarly, 
one might invoke triplet leptoquarks with carefully chosen mass 
splittings~\cite{Davidson:2001ji}. A $\nu_\mu$-mixing with an extra heavy 
neutrino could also be the cause, but this would imply shifts in other 
observables, as well, so that this scenario~\cite{Babu:2002en,Loinaz:2002ep} 
requires a conspiracy of effects to obtain an overall consistent picture.

\begin{table}[t]
\begin{center}
\begin{minipage}[t]{16.5 cm}
\caption[]{Coefficients defining the linear combinations in Eq.~(\ref{lc}).
CHARM has been adjusted to be directly comparable to CDHS. The average momentum
transfer, $Q^2 = -q^2$, is also shown.  In the case of NuTeV it corresponds to 
the geometric mean between the average $\log Q^2$-values of neutrino and 
anti-neutrino reactions.  In the case of the CCFR~\cite{McFarland:1997wx} and 
NuTeV experiments, $\delta$ has been absorbed into the $\epsilon$ parameters.}
\label{tab:dis2}
\vspace*{4pt}
\end{minipage}
\begin{tabular}{|l|c|c|c|c|c|c|r|}
\hline
&&&&&&&\\[-8pt]
Quantity & Group(s) & $\delta$ & $Q^2$~[GeV$^2$] & $a_L(u)$ & $a_L(d)$ & 
$a_R(u)$ & $a_R(d)$ \\[4pt]
\hline
&&&&&&&\\[-8pt]
$g_L^2$      &NuTeV        & 0     & 12 & 1     & 1     & 0     & 0    \\
$g_R^2$      &NuTeV        & 0     & 12 & 0     & 0     & 1     & 1    \\
$R^\nu$      &CCFR         & 0     & 35 & 1.698 & 1.881 & 1.070 & 1.226\\
$R^\nu$      &CDHS + CHARM & 0.023 & 21 & 0.936 & 1.045 & 0.379 & 0.453\\
$R^{\bar\nu}$&CDHS + CHARM & 0.026 & 11 & 0.948 & 1.134 & 2.411 & 2.690\\
$R^{\bar\nu}$&CDHS (1979)  & 0.024 & 11 & 0.944 & 1.126 & 2.295 & 2.563\\[-8pt]
&&&&&&&\\
\hline
\end{tabular}
\end{center}
\end{table}

The other results in Table~\ref{tab:dis1} are only very approximately 
proportional to $R_\nu$ or $R_{\bar\nu}$. In practice, the measured quantities
are the linear combinations of the $\epsilon_{L,R}^2(q)$~\cite{Blondel:1989ev},
\be
   R^{\nu(\bar\nu)} = (1 - \delta) \left[
                      a_L(u) \epsilon_L^2(u) + a_L(d) \epsilon_L^2(d)
                    + a_R(u) \epsilon_R^2(u) + a_R(d) \epsilon_R^2(d) \right],
\label{lc}
\ee
where $\delta$ accounts for the purely photonic (QED) radiative corrections. 
In addition to the charm threshold and radiative corrections, 
the interpretation of $\nu$-nucleus DIS experiments is complicated by 
corrections for and uncertainties from non-isoscalarities, structure functions,
sea quark effects, {\em etc.}, which also gives rise to theoretical 
correlations among the results. The residual correlation between $g_L^2$ and 
$g_R^2$ from NuTeV is accidentally very small ($-0.017$\%). This and 
the correlations between the older results are taken into account, while 
the correlations between NuTeV and the other experiments are currently 
neglected. The results of the older experiments should also be updated with 
the more precise NuTeV results on $m_c^{\rm eff}$ and the quark sea, and if 
possible, with more recent structure functions.

Initially, elastic $\nu$-$p$ scattering~\cite{Mann:qh} also served as a SM 
test and as a measurement of $\sin^2\theta_W$. The most precise 
result~\cite{Ahrens:xe}, $\sin^2\theta_W = 0.218^{+0.039}_{-0.047}$, is no 
longer competitive with present day determinations, but elastic $\nu$
scattering continues to play a crucial role in form factor measurements (see 
Section~\ref{strangeness}). 

NC $\nu$-$e$ scattering~\cite{Panman:rg} is described by 
the Lagrangian~(\ref{nuf}) with $f = e$, and one usually extracts,
\be
   g_V^{\nu e} = \epsilon_L(e) + \epsilon_R(e) = -{1\over 2} + 2\sin^2\theta_W,
\hspace{50pt}
   g_A^{\nu e} = \epsilon_L(e) - \epsilon_R(e) = -{1\over 2}.
\ee
EW radiative corrections to $g_{V,A}^{\nu e}$ have been obtained in 
Refs.~\cite{Sarantakos:1982bp,Bardin:1983yb}. The results are summarized in 
Table~\ref{tab:nue}. $\nu_e$ scattering has been studied at 
\begin{table}
\begin{center}
\begin{minipage}[t]{16.5 cm}
\caption[]{$\nu_\mu$-$e$ scattering results from CHARM~\cite{Dorenbosch:1988is}
and CHARM~II~\cite{Vilain:1994qy} at CERN and from E734~\cite{Ahrens:fp} at
BNL, obtained from two-parameter fits.  The E734 Collaboration reports a 16.3\%
statistical anti-correlation and a 41.3\% systematic correlation between 
$g_V^{\nu e}$ and $g_A^{\nu e}$, for a total 11.9\% correlation. Only the world
averages (which are updated from Ref.~\cite{Erler:ew}) are used as constraints
in the fits.  These also include results from $\nu_e$-$e$ scattering, as well 
as older $\nu_\mu$-$e$ experiments. The overall correlation of the world 
averages is $-5$\%.}
\label{tab:nue}
\vspace*{4pt}
\end{minipage}
\begin{tabular}{|c|c|c|c|r|}
\hline
&&&&\\[-8pt]
Quantity & Group(s) & Value & Standard Model & Pull \\[4pt]
\hline
&&&&\\[-8pt]
$g_V^{\nu e}$ & CHARM    & $-0.060 \pm 0.073$ & $-0.0398 \pm 0.0003$ & --- \\
$g_V^{\nu e}$ & E 734    & $-0.107 \pm 0.046$ &                      & --- \\
$g_V^{\nu e}$ & CHARM II & $-0.035 \pm 0.017$ &                      & --- \\
$g_V^{\nu e}$ & world average & $-0.041 \pm 0.015$ &          &$-0.1$ \\[4pt]
\hline
&&&&\\[-8pt]
$g_A^{\nu e}$ & CHARM    & $-0.570 \pm 0.072$ & $-0.5065 \pm 0.0001$ & --- \\
$g_A^{\nu e}$ & E 734    & $-0.514 \pm 0.036$ &                      & --- \\
$g_A^{\nu e}$ & CHARM II & $-0.503 \pm 0.017$ &                      & --- \\
$g_A^{\nu e}$ & world average & $-0.507 \pm 0.014$ &          &$0.0$ \\[-8pt]
&&&&\\
\hline
\end{tabular}
\end{center}
\end{table}     
LANL~\cite{Allen:qe,Auerbach:2001wg} and $\bar\nu_e$ scattering at the Savannah
River plant~\cite{Reines:pv}. In these cases one has to add the CC Lagrangian 
so that effectively, $\epsilon_L(e) \rightarrow \epsilon_L(e) + 1$, and, 
$g_{V,A}^{\nu e} \rightarrow g_{V,A}^{\nu e} + 1$.  The basic observables are 
the cross sections, which in the limit of large incident $\nu$ energies, 
$E_\nu \gg m_e$, read,
\be
   \sigma = \frac{G_F^2 m_e E_\nu}{2\pi} \left[ (g_V^{\nu e} \pm g_A^{\nu e})^2
            + {1\over 3} (g_V^{\nu e} \mp g_A^{\nu e})^2 \right],
\ee
where the upper (lower) sign refers to (anti-)neutrinos.  Some experiments
achieved slight improvements by also including differential cross section 
information.  The NC-CC interference in $\nu_e$-$e$ scattering resolves 
a sign ambiguity, $g_{V,A}^{\nu e} \rightarrow - g_{V,A}^{\nu e}$, relative
to the CC coupling, and is found in agreement with the SM~\cite{Allen:qe}. 
A new reactor-based elastic $\bar{\nu}_e$-$e^-$ scattering experiment has been
suggested in Ref.~\cite{Conrad:2004gw}, aiming at an improvement by a factor of
four in $\sin^2\theta_W$ relative to the results in Table~\ref{tab:nue}.

\subsection{\it Charged Lepton Scattering}
\label{escatt}
The parity (P) or charge-conjugation\footnote{We refer here to the conjugation
of the lepton charge, and not the fundamental charge-conjugation operation 
which would also require replacing the target by anti-matter.} (C) violating NC
Lagrangian for charged lepton-hadron scattering is given by (assuming 
lepton-universality),
\be
  {\cal L}_{\rm NC}^{\nu f} = \frac{G_F}{\sqrt 2} \sum\limits_q \left[
  C_{1q} \bar{\ell} \gamma^\mu \gamma_5 \ell \bar{q} \gamma_\mu          q +
  C_{2q} \bar{\ell} \gamma^\mu          \ell \bar{q} \gamma_\mu \gamma_5 q +
  C_{3q} \bar{\ell} \gamma^\mu \gamma_5 \ell \bar{q} \gamma_\mu \gamma_5 q 
\right],
\label{ellf}
\ee
where the effective couplings at the SM tree level are again obtained from 
Eqs.~(\ref{Lagrangian}) and (\ref{zcouplings}),
\be
   C_{1q} = - T_3^q + 2 Q_q \sin^2\theta_W, \hspace{40pt}
   C_{2u} = - C_{2d} = - {1\over 2} + 2 \sin^2\theta_W, \hspace{40pt}
   C_{3u} = - C_{3d} = {1\over 2}.
\label{cij}
\ee
The $C_{3q}$ are P conserving, but C violating couplings affecting asymmetries 
involving charge reversal. Alternatively, the following linear 
combinations~\cite{Hung:yg} motivated by isospin symmetry are also used,
\be
\ba{l}
   \tilde{\alpha} = C_{1u} - C_{1d} =  - 1  +  2 \sin^2\theta_W, \hspace{50pt}
   \tilde{\gamma} = C_{1u} + C_{1d} = {2\over 3} \sin^2\theta_W, \\[4pt]
   \tilde{\beta } = C_{2u} - C_{2d} =  - 1  +  4 \sin^2\theta_W, \hspace{50pt}
   \tilde{\delta} = C_{2u} + C_{2d} = 0.
\label{cij_alt}
\ea
\ee
EW radiative corrections to the couplings in Eq.~(\ref{cij}) other than 
the $C_{3q}$ have first been obtained in 
Refs.~\cite{Wheater:ym,Marciano:1982mm} and updated in 
Ref.~\cite{Marciano:1993ep}.

The terms in Eq.~(\ref{ellf}) give rise to interference 
effects~\cite{Derman:sp,Berman:1973pt} with the parity conserving QED terms in
Eq.~(\ref{Lagrangian}), which can be isolated by measuring parity violating LR
asymmetries or by comparing processes related by charge-conjugation in charged 
lepton-hadron scattering. Currently the most precise results are obtained from
APV which will be discussed in the next Subsection~\ref{APV}.

Historically, the SLAC $e$-D scattering experiment~\cite{Prescott:1978tm}
was crucial in establishing the SM.  It measured~\cite{Prescott:1979dh}
the cross section asymmetry~\citer{Cahn:1977uu,Fritzsch:1978ku},
\be
   \frac{\sigma_R - \sigma_L}{\sigma_R + \sigma_L} = 
   {3 G_F Q^2 \over 10\sqrt{2}\pi\alpha} 
   \left[ (2 C_{1u} - C_{1d}) + g(y) (2 C_{2u} - C_{2d}) \right],
\label{asymmetrya}
\ee
where $y = (E_0 - E')/E_0$, where $E_0$ ($E'$) is the incident (scattered) 
electron energy, and where\footnote{For more precise measurements, $g(y)$ needs
to be corrected for the longitudinal contributions to $\gamma$ and $Z$ 
exchange.},
\be
   g(y) = \frac{1 - (1 - y)^2}{1 + (1 - y)^2}.
\ee
Eq.~(\ref{asymmetrya}) is valid for isoscalar targets and neglects $s$ quarks,
$c$ quarks, and anti-quarks. The only experiment to date which has measured 
a charge-conjugation cross section asymmetry~\cite{Kim:sa,Hung:1981nv},
\be
   \frac{\sigma^+(-|\lambda|) - \sigma^-(+|\lambda|)}
        {\sigma^+(-|\lambda|) + \sigma^-(+|\lambda|)} = 
   - {G_F Q^2 \over 10 \sqrt{2}\pi\alpha} g(y) \left[ (2 C_{3u} - C_{3d}) + 
     \lambda (2 C_{2u} - C_{2d}) \right],
\ee
was the CERN $\mu^\pm$C scattering experiment~\cite{Argento:1982tq}. 
In this experiment, the $\mu$-polarization, $\lambda$, was reserved 
simultaneously with the $\mu$-charge. A linear combination of the $C_{1q}$ and 
$C_{2q}$ different from those entering DIS was obtained in an experiment at 
Mainz~\cite{Heil:dz} in the quasi-elastic (QE) kinematic regime. The asymmetry
is a superposition of various distinct contributions and described by nuclear 
form factors~\cite{Hoffmann:1978xy} which were taken from other experiments. 
Scattering off carbon at even lower energies needs only two elastic form 
factors, $G_E^{T=0}$ and $G_E^S$, the isoscalar electromagnetic and strange 
quark electric form factors. The dependence on $G_E^{T=0}$ cancels in 
the asymmetry~\cite{Feinberg:cg} of 
the form~\cite{Walecka:1977us,Musolf:1993tb},
\be
   \frac{\sigma_R - \sigma_L}{\sigma_R + \sigma_L} = {3 G_F Q^2 \over 2\sqrt{2}
   \pi\alpha} \left[(C_{1u} + C_{1d})+{G_E^S\over 4 G_E^{T=0}}\right],
\ee
which has been measured in elastic $e$-C scattering~\cite{Souder:ia} at
the MIT-Bates accelerator. Measurements of PV elastic $e$-$p$ and QE $e$-D
scattering~\cite{Hasty:2001ep} at the same facility yielded a value for 
$C_{2u} - C_{2d}$. However, there is uncertainty in the SM prediction due to 
the presence of the proton anapole moment~\cite{Musolf:ts,Zhu:2000gn}.

Table~\ref{tab:ellhad} summarizes the lepton-hadron scattering experiments
\begin{table}
\begin{center}
\begin{minipage}[h]{16.5 cm}
\caption[]{Observables sensitive to the $P$ or $C$ violating coefficients 
$C_{iq}$. The errors are the combined (in quadrature) statistical, systematic 
and theoretical uncertainties. The first two lines result from a fit to 
11~different kinematic points (a 5\% uncertainty in the polarization was common
to all points) and have a $-92.7\%$ correlation. Including a 7\% theory 
uncertainty~\cite{Souder:qk} increases the error in the first line to 
$\pm 0.18$ and decreases the correlation to $-86.6\%$. The two CERN entries are
for muon beam energies (polarizations) of 120~GeV (66\%) and 200~GeV (81\%), 
respectively. Assuming 100\% correlated systematic errors yields a correlation 
of 17.4\% between them. The second line (SLAC) contains a 31.6\% correction 
to account for sea quarks, while the corresponding correction is 7.5\% for 
CERN~\cite{Souder:qk}. The Mainz result includes a 10\% theory 
error~\cite{Souder:qk}.} 
\label{tab:ellhad}
\vspace*{4pt}
\end{minipage}
\begin{tabular}{|c|c|c|c|c|c|c|c|c|}
\hline
&&&&&\\[-8pt]
Beam & Process & $\overline{Q^2}$~[GeV$^2$] & Combination & Result/Status & 
SM \\[4pt] 
\hline
&&&&&\\[-8pt]
SLAC & $e^-$-D DIS    & 1.39 & $2C_{1u}-C_{1d}$ & $-0.90\pm 0.17$ & $-0.7185$\\
SLAC & $e^-$-D DIS    & 1.39 & $2C_{2u}-C_{2d}$ & $+0.62\pm 0.81$ & $-0.0983$\\
CERN & $\mu^\pm$-C DIS&  34  & $0.66(2C_{2u}-C_{2d})+2C_{3u}-C_{3d}$
                                                & $+1.80\pm 0.83$ & $+1.4351$\\
CERN & $\mu^\pm$-C DIS&  66  & $0.81(2C_{2u}-C_{2d})+2C_{3u}-C_{3d}$
                                                & $+1.53\pm 0.45$ & $+1.4204$\\
Mainz& $e^-$-Be QE    & 0.20 & $2.68C_{1u}-0.64C_{1d}+2.16C_{2u}-2.00C_{2d}$ 
                                                & $-0.94\pm 0.21$ & $-0.8544$\\
Bates& $e^-$-C elastic&0.0225& $ C_{1u}+C_{1d}$ & $0.138\pm 0.034$& $+0.1528$\\
Bates& $e^-$-D QE     & 0.1  & $ C_{2u}-C_{2d}$ & $0.015\pm 0.042$& $-0.0624$\\
JLAB & $e^-$-$p$ elastic& 0.03 & $2C_{1u}+C_{1d}$ & approved      & $+0.0357$\\
SLAC & $e^-$-D DIS    &  20  & $2C_{1u}-C_{1d}$ & to be proposed  & $-0.7185$\\
SLAC & $e^-$-D DIS    &  20  & $2C_{2u}-C_{2d}$ & to be proposed  & $-0.0983$\\
SLAC & $e^\pm$-D DIS  &  20  & $2C_{3u}-C_{3d}$ & to be proposed  & $+1.5000$\\[4pt] 
\hline
&&&&&\\[-8pt]
 --- & $^{133}$Cs APV & 0 &$-376C_{1u}-422C_{1d}$&$-72.69\pm 0.48$& $-73.16$ \\
 --- & $^{205}$Tl APV & 0 &$-572C_{1u}-658C_{1d}$&$-116.6\pm 3.7$ & $-116.8$ \\[-8pt] 
&&&&&\\
\hline
\end{tabular}
\end{center}
\end{table}     
described above. We applied corrections for $\alpha (Q^2) \neq\alpha$. The most
precise results from APV (discussed in the next Subsection) are also shown. 
Furthermore, the weak charge of the proton,
\be
   Q_W (p) = 2 C_{1u} + C_{1d},
\label{qweak}
\ee
will be measured at the Jefferson Lab~\cite{Armstrong:2001} in elastic $e$-$p$ 
scattering. Ward identities associated with the weak NC protect $Q_W (p)$ 
(defined at $Q^2=0$) from incalculable strong interaction effects and $Q_W(p)$ 
can be computed with a high degree of reliability~\cite{Erler:2003yk}. 
Approximately linear effects due to $Q^2 \approx 0.03 \mbox{ GeV}^2 \neq 0$ can
be accounted for experimentally by extrapolating across various $Q^2$ points 
performed by other experiments (see Section~\ref{strangeness}). A new and more
precise deep inelastic $e$-D scattering experiment at SLAC has also been 
suggested~\cite{Bosted:2003} conceivably including an electron charge asymmetry
as a separate observable. 

We performed a simultaneous fit to the couplings defined in Eq.~(\ref{cij_alt})
in addition to the combination $\tilde\epsilon \equiv 2 C_{3u} - C_{3d}$, 
including all theoretical and experimental uncertainties and correlations. 
The result is,
\be
\ba{rlllll}
   \tilde\alpha      &=& - 0.67  &\pm& 0.10  & ({\rm SM} = - 0.530), \\
   \tilde\beta       &=& + 0.007 &\pm& 0.040 & ({\rm SM} = - 0.062), \\
   \tilde\gamma      &=& + 0.144 &\pm& 0.006 & ({\rm SM} = + 0.153), \\
   \tilde\delta      &=& + 1.4   &\pm& 1.5   & ({\rm SM} = - 0.009), \\
   \tilde\epsilon    &=& + 0.87  &\pm& 0.88  & ({\rm SM} = + 1.500),
\ea
\ee
with the correlation matrix\footnote{The individual $C_{1q}$ and $C_{2q}$ are 
even stronger correlated than the combinations chosen here.} given by:

\be
\ba{|c|rrrrr|}
\hline
&&&&&\\[-8pt]
& \tilde\alpha & \tilde\beta & \tilde\gamma & \tilde\delta & \tilde\epsilon 
\\[4pt]
\hline
&&&&&\\[-8pt]
\tilde\alpha      &  1.00 & -0.09 &  0.98 & -0.92 &  0.81 \\
\tilde\beta       & -0.09 &  1.00 & -0.09 & -0.02 & -0.05 \\
\tilde\gamma      &  0.98 & -0.09 &  1.00 & -0.90 &  0.80 \\
\tilde\delta      & -0.92 & -0.02 & -0.90 &  1.00 & -0.88 \\
\tilde\epsilon    &  0.81 & -0.05 &  0.80 & -0.88 &  1.00 \\[-8pt]
&&&&&\\
\hline
\ea
\ee

This concludes the discussion of PV charged lepton scattering off hadrons.
Another possibility is to have leptons in both the initial and final states. 
The first PV experiment of this type is being performed at SLAC. The E158 
Collaboration~\cite{Anthony:2003ub} measures the weak charge of the electron,
$Q_W(e)$, from the left-right asymmetry in polarized M\o ller scattering, 
$e^- e^- \rightarrow e^- e^-$. A precision of better than $\pm 0.001$ in 
$\sin^2 \theta_W$ at $Q^2 \sim 0.03$ GeV$^2$ is anticipated. The result of 
the first of three runs yields for the mixing angle in the $\overline{\rm MS}$ 
renormalization scheme at the $Z$ scale, 
$\sin^2\hat\theta_W (M_Z) = 0.2279\pm 0.0032$. 

Since the SM tree level expressions for $Q_W(e)$ and $Q_W(p)$ are suppressed by
a factor $1 - 4 \sin^2 \theta_W \ll 1$, these observables are particularly 
useful to search for a variety of NP 
scenarios~\cite{Erler:2003yk,Ramsey-Musolf:1999qk}.

\subsection{\it Parity Violation in Atoms}
\label{APV}
The experimental study of PV in atoms generally follows one of two approaches.
The first is to measure the rotation of the polarization plane of linearly
polarized light as it passes through a vapor of atoms. Such a rotation occurs 
because PV interactions between the atomic electrons and the nucleus induce 
a difference in the absorption cross section for the left- and right-handed 
components of the polarized light. A second method involves applying 
an external electric field to a vapor, thereby inducing Stark-mixing of 
the atomic levels and leading to parity forbidden atomic transitions. The PV
electron-nucleus interactions lead to small modulations of the Stark induced
transitions, and their effect may be isolated by the appropriate combination of
field reversals. The most precise measurement of an APV effect has been
performed by the Boulder group exploiting the Stark interference method and 
a beam of cesium atoms~\cite{Wood:zq}. A summary of APV results obtained by 
either of these methods appears in Table~\ref{tab:apv}.
\begin{table}
\begin{center}
\begin{minipage}[t]{16.5 cm}
\caption[]{Representative results of various APV experiments. Quoted atomic 
theory uncertainties for the first eight entries are taken from 
Ref.~\cite{Masterson:qi}. The final entry for francium indicates a measurement
only of the nuclear anapole moment.}
\label{tab:apv}
\vspace*{4pt}
\end{minipage}
\begin{tabular}{|l|l|l|r|r|}
\hline
&&&&\\[-8pt]
Element & Method & Group & Experimental & Atomic Theory \\
        &        &       & Uncertainty  & Uncertainty   \\[4pt] 
\hline
&&&&\\[-8pt]
Pb & spin rotation & Seattle~1983~\cite{Emmons:vn}     &$\pm 28\%$&$\pm 10\%$\\
Bi & spin rotation & Seattle~1981~\cite{Hollister:1981}&$\pm 18\%$&$\pm 15\%$\\
   &               & Oxford~1991~\cite{MacPherson:1991}&$\pm  2\%$&$\pm 15\%$\\
   &               & Seattle~1993~\cite{Meekhof:1993}  &$\pm  1\%$&$\pm 15\%$\\
Tl & spin rotation & Oxford~1991~\cite{Wolfeden:1991}  &$\pm 15\%$&$\pm  3\%$\\
Tl & Stark interference&Berkeley~1985~\cite{Drell:mx}  &$\pm 28\%$&$\pm  6\%$\\
   &               & Oxford~1995~\cite{Edwards:1995}   &$\pm  3\%$&$\pm  6\%$\\
   &               & Seattle~1995~\cite{Vetter:vf}     &$\pm  1\%$&$\pm  6\%$\\
Cs & Stark interference&Boulder~1985~\cite{Gilbert:ki} &$\pm 12\%$&$\pm  1\%$\\
   &               & Paris~1986~\cite{Bouchiat:1986}   &$\pm 12\%$&$\pm  1\%$\\
   &               & Boulder~1988~\cite{Noecker:ys}    &$\pm 2 \%$&$\pm  1\%$\\
   &               & Boulder~1998~\cite{Wood:zq}    &$\pm 0.35\%$&$\pm 0.5\%$\\[4pt]
\hline
&&&&\\[-8pt]
Yb & isotope ratios (${\cal R}_1$) & Berkeley~(in progress)~\citer{DeMille:1995,Stalnaker:2004} & $\pm 0.1\%$ & \\
Ba$^+$ & ion trap & Seattle~(in progress)~\cite{Fortson:2004} & $\sim\pm 0.35\%$&$\lsim\pm 1\%$\\
Fr & atom trap & Stony Brook~(in progress)~\cite{Orozco:2004} & $\lsim\pm  10\%$&$\sim\pm 1\%$ \\[-8pt] 
&&&&\\
\hline
\end{tabular}
\end{center}
\end{table}     

Theoretically, APV effects are described by the effective, PV atomic 
Hamiltonian,
\be
   {\hat H}_{\rm PV}^{\rm atom} = {\hat H}_{\rm PV}^{\rm atom}({\rm NSID}) + 
   {\hat H}_{\rm PV}^{\rm atom}({\rm NSD}),
\ee
where the second term depends on the nuclear spin (NS) while the first term
is nuclear spin independent. The former is given by,
\be
\label{eq:apvnsid}
  {\hat H}_{\rm PV}^{\rm atom}({\rm NSID}) ={G_F\over 2\sqrt{2}}
  \int d^3x {\hat\psi}^{\dag}_e({\vec x}) \gamma_5 \psi_e({\vec x})
  \rho^{\rm NC}({\vec x}) + \cdots,
\ee
where $\rho^{\rm NC}({\vec x})$ is the nuclear matrix element of the NC charge
operator and where the dots indicate contributions from the spatial components 
of the nuclear NC. The analogous form for 
${\hat H}_{\rm PV}^{\rm atom}({\rm NSD})$ is,
\be
\label{eq:apvnsd}
   {\hat H}_{\rm PV}^{\rm atom}({\rm NSD}) ={G_F\over \sqrt{2}}\kappa
   \int d^3x {\hat\psi}^{\dag}_e({\vec x}) {\vec\alpha} \psi_e({\vec x}) \cdot
   {\vec I}{\tilde\rho}({\vec x}),
\ee
where ${\vec\alpha}$ are Dirac matrices, ${\vec I}$ is the nuclear spin,
and ${\tilde\rho}({\vec x})$ is the nuclear density~\cite{Haxton:2001ay}. 
The quantity ${\kappa}$ receives contributions from the hadronic axial vector 
NC as well as from the anapole moment (see below). The effect of 
${\hat H}_{\rm PV}^{\rm atom}$ is to mix atomic states of opposite parity.
In cesium, the relevant mixing involves $n S_{1/2}$ and $n' P_{1/2}$ states, 
with a mixing matrix element given by,
\be
\label{eq:APVmix}
   \bra{P} {\hat H}_{\rm PV}^{\rm atom}({\rm NSID})\ket{S} =
   i {G_F\over 2\sqrt{2}} C_{SP}(Z) Q_W(Z,N) + \cdots,
\ee
where $C_{SP}$ is an atomic structure dependent coefficient that must be
computed theoretically and where the dots denote small corrections arising from
the spatial components of the NC, finite nuclear size, nucleon substructure, 
and the NSD term. Since the weak charge [cf.\ Eq.~(\ref{qweak})] is given by,
\be
   Q_W(Z,N) = (2 Z + N) C_{1u} + (Z + 2 N) C_{1d} 
      \approx Z (1 - 4\sstw) - N \approx - N,
\ee
the finite nuclear size corrections are dominated by the spatial dependence of
the neutron density. The uncertainties associated with this correction have 
been estimated in Ref.~\cite{Pollock:1999ec} to be about $\pm 0.15\%$ for 
cesium. The uncertainties associated with nucleon structure are comparably
small~\cite{Musolf:1993tb,Erler:2003yk}.

A more significant source of theoretical uncertainty arises from computations 
of the atomic structure dependent constants $C_{SP}(Z)$. In practical 
calculations, however, one must compute not only the mixing matrix elements of
Eq.~(\ref{eq:APVmix}) for a tower of $P$-states that are mixed into the cesium
ground state and first excited state, but also those of the electric dipole 
operator between the various $S$ and $P$ states as well as the energy 
differences appearing in the denominator in the perturbation series. Prior to 
the most recent measurement of cesium APV, {\em ab initio\/} calculations of 
these quantities had been carried out by the Notre 
Dame~\cite{Blundell:1990ji,Blundell:ss} and Novosibirsk~\cite{Dzuba:yu} groups.
The uncertainty in $Q_W(Z,N)$ associated with these computations were estimated
to be about one percent. Following their 0.35\% measurement of the PV 
transition in cesium~\cite{Wood:zq}, the members of the Boulder group performed
additional measurements of transition dipole amplitudes and argued that 
the results considerably reduced the theoretical, atomic structure uncertainty 
in $Q_W$ for cesium, with a combined experimental and theoretical error of 
0.6\%~\cite{Bennett:1999pd}. With this reduced uncertainty, the measurement 
implied a 2.5~$\sigma$ deviation from the SM prediction. 

The report of this deviation stimulated considerable theoretical activity. From
the standpoint of particle physics, various studies argued that it suggested 
the presence of a light 
$Z'$~\cite{Erler:1999nx,Rosner:1999cy,Casalbuoni:1999mw}, 
leptoquarks~\cite{Barger:2000gv}, or RPV SUSY 
interactions~\cite{Ramsey-Musolf:2000qn}. At the same time, atomic structure 
theorists scrutinized previous calculations and discovered several 
${\cal O}(1\%)$ effects that had not been properly included. Among 
these effects were correlation-enhanced contributions from the Breit
interaction~\cite{Derevianko:2000dt,Dzuba:2001}, contributions from the Uehling
potential~\cite{Johnson:2001nk,Milstein:2002ai}, and QED vertex and self-energy
corrections that are amplified in the presence of the nuclear
field~\citer{Dzuba:2002kx,Kuchiev:2003pk}. Inclusion of the Uehling potential 
contribution tends to increase the disagreement between the experimental and SM
values for $Q_W$, whereas the Breit correction and QED vertex and self-energy 
contributions reduce it. The net result is the value,
\be
\label{eq:csexp}
   Q_W^{\rm Cs} ({\rm exp.}) = - 72.69 \pm 0.48,
\ee
in agreement with the SM prediction~\cite{Erler:2003yk},
\be
\label{eq:cssm}
   Q_W^{\rm Cs} ({\rm SM}) = - 73.16.
\ee
The error in~(\ref{eq:csexp}) includes the uncertainty from atomic 
structure calculations. Note that this error (obtained from the most recent 
atomic structure publications) has been reduced from the 1\%~uncertainty 
associated with the previous calculations in 
Refs.~\citer{Blundell:1990ji,Dzuba:yu}. When used to extract a value for 
$\sstw$ at $Q^2 \approx 0$, one obtains from~(\ref{eq:csexp}) the result 
indicated in Figure~\ref{s2w}. 

The importance of the cesium result~(\ref{eq:csexp}) is underlined by 
on-going experimental work in the field. The Paris group is attempting 
to perform a more precise version of their earlier cesium measurement, applying
the Stark induced mixing technique to a cell of cesium gas. The Seattle group
has undertaken a measurement of APV with trapped Ba$^+$ ions that involves 
looking for frequency shifts associated with parity forbidden 
transitions~\cite{Fortson:2004}. The latter involve a study of 
the $6S_{1/2}~(\mbox{ground state})\to 5D_{3/2}$ transition that will contain 
a parity forbidden component due to mixing of P-states into the ground state. 
Ref.~\cite{Ginges:2001} argued that the atomic structure computations for 
Ba$^+$ isotopes should achieve the same level of precision as for cesium. 

At Berkeley efforts are underway to measure APV effects for different isotopes
of Yb. The latter approach was motivated by the observation that a comparison 
of APV effects in different isotopes would eliminate the large atomic structure
theory uncertainties. In particular, if one forms the quantity,
\be
   {\cal R}_1 = {A_{\rm PV}^{\rm NSID}(N')-A_{\rm PV}^{\rm NSID}(N) \over 
   A_{\rm PV}^{\rm NSID}(N') + A_{\rm PV}^{\rm NSID}(N)},
\ee
where $A_{\rm PV}^{\rm NSID}(N)$ is a NS independent atomic PV observable, and 
if the atomic structure effects (governed largely by the nuclear Coulomb 
field) do not vary appreciably along the isotope chain, then one has,
\be
   {\cal R}_1 = {Q_W(N')-Q_W(N)\over Q_W(N')+ Q_W(N)} \approx {N'-N\over N'+N},
\ee
where the dependence on atomic structure has largely canceled from the ratio.
An analogous result occurs for the ratio 
${\cal R}_2 = A_{\rm PV}^{\rm NSID}(N')/A_{\rm PV}^{\rm NSID}(N)$.

Corrections to these ratios are generated by nuclear structure since 
the neutron distribution [and thus, the quantity $\rho^{\rm NC}({\vec x})$
appearing in Eq.~(\ref{eq:apvnsid})] vary along the isotope chain. At present, 
the theoretical uncertainties associated with this effect appear to be larger 
than one would like for isotope measurements to provide meaningful probes of
NP~\cite{Pollock:1999ec}. In principle, a new measurement of the neutron 
distribution in Pb using elastic PV electron scattering at the Jefferson 
Lab~\cite{Michaels:1999} may help to reduce the nuclear structure uncertainties
associated with this and similar measurements. Moreover, the NP sensitivity of 
the ${\cal R}_i$ is dominated by the possible effects of NP on 
the proton~\cite{Ramsey-Musolf:1999qk}, making a direct measurement with, 
{\em e.g.,} PV $e$-$p$ scattering, a cleaner probe. Nonetheless, work is 
proceeding to carry out isotope comparisons with Yb. A combined experimental 
and theoretical uncertainty of $\sim 0.2\% $ in ${\cal R}_1({\rm Yb})$ would 
yield a similar sensitivity to NP as the determination of $Q_W(p)$ planned at 
Jefferson Lab.

While the primary focus of atomic PV measurements has been on testing the SM 
{\em via\/} the NS independent weak charge interaction, the NS dependent 
contribution has also received considerable attention. The latter is dominated
by the nuclear anapole moment, which gives the leading PV coupling of a photon
to a nucleus. For a spin-1/2 system, the anapole interaction has the form,
\be
\label{eq:anapoledef}
   {\cal L}_{\rm anapole} = {a\over M^2} 
   {\psibar}(x)\gamma_\mu\gamma_5 \psi \partial_\nu F^{\mu\nu},
\ee
where the coefficient, $a$, is the anapole moment. The interaction in 
Eq.~(\ref{eq:anapoledef}) vanishes for real photons, while for virtual photons
it has the same contact interaction character as the low energy $Z$ exchange
amplitude. As discussed in Section~\ref{sec:HPV}, the largest contributions to 
$a$ arise from PV nucleon-nucleon ($NN$) interactions, whose effects grow as 
$A^{2/3}$ in nuclei. A study of the anapole moment, then, provides a probe of
the $\Delta S = 0$ hadronic weak interaction in nuclei. The first non-zero 
determination of a nuclear anapole moment was carried out for cesium at 
Boulder~\cite{Wood:zq}, and a limit on the anapole moment of thallium
has been obtained at Seattle. On-going atomic PV experiments involving Yb, 
Ba$^+$ ions, and Fr seek to isolate the anapole effect in those nuclei (for 
a recent review, see Ref.~\cite{Haxton:2001ay}).

\section{Charged Current Phenomena}
\subsection{\it $\mu$ Decay}
The study of heavy lepton decays continues to provide important input into 
the SM and constrain various SM extensions. Indeed, the muon lifetime,
$\tau_\mu$, remains one of the most precisely measured weak interaction 
observables and yields, {\em via\/} the Fermi constant, one of the three inputs
needed to determine properties of the EW gauge sector of the SM. Taking into 
account QED radiative corrections up to ${\cal O}(\alpha^2)$, the lifetime and 
Fermi constant are related through~\cite{Marciano:1999ih},
\be
   {1\over \tau_\mu} = {G_\mu^2 m_\mu^2\over 192\pi^3} 
   f \left( {m_e^2\over m_\mu^2} \right) (1 + R) 
   \left(1 + \frac{3}{5}\frac{m_\mu^2}{M_W^2}\right),
\ee
where,
\be
   R = {\alpha\over 2\pi} \left({25\over 4} - \pi^2 \right) \Biggl[ 1 + 
   \frac{\alpha}{\pi} \left( \frac{2}{3} \ln\frac{m_\mu}{m_e} - 3.7 \right) +
   \left( \frac{\alpha}{\pi}\right)^2 \left(\frac{4}{9} 
   \ln^2 \frac{m_\mu}{m_e} - 2.0 \ln \frac{m_\mu}{m_e} + C \right) + \cdots
   \Biggr],
\ee
are the QED radiative corrections and where $f(x) = 1-8x+8x^3-x^4-12x^2\ln x$.
The ${\cal O}(\alpha)$ contribution was first computed in 
Refs.~\cite{Berman:1958ti,Kinoshita:1958ru}. The ${\cal O}(\alpha^2,\alpha^3)$
terms containing $\ln m_\mu/m_e$ terms have been obtained using renormalization
group methods in Ref.~\cite{Roos:mj}, while the non-logarithmic 
${\cal O}(\alpha^2)$ contribution was worked out in 
Ref.~\cite{vanRitbergen:1998yd}. The constant $C$, which describes 
the non-logarithmic ${\cal O}(\alpha^3)$ corrections has not been computed.

The present value for the lifetime is $\tau_\mu =2.197035(40)\times 10^{-6}$~s,
which leads to $G_\mu= 1.16637 (1)\times 10^{-5}$~GeV$^{-2}$ (the subscript,
$\mu$, indicates a value for the Fermi constant taken from the muon lifetime).
The dominant contribution to the uncertainty arises from the experimental error
in $\tau_\mu$ (18 ppm), with small errors arising from the uncertainty in
the neutrino mass (10 ppm), the muon mass (0.38 ppm) and from higher order QED 
contributions (0.50 ppm).

By itself, the value for $G_\mu$ cannot be used to constrain NP. However, 
requiring consistency between $G_\mu$ and other SM quantities can lead 
to constraints~\cite{Marciano:1999ih}. For example, given values for the fine
structure constant, the $Z$ boson mass, and $G_\mu$, the SM predicts a value 
for the weak mixing angle as a function of all the other parameters in 
the theory,
\be
\label{eq:Gfermi}
  {\hat s}^2{\hat c}^2={\pi\alpha\over\sqrt{2} G_\mu M_Z^2(1-\Delta\hat{r})},
\ee
where the hat indicates quantities renormalized in the $\overline{\rm MS}$ 
scheme, ${\hat s}^2 = \sin^2{\hat\theta}_W (\mu=M_Z)$, and $\Delta\hat{r}$ is
a radiative correction parameter. Alternately, if one treats all the SM 
parameters (including the value of $\hat{s}^2$) as independent quantities to be
taken from experiment, Eq.~(\ref{eq:Gfermi}) is a self-consistency test 
constraining NP that could affect the value of 
$\Delta\hat{r} =\Delta\hat{r}^{\rm SM} + \Delta\hat{r}^{\rm NEW}$. Updating
Ref.~\cite{Marciano:1999ih} we find,
\be
\label{eq:deltarhat}
   \Delta{\hat r}^{\rm NEW} = 0 \pm 0.0006,
\ee
This can be used to constrain a variety of NP scenarios~\cite{Marciano:1999ih}.
For example, if an excited $W$ boson contributed to the decay rate, then its 
mass would have to satisfy,
\be
  M_{W^\ast} > 3.3 \sqrt{C} {g^\ast\over g}~{\rm TeV},
\ee
where $C$ is a model dependent constant expected to be of ${\cal O}(1)$, and 
$g^\ast$ is the gauge coupling of the $W^\ast$. In a similar way, 
Eq.~(\ref{eq:Gfermi}) can be used to derive constraints for SUSY 
scenarios~\cite{Ramsey-Musolf:2000qn,Kurylov:2001zx} (see 
Refs.~\citer{Drees:1991zk,Erler:1998ur} for earlier analyses).

At present, the range~(\ref{eq:deltarhat}) is determined by 
the uncertainties in the values of $m_t$ and $\alpha(M_Z)$ that appear in 
$\Delta\hat{r}^{\rm SM}$, and the value of $\hat{s}^2$ entering 
Eq.~(\ref{eq:Gfermi}). Substantial improvements in the precision for 
these quantities must be achieved before the uncertainty in $G_\mu$ will 
present a serious limitation to future improvements in NP sensitivity. 
Nevertheless, a more precise measurement of $\tau_\mu$ is being carried out at
PSI by the FAST~\cite{FAST} and $\mu$Lan~Collaborations~\cite{mulan}. The goal
for both measurements is a 2~ps (1~ppm) uncertainty in the lifetime.

In addition to precision measurements of $\tau_\mu$, studies of the $\mu$ decay
spectral shape and $\beta$-asymmetry provide tests of EW theory\footnote{Studies
of the $\beta$-polarization may also lead to constraints on NP (see, {\em e.g.},
Ref.~\cite{Aysto:2001zs}).}. These properties have historically been described 
by the Michel parameters~\cite{Michel:1950,Bouchiat:1957}, which appear in 
the partial decay rate for a $\mu^{\pm}$:
\bea
\label{eq:michel1}
   d\Gamma& = & {G_\mu^2 m_\mu^5\over 192\pi^3} {d\Omega\over 4\pi} x^2\ dx
  \times \Biggl\{ {1+h(x)\over 1 + 4\eta(m_e/m_\mu)}\left[
  12(1-x)+\frac{4}{3}\rho(8x-6)+ 24\frac{m_e}{m_\mu}{(1-x)\over x}\eta\right]\\
\nonumber
   && \pm P_\mu\; \xi\cos\theta \left[ 4 (1-x) + \frac{4}{3}\delta(8x - 6) +
   {\alpha\over 2\pi}{g(x)\over x^2}\right]\Biggl\},
\eea
where $x=|{\vec p}_e|/|{\vec p}_e|_{\rm max}$, 
$\theta=\cos^{-1}({\hat p}_e\cdot{\hat s}_\mu)$, $P_\mu$ is the $\mu^{\pm}$ 
polarization, and $h(x)$ and $g(x)$ are momentum dependent radiative 
corrections. The SM predictions for the Michel parameters, $\rho$, $\delta$, 
$\xi$, and $\eta$, along with the present experimental limits are listed in 
Table~\ref{tab:michel}. 
\begin{table}
\begin{center}
\begin{minipage}[t]{16.5 cm}
\caption[]{Present experimental values for the Michel parameters, compared with
SM predictions. Experimental errors have been combined in quadrature. Projected
uncertainties for the TWIST measurement are shown in the last column.}
\label{tab:michel}
\vspace*{4pt}
\end{minipage}
\begin{tabular}{|c|c|c|c|}
\hline
&&&\\[-8pt]
Parameter & Present  & SM  &  TWIST (projected) \\[4pt] 
\hline
&&&\\[-8pt]
$\rho$ & $0.7518\pm 0.0026$~\cite{Derenzo:za}   & $3/4$ & $\pm 0.0001$    \\
$\delta$ & $0.7486\pm 0.0040$~\cite{Balke:1987vr} & $3/4$ & $\pm 0.00014$ \\
$P_\mu\; \xi$ & $1.0027\pm 0.0085 $~\cite{Beltrami:ne} & 1 & $\pm 0.00013$\\
$\eta$ & $-0.007\pm 0.013$~\cite{Burkard:kf}    &   0   & $\pm0.003 $     \\[-8pt] 
&&&\\
\hline
\end{tabular}
\end{center}
\end{table}     

Note that the effect of $\eta$ on the differential rate is suppressed by 
$m_e/m_\mu$, making this quantity more difficult to measure than the other 
Michel parameters. Indeed, the level of agreement of the SM predictions for 
the shape parameters, $\rho$ and $\delta$, compared to experiment is quite high
(per mille), while the precision for $\eta$ and the asymmetry parameter 
$P_\mu\; \xi$ is presently a factor of ten weaker. One expects significant 
improvements in these limits from the TWIST~Collaboration~\cite{Poutissou:kg}, 
which has undertaken a new measurement of polarized $\mu^+$ decay at TRIUMF. 
As indicated in Table~\ref{tab:michel}, the Collaboration expects to decrease 
the experimental errors by factors ranging from 60 for $P_\mu\; \xi$ to 4 for 
$\eta$. 

With the expected improvement in precision, the results of the TWIST experiment
could have significant implications for NP that might affect muon decay. 
Historically, the effects of non-SM interactions on the Michel parameters have
been characterized by a general set of four-fermion contact operators,
\be
\label{eq:michel2}
  {\cal L}^{\rm eff}_{\mu\ {\rm decay}} ={G_\mu\over\sqrt{2}}\sum_j{\bar\psi}_e
  \Gamma_j\psi_\mu 
  {\bar \psi}_{\nu_\mu}\Gamma_j(C_j+C_j^\prime\gamma_5)\psi_{\nu_e},
\ee
where the sum runs over all of the independent Dirac matrices $\Gamma_j$. While
the operators appearing in Eq.~(\ref{eq:michel2}) are non-renormalizable, they 
may arise in the low energy limit of renormalizable gauge theories. In theories
such as the SM that contain purely left- or right-handed gauge interactions, 
all but the vector and axial vector type couplings, $C_{V,A}$ and $C_{V,A}'$, 
vanish. Scalar and pseudoscalar interactions can be induced, for example, in 
the presence of mixing between left- and right-handed gauge bosons (see below).
General expressions for the Michel parameters in terms of the $C_j$ and $C_j'$
can be found, for example, in Ref.~\cite{Commins:ns}.

To illustrate, we consider the effects of right-handed gauge interactions on 
the Michel parameters. An extensive analysis of such effects has been carried
out in Ref.~\cite{Herczeg:cx}. For a situation involving an additional, light 
right-handed gauge boson that mixes with the $SU(2)_L$ gauge boson, one has,
\bea
\label{eq:lrmix}
   W_1 & = & \cos\zeta W_L  - \sin\zeta e^{-i\omega}W_R, \\
   W_2 & = & \sin\zeta W_L + \cos\zeta e^{-i\omega}W_R,
\eea
for the two mass eigenstates with $M_2 > M_1$. The resultant effective, 
low energy interaction for muon decay is given by,
\be
\label{eq:mueffective}
  {\cal L}_{\rm  eff}^{\mu\ {\rm decay}} = 4 \sum_{ij} c_{ij}
  {\bar e}_i\gamma^\lambda \nu_{ei} {\bar\nu}_{\mu j}\gamma_\lambda \mu_j,
\ee
where the sum runs over all chiralities $i,j = L,R$ and where the neutrino 
flavor states may be mixtures of mass eigenstates. In terms of the couplings, 
masses, and mixing angles, one has~\cite{Herczeg:cx},
\bea
\label{eq:mucouplings}
   c_{LL} &=&{g_L^2\over 8 M_1^2}\cos^2\zeta+{g_R^2\over 8 M_2^2}\sin^2\zeta,\\
\nonumber
   c_{RR} &=&{g_L^2\over 8 M_1^2}\sin^2\zeta+{g_R^2\over 8 M_2^2}\cos^2\zeta,\\
\nonumber
   c_{LR}=c_{RL}^\ast &=& -{g_L g_R\over 8 M_1^2} \left( 1-{M_1^2\over M_2^2}
   \right)\sin\zeta\cos\zeta e^{i\omega}.
\eea
In order to translate these effective couplings into the Michel parameters, one
must consider various scenarios for the neutrino sector. For example, assuming
massive Dirac neutrinos leads to,
\bea
\label{eq:rho}
   \rho & = & {3\over 4} {1+|\kappa_{RR}|^2{\tilde v_e}{\tilde v_\mu}\over
   1+|\kappa_{RR}|^2 {\tilde v_e}{\tilde v_\mu}+|\kappa_{RL}|^2({\tilde v_\mu}
   + {\tilde v_e})},\\
\label{eq:eta}
   \eta &=& 0, \\
\label{eq:xi}
   \xi &=& {1-|\kappa_{RR}|^2{\tilde v_e}{\tilde v_\mu}\over
   1+|\kappa_{RR}|^2 {\tilde v_e}{\tilde v_\mu}+|\kappa_{RL}|^2({\tilde v_\mu}
   +{\tilde v_e})},\\
\label{eq:delta}
   \delta &=& {1-|\kappa_{RR}|^2{\tilde v_e}{\tilde v_\mu}\over
   1-|\kappa_{RR}|^2 {\tilde v_e}{\tilde v_\mu}+3|\kappa_{RL}|^2({\tilde v_\mu}
   -{\tilde v_e})},
\eea
for the Michel parameters and,
\be
\label{eq:mupol}
   P_\mu = {|1-\kappa_{LR}\lambda e^{i\alpha}|^2 - |\kappa_{RL} - \kappa_{RR}
   \lambda e^{i\alpha}|^2{\tilde v_\mu} \over |1-\kappa_{LR}\lambda 
  e^{i\alpha}|^2+|\kappa_{RL}-\kappa_{RR}\lambda e^{i\alpha}|^2{\tilde v}_\mu},
\ee
for the polarization. The quantities ${\tilde v_\ell}$ are ratios of sums over
the mixing angles for the right- and left-handed neutrinos,  
\be
   {\tilde v} = {\sum_i^\prime\ |V_{li}|^2\over \sum_i^\prime\ |U_{li}|^2},
\ee
with the $U_{li}$ ($V_{li}$) being the analogs of the CKM matrix for 
left-handed (right-handed) Dirac neutrinos and the prime indicating that only
mass eigenstates produced in the decay are included, 
$\lambda=\cos\theta_1^R/\cos\theta_1^L$ is the ratio of (1,1) entries in 
the right- and left-handed CKM matrices, $\alpha$ is the CP violating phase in 
the right-handed CKM matrix, and $\kappa_{ij}=c_{ij}/c_{LL}$.

As Eqs.~(\ref{eq:mucouplings}-\ref{eq:mupol}) make evident, the Michel 
parameters depend in a complicated way on the couplings, mixing angles, and 
masses that appear in the simplest, but most general, left-right symmetric
model with massive Dirac neutrinos. Allowing for Majorana mass terms introduces
additional contributions to expression for the Michel spectrum. Thus, 
the analysis of the Michel spectrum must occur in the context of a more general
study of CC processes, including direct searches for a $W_R$ in collider 
experiments, light quark $\beta$ decay, neutrino oscillations, neutrinoless 
$\beta\beta$ decay, {\em etc.} Such a comprehensive study has yet to be 
performed.

Some simplifications occur by considering the combination of Michel parameters
\be
\label{eq:muR}
   R=1-{\delta\xi\over\rho} P_\mu.
\ee
As a practical matter, only the combination $\xi P_\mu$ can be accessed by 
experiment, making $R$ an appropriate quantity to constrain experimentally. 
Expanding the foregoing expressions for the Michel parameters and $P_\mu$ 
to second order in small quantities, one obtains~\cite{Herczeg:cx},
\be
  R \approx 2 t^2 {\tilde v_e}{\tilde v_\mu} + 2 t_\theta^2 {\tilde v}_\mu + 2
  \zeta_g^2{\tilde v_\mu} + 4t_\theta\zeta_g{\tilde v_\mu}\cos(\alpha_+\omega),
\label{eq:muRapprox}
\ee
where $t=(g_R/g_L)^2(M_1/M_2)^2$, $t_\theta=t\lambda$, and 
$\zeta_g=(g_R/g_L)\zeta$. Information on the combination of phases 
$\alpha+\omega$ can be derived from searches for the electric dipole moment of
the electron, neutron, or neutral atoms, which are sensitive to the combination
$\zeta_g\lambda\sin(\alpha+\omega)$, while neutron, pion, and nuclear 
$\beta$ decays provide independent constraints on $\zeta_g$. Studies of 
neutrino properties are clearly required for ${\tilde v_\ell}$. Collider 
experiments presently provide upper bounds on the mass ratios $M_1^2/M_2^2$, 
though the extraction of these bounds depends to some degree on assumptions 
about the right-handed CKM matrix and the relative strengths of the left- and 
right-handed couplings.

\subsection{\it Pion Decay}
The decay modes of the $\pi^\pm$ have long been a subject of study in EW 
physics. The dominant decay mode, $\pi^-\to \mu^ -{\bar\nu}_\mu $ provides 
a value for the pion decay constant, $F_\pi$, that encodes the effects of 
non-perturbative strong interactions involving the light quarks in the decay. 
Since these effects cannot be computed at present with high precision, 
the dominant decay mode does not provide a useful testing ground for the SM EW
interaction. However, the value of $F_\pi$ obtained from this decay plays 
an important role in the analysis of chiral dynamics in strong interactions. 
In contrast, a comparison of the rates
$\Gamma[\pi^+ \to \mu^- \bar\nu_\mu (\gamma)]$ and 
$\Gamma[\pi^- \to   e^- \bar\nu_e   (\gamma)]$ is insensitive to $F_\pi$ at 
leading order and can be used to study the underlying EW interaction.
Similarly, the pion $\beta$ decays $\pi^+ \to \pi^0 e^+{\nu}_e$
($\pi^- \to \pi^0 e^- \bar\nu_e$) and their radiative counterparts are also 
quite insensitive to strong interaction uncertainties, making them in principle
an interesting SM testing ground.

When extracting the value of $F_\pi$ from 
$\Gamma[\pi^- \to \mu^- \bar\nu_\mu (\gamma)]$, one must take into 
consideration EW radiative corrections. Doing so ensures that they are not 
inadvertently included ({\em via\/} $F_\pi$) in strong interaction processes 
such as $\pi$-$N$ scattering. In contrast to the situation with muon decay, 
however, the treatment of these radiative corrections is convention dependent 
and entails some degree of theoretical uncertainty. These features arise 
because some of the ${\cal O}(\alpha)$ contributions involve loops containing 
light quarks that interact non-perturbatively. The most widely used convention
for treating the radiative corrections has been given by Marciano and 
Sirlin~\cite{Marciano:1993sh}. Including all effects through order 
$G_\mu^2\alpha$ yields,
\begin{eqnarray}
\label{eq:piona}
  \Gamma[\pi^- \to \ell^- \bar\nu_\ell (\gamma)] = {G_\mu^2 |V_{ud}|^2 \over 
  4\pi} F_\pi^2 m_\pi m_\ell^2 \left[ 1 - {m_\ell^2\over m_\pi^2} \right]
  \left[ 1 + {2\alpha\over\pi}\ln\frac{M_Z}{\mu} \right] \\
\nonumber
  \times \left[ 1 - {\alpha\over\pi} \left \{ \frac{3}{2} \ln\frac{\mu}{m_\pi}
  + \bar{C}_1(\mu) + \bar{C}_2(\mu) \frac{m_\ell^2}{\Lambda_\chi^2}
  \ln\frac{\mu^2}{m_\ell^2} + \bar{C}_3(\mu) \frac{m_\ell^2}{\Lambda_\chi^2}
  + \cdots \right\} \right] \left[ 1 + \frac{\alpha}{\pi} F(x) \right],
\end{eqnarray}
where ${\bar C}_i$ are {\em a priori\/} unknown constants that parameterize 
presently incalculable non-perturbative QCD effects, $\Lambda_\chi =4\pi F_\pi$
is the chiral scale, the dots denote terms suppressed by additional powers of
the lepton mass square, $m_\ell^2/\Lambda_\chi^2$, and $x = m_\ell^2/m_\pi^2$. 

The function $F(x)$, along with the terms containing the $\bar{C}_i$, arise 
from QED corrections to the decay of a point-like pion. The first 
${\cal O}(\alpha)$ correction containing the $\ln M_Z$ is a short-distance 
contribution. Symmetry considerations protect part of this term from receiving 
any perturbative corrections, while another component receives corrections. 
In addition, Ref.~\cite{Marciano:1993sh} summed the contributions of the form 
$[(\alpha/\pi)\ln(M_Z/\mu)]^n$ for all $n$, using the RG to produce an improved
estimate of the short distance correction factor, $S_{EW}(\mu, M_Z)$ that 
replaces $1 + 2 (\alpha/\pi)\ln(M_Z/\mu)$. Choosing $\mu = m_\rho$, one has 
$S_{EW}(\mu, M_Z)=1.0232$.

More serious theoretical uncertainties arise from the terms proportional to 
the $\bar{C}_i$. In general, they depend on the choice of scale associated with
matching short- and long-distance contributions\footnote{Our conventions differ
slightly from those of Ref.~\cite{Marciano:1993sh}, since we have normalized 
the terms containing powers of $m_\ell^2$ to $\Lambda_\chi^2$ rather than to
$m_\rho^2$.}. In particular, the $\mu$ dependence of $\bar{C}_1(\mu)$ must 
cancel the $\mu$ dependence of the short-distance, RG-improved correction 
factor. The authors have estimated the uncertainty in $\bar{C}_1$ by varying 
$\mu$ from $m_\rho$ by a factor of two and requiring a corresponding variation
in $\bar{C}_1(\mu)$. Taking $\bar{C}_1(m_\rho) = 0$ they estimate
$\delta\bar{C}_1 = \pm 2.4$, corresponding to a $\pm 0.56\% $ correction to 
the rate. 

An estimate for $\bar{C}_2$ can be obtained using PCAC and the ratio of axial 
and vector form factors in radiative pion decay. The uncertainty associated 
with $\bar{C}_3$ should be small as its magnitude is $(\alpha/\pi)
(m_\mu^2/\Lambda_\chi^2)\bar{C}_3\approx 1.9\times 10^{-5}\bar{C}_3$.
Including these effects, using the latest value for the lifetime and branching
ratio~\cite{Hagiwara:fs},
\be
   \tau_{\pi^\pm} = (2.6033 \pm 0.0005) \times 10^{-8}~s, \hspace{50pt}
{\Gamma(\pi^+\to\mu^+\nu_\mu)\over\Gamma_{\rm tot}}=(99.98770\pm 0.00004)\%,
\ee
and the values of $G_\mu$ and $|V_{ud}|$ from muon and super-allowed, 
nuclear $\beta$ decays (see below), one obtains,
\begin{equation}
   F_\pi = 92.4 \pm 0.07\pm 0.25~{\rm MeV},
\end{equation}
where the first error is from the experimental uncertainty in $|V_{ud}|$ and 
the second is associated with $\bar{C}_1$.

In contrast to $\Gamma[\pi^-\to \ell^- \bar\nu_\ell (\gamma)]$, the ratio 
$R_{e/\mu}$ of electronic to muonic widths is fairly insensitive to strong 
interaction uncertainties and can provide an interpretable test of 
the electron-muon universality of the SM. In particular,
\begin{eqnarray}
   R_{e/\mu} = {\Gamma[\pi^-\to e^- \bar\nu_e (\gamma)]\over 
                \Gamma[\pi^-\to \mu^- \bar\nu_\mu (\gamma)]} = 
   \frac{m_e^2}{m_\mu^2} \left[ {m_\pi^2-m_e^2 \over m_\pi^2-m_\mu^2} \right]^2
   \left\{ 1 +\frac{\alpha}{\pi} \left[ F({m_e\over m_\pi}) - 
   F({m_\mu\over m_\pi}) + \frac{m_\mu^2}{\Lambda_\chi^2} ( \bar{C}_2 
   \ln {m_\mu^2\over \Lambda_\chi^2} + \bar{C}_3) \right] \right\},
\end{eqnarray}
where terms proportional to $\alpha m_e^2$ are negligible and have been 
dropped. After including structure dependent bremsstrahlung corrections and 
re-summing terms of the form $[\alpha/\pi\ln(m_e/m_\mu)]^n$ that arise in 
the difference $F(m_e/m_\pi)-F(m_\mu/m_\pi)$ one 
obtains~\cite{Marciano:1993sh}, 
\begin{equation}
   R_{e/\mu}^{\rm SM} = (1.2352\pm 0.0005)\times 10^{-4},
\end{equation}
where the error originates dominantly from the structure dependent 
bremsstrahlung contributions. 

Precise determinations of $R_{e/\mu}$ have been carried out at 
PSI~\cite{Czapek:kc} and TRIUMF~\cite{Britton:1992pg}, averaging to,
\begin{eqnarray}
\label{eq:remuresult}
   {R_{e/\mu}^{\rm exp}\over R_{e/\mu}^{\rm SM}}=0.9966\pm 0.0030\pm 0.0004,
\end{eqnarray}
where the first error is experimental and the second is the estimated 
theoretical uncertainty. At present, one has no indication of any disagreement
with the SM, though the ratio~(\ref{eq:remuresult}) does provide rather stringent 
constraints on possible universality violating, NP contributions. For example,
the exchange of squarks (${\tilde q}$) in RPV SUSY may lead to a non-universal
contribution~\cite{Ramsey-Musolf:2000qn,Barger:1989rk},
\begin{equation}
   {R_{e/\mu}\over R_{e/\mu}^{\rm SM}} = 1 + 2\left[ \Delta^\prime_{11k}
   ({\tilde d}_R^k) - \Delta^\prime_{21k} ({\tilde q}_R^k) \right],
\end{equation}
where $\Delta^\prime_{11k}({\tilde d}_R^k) = 
|\lambda^\prime_{11k}|^2/4\sqrt{2}G_F M^2_{{\tilde d}_R^k}$, {\em etc.}, with
$\lambda^\prime_{ijk}$ denoting the RPV coupling of various generations of 
squarks with masses $M_{{\tilde q}_k}$. The experimental agreement with the SM
prediction illustrated by the ratio~(\ref{eq:remuresult}) places substantial 
constraints on the possible size of RPV effects in other low energy weak 
processes that also depend on 
the $\Delta^{\prime}_{ijk}$~\cite{Kurylov:2003by,Ramsey-Musolf:2000qn}.

\begin{figure}[t]
\begin{center}
\begin{minipage}[t]{12 cm}
\epsfig{file=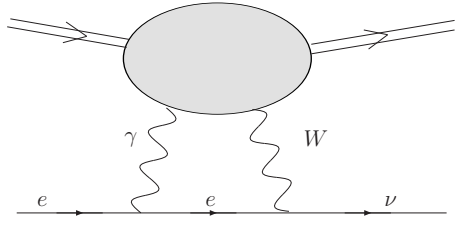,scale=0.60}
\end{minipage}
\vspace*{-310pt}
\begin{minipage}[t]{16.5 cm}
\vspace*{-370pt}
\caption{The ${\cal O}(\alpha)$ box graph correction to pion, neutron, and 
nuclear $\beta$ decay generated by the exchange of one $W$ boson and 
one $\gamma$. The double lines indicate initial and final hadronic states, 
respectively.
\label{diagram}}
\end{minipage}
\end{center}
\end{figure}

In addition, CC universality tests can be performed by the study of pion 
$\beta$ decays. In this case, the transition involves hadronic matrix elements
of the charged vector current, so the theoretical prediction for the rate is 
protected from large and theoretically uncertain strong interaction corrections
of the type encoded by $F_\pi$. However, some uncertainties do appear at 
${\cal O}(\alpha)$ due to the diagrams of the type in Figure~\ref{diagram}. 
The benchmark study of the SM, EW radiative corrections to pion $\beta$ decay 
was performed by Sirlin~\cite{Sirlin:1977sv}, who used current algebra 
techniques. A more recent analysis within the context of chiral perturbation 
theory ($\chi$PT) has been carried out by Cirigliano 
\etal~\cite{Cirigliano:2002ng}. The latter analysis also includes 
the corrections due to pion form factors that were not included in the earlier
studies. 

Experimentally, the study of pion $\beta$ decay is a challenging enterprise, 
given that the branching ratio for this decay mode is of the order 
$\sim 10^{-8}$. The most accurate result has been published in 
Ref.~\cite{McFarlane:1985}, leading to the Particle Data Group 
value~\cite{Hagiwara:fs},
\begin{equation}
  {\cal B}(\pi^+\to\pi^0 e^+\nu_e) = (1.025\pm 0.034) \times 10^{-8}.
\end{equation}
After inclusion of EW radiative corrections~\cite{Cirigliano:2002ng}, one can 
extract,
\begin{equation}
   |V_{ud}|^2 = 0.9675\pm 0.0160\pm 0.0005,
\end{equation}
where the first error is experimental and the second theoretical. The latter is
dominated by extrapolation of the pion form factor $f_+$ to $Q^2=0$ and is
negligible compared to the experimental uncertainty.

A reduction in the experimental error bar is a primary goal of the PIBETA
experiment~\cite{Pocanic:2003jp} underway at PSI. The experiment relies on
a 113~MeV beam of $\pi^+$ stopped in a plastic scintillator target. Energy from
the decay products is deposited in a CsI shower calorimeter. The ultimate goal
of the measurement is to determine ${\cal B}(\pi^+\to\pi^0 e^+\nu_e)$ to 0.5\%
precision, representing a factor of 7 improvement over the present world
average. A new result with 0.8\% uncertainty, based on a partial data set, has
recently been reported by the Collaboration~\cite{Pocanic:2003pf}.

The PIBETA Collaboration is also carrying out a study of radiative pion
decay (RPD). The branching ratio, ${\cal B}(\pi^+\to e^+\nu_e\gamma)$, is
sensitive to the ratio of axial vector and vector pion form factors, $F_A$ and 
$F_V$, respectively, defined as~\cite{Poblaguev:1990tv,Herczeg:ur},
\begin{eqnarray}
   \langle\gamma(q,\varepsilon)|\bar{d}\gamma_\lambda u|\pi^+(p)\rangle &=&
   e F_V\epsilon_{\mu\lambda\rho\sigma}\varepsilon^\mu
   {p^\rho q^\sigma\over m_\pi}, \\ \langle\gamma(q,\varepsilon)
   |\bar{d}\gamma_\lambda\gamma_5 u|\pi^+(p)\rangle &=& ieF_A
   \varepsilon^\mu{p\cdot q g_{\mu\lambda}-p_\mu q_\lambda\over m_\pi},
\end{eqnarray} 
where $p$ is the pion momentum, and $q$ and $\varepsilon$ are the photon
momentum and polarization, respectively. Present constraints on RPD
give~\cite{Hagiwara:fs}, $F_V=0.0259\pm 0.0005$, and $F_A\approx 0.5 F_V$.
Recently, the PIBETA Collaboration has performed a precise determination of
the ratio, $\gamma=F_A/F_V$, yielding~\cite{Frlez:2003pe},
\begin{equation}
   \gamma = 0.443 \pm 0.014.
\end{equation}
The presence of only these two form factors ($F_{V,A}$) implies a spectral
shape for the RPD as a function of 
$(2E_{e^+}/m_{\pi^+})\sin^2(\theta_{e\gamma}/2)$. An analysis of the PIBETA 
data set indicates a departure from this expectation for hard photons and soft
positrons~\cite{Frlez:2003pe}. If one ultimately does not find a refined SM 
prediction that is consistent with the observed spectral shape, the explanation
could lie in the presence of a non-vanishing tensor form 
factor~\cite{Herczeg:ur} arising from a lepton-quark interaction producing 
an amplitude,
\begin{equation}
   {\cal M}(\pi^+\to e^+\nu_e\gamma) = {G_\mu |V_{ud}|\over\sqrt{2}} i e F_T
   \varepsilon^\lambda q^\mu \bar{e}\sigma_{\lambda\mu}(1-\gamma_5) \nu_e.
\end{equation}
Including such a non-vanishing tensor amplitude with $F_T \approx 0.0016$ 
appears to produce better agreement with the preliminary experimental PIBETA 
spectrum~\cite{Pocanic:2003jp}. While the value of $F_T$ vanishes at tree level
in the SM, one can presently not exclude an induced $F_T$ arising from 
radiative corrections. Alternatively, this type of term may arise in various 
leptoquark scenarios or from sfermion exchange in RPV SUSY.

\subsection{\it Beta Decay, Kaon Decay, and CKM Unitarity}
\label{CKM}
As illustrated in the previous Subsection, one of the arenas in which 
the predictions of the SM have been tested most 
precisely at low energies is in the weak decays of light quarks. Of particular
interest is the first row of the CKM matrix and the SM unitarity requirement,
\be
\label{eq:ckm1}
   |V_{ud}|^2 + |V_{us}|^2 + |V_{ub}|^2 = 1.
\ee
The largest contributor to the unitarity sum is $|V_{ud}|$, which can be 
determined from the $\beta$ decay of the pion, neutron, or nuclei. The current
experimental work on pion $\beta$ decay has been discussed above, so we 
concentrate here on neutron and nuclear $\beta$ decays. In both cases, 
the relevant quantity that characterizes the total decay rate is the so-called 
$ft$ value,
\be
\label{eq:ft1}
  ft = {K\over (G_F^\beta)^2\langle M_V\rangle^2 + 
       (G_A^\beta)^2\langle M_A \rangle^2},
\ee
where $t$ is the half-life, $f$ is a factor that corrects for the outgoing 
$\beta$ particle wave-function, $G_A = G_\mu V_{ud} g_A$ with $g_A$ being
the axial vector coupling of the nucleon to the $W^\pm$ boson, $M_V$ ($M_A$) 
are vector (axial vector) matrix elements, and $K$ is the combination of
constants,
\be
   K=\hbar (2\pi^3\ln2 )(\hbar c)^6 /(m_e c^2)^5.
\ee
The quantity $G_F^\beta$ can be expressed in terms of $G_\mu$, $V_{ud}$, and 
radiative corrections to the $\beta$ decay ($\Delta r_\beta$) and $\mu$ decay 
($\Delta r_\mu$) amplitudes,
\be
\label{eq:ckm2}
   {G_F^\beta\over G_\mu}=V_{ud}(1+\Delta r_\beta-\Delta r_\mu).
\ee
The quantity $\Delta r_\beta$ is sensitive to various non-perturbative strong 
interaction effects that contribute to the EW radiative corrections. The matrix
elements $M_{V,A}$ are similarly dependent on hadronic structure. For neutron 
decay, for example, one has $\langle M_V\rangle^2=1$ and 
$\langle M_A\rangle^2=3$.

The neutron lifetime, $\tau_n$, depends on both $G_F^\beta$ and $G_A^\beta$ and
both of these quantities are proportional to $V_{ud}$. Unlike $G_F^\beta$, 
which arises from the vector coupling of the $W^\pm$ boson to light quarks, 
$G_A^\beta$ arises from the axial vector charge changing quark current and, 
thus, is susceptible to significant strong interaction renormalization. 
The quantity $g_A =-1.2670\pm 0.0030$ encodes this renormalization. At present,
one cannot compute it directly from QCD with the precision needed to perform 
a significant determination of $V_{ud}$ from $\tau_n$. Instead, an experimental
separation of $G_F^\beta$ and $G_A^\beta$ is needed. In practice, such 
a separation can be obtained by combining a determination of $\tau_n$ with
a measurement of any one of several $\beta$ decay correlation coefficients.
The latter appear in the partial rate~\cite{Jackson:1957},
\be
\label{eq:betacor}
   d\Gamma\propto N(E_e)\left[ 1+a {{\vec p}_e\cdot{\vec p}_\nu\over E_e E_\nu}
   + b{m_e\over E_e} + A{{\vec p}_e\cdot\langle {\vec\sigma}_n\rangle\over E_e}
   + B{{\vec p}_\nu\cdot \langle {\vec\sigma}_n\rangle\over E_\nu} + D \langle
   {\vec\sigma}_n\rangle{{\vec p}_e\times {\vec p}_\nu \over E_e E_\nu}\right]
   d\Omega_e d\Omega_\nu d E_e,
\ee
where $N(E_e)=p_e E_e(E_0-E_e)^2$, $E_e$ ($E_\nu$) and ${\vec p}_e$ 
(${\vec p}_\nu$) are the $\beta$ (neutrino) energy and momentum, respectively,
and ${\vec\sigma}_n$ is the neutron polarization. The coefficients $a$, $A$, 
and $B$ can be expressed in terms of the ratio, 
$\lambda = {G_A^\beta/G_F^\beta}$, as, 
\be
\label{eq:corcoeff}
a = {1-\lambda^2\over 1+3\lambda^2}, \hspace{80pt}
A = -2{\lambda(1+\lambda)\over 1+3\lambda^2}, \hspace{80pt}
B =  2{\lambda(\lambda - 1)\over 1+3\lambda^2}.
\ee
The quantity $b$ appearing in the so-called Fierz interference term is zero for
purely vector and axial vector interactions, while the triple correlation
parameterized by $D$ is time-reversal odd. A non-zero value for the latter can
be induced by electromagnetic final state interactions between the outgoing 
$\beta$ particle and the proton or by possible new T-odd, parity conserving 
interactions. A search for the latter {\em via\/} a measurement of the $D$ 
coefficient has been completed by the emiT Collaboration~\cite{Lising:2000pa},
who obtained the null result,
\be
\label{eq:emit}
   D = [-0.6\pm 1.2 {\rm~(stat.)} \pm 0.5 {\rm~(syst.)}] \times 10^{-3}.
\ee

The neutron lifetime has been obtained from both neutron beam experiments and 
stored ultra-cold neutrons (UCNs). An average of the five most recent 
experiments yields the current PDG value, $\tau_n = 885.7\pm 0.7$~s.  A more 
precise measurement of $\tau_n$ is being pursued at NIST using magnetically 
trapped ultra-cold neutrons~\cite{Huffman:2002bb}. Historically, determinations
of $\lambda$ from correlation measurements have proved more challenging due to
a host of systematic effects. The quoted uncertainty for each of the four most 
recent cold neutron measurements of $A$ is roughly one percent. However, 
the central values form the different measurements are in rather poor 
agreement, leading to the PDG average, $A = -0.1162\pm 0.0013$. When combined 
with the current uncertainty in $\tau_n$, this average implies a sizable 
uncertainty in $V_{ud}$ obtained from neutron decay. 

Recent technological advances involving cold and ultra-cold neutron methods 
should lead to improved determinations of $\lambda$. The PERKEO Collaboration 
has performed a determination of $A$ with $\sim 0.6\%$ uncertainty using cold 
neutrons at the ILL reactor, leading to 
$\lambda = -1.2739\pm 0.0019$~\cite{Abele:2002wc}. In contrast to previous
determinations of $A$, which entailed application of large, ${\cal O}(25\%)$, 
corrections to the raw data, the subtractions applied in the latest experiment
are roughly ten times smaller than the quoted uncertainty. When combined with 
the current world average for $\tau_n$, the result implies 
$|V_{ud}| = 0.9713\pm 0.0013$. Using the current PDG values for $|V_{us}|$ (see
below) and $|V_{ub}|$ one obtains,
\be
\label{eq:perkeo}
   |V_{ud}|^2 + |V_{us}|^2 + |V_{ub}|^2 = 0.9917\pm 0.0028,
\ee
indicating a deviation from unitarity by more than three standard deviations.

A new measurement of $A$ presently underway at LANSCE seeks to reduce 
the uncertainty in $\lambda$ to the $0.1\%$ level~\cite{Saunders:2003jg}.
The experiment relies on 
a solid D$_2$ moderator and the LANSCE beam of cold neutrons to create 
a source of ultra cold, polarized neutrons. Cold neutrons ($T < 100$~K) are 
first produced using conventional cold moderators and subsequently cooled to 
$T < 4$~mK {\em via\/} phonon interactions in D$_2$. The experiment has thus 
far produced the world's record density of stored ultra cold neutrons 
(140~UCN/cm$^3$). The advantage of UCNs for a measurement of $A$ is two-fold: 
(a) the neutrons can be stored in bottles for a duration comparable to $\tau_n$
and (b) a polarization of $P \sim 100\%$ can be achieved.   

A variety of future possibilities for both more precise $\tau_n$ determinations
as well as correlation coefficient measurements have been discussed. The use of
magnetically trapped UCNs in super-fluid $^4$He has been demonstrated at NIST, 
and a $\tau_n$ measurement using this method with $\pm 1$~s uncertainty could 
be carried out. The construction of an UCN beam line for fundamental neutron 
physics at the Spallation Neutron Source (SNS) at Oak Ridge National Laboratory
would provide the capability for further improvements in precision. The present
uncertainty in $\tau_n$ contributes an error of $\sim 0.05\%$ to the value of 
$|V_{ud}|$ obtained from neutron decay, and as the precision on $\lambda$ is 
increased with new UCN measurements, corresponding reductions in the error on 
$\tau_n$ will become important. A comprehensive determination of the complete 
set of neutron decay correlation coefficients using a beam of pulsed cold 
neutrons at the SNS has also been considered. Such an experiment would allow 
a redundant determination of $\lambda$, with multiple cross checks on 
systematic effects, thereby providing additional confidence in the value of 
this quantity. For a recent theoretical study of neutron decay correlation 
coefficients, see Ref.~\cite{Gardner:2000nk}. For other theoretical 
considerations, see Ref.~\cite{Ando:2004rk}.

To date, the most precise determination of $|V_{ud}|$ has been obtained from 
analysis of \lq\lq super-allowed" Fermi nuclear $\beta$ decays. These decays 
involve transitions from spin-parity $J^\pi=0^+$ initial states to $0^+$ 
daughter nuclei and, thus, only matrix elements of the vector part of 
the charge changing weak current may contribute. In the limit of zero momentum
transfer squared, the latter is simply the isospin raising or lowering 
operator, and to the extent that the states involved in the transition are part
of an isospin multiplet, the matrix element $\langle M_V \rangle$ of the time
component of the charge changing vector current is
\be
  \langle I,I_Z\pm 1|J_0^\pm |I, I_Z 1\rangle = [(I\mp I_Z)(I\pm I_Z+1)]^{1/2},
\ee
independently of nuclear structure. Small deviations from this exact isospin
symmetric result are induced by configuration mixing as well as small 
differences in the neutron and proton radial wave-functions in the initial and 
final nuclear states. The corresponding corrections, along with those due to 
nuclear structure dependent effects in EW radiative corrections, must be 
computed and applied to the $ft$ values before a nucleus independent value of 
$G_F^\beta$ is obtained. The resulting expression for the corrected $ft$ 
values, ${\cal F}t$, is given by,
\be
\label{eq:super1}
   {\cal F}t (1+\delta_R)(1-\delta_C) = {K\over 2 (G_F^\beta)^2},
\ee
where $\delta_R$ is a nucleus dependent component of the ${\cal O}(\alpha)$ EW
radiative corrections, and $\delta_C$ is a nuclear structure correction that 
enters because the initial and final nuclear states are not perfect isospin 
multiplet partners. The latter correction accounts for two effects, 
configuration mixing in the nuclear states and small differences in the initial
and final radial wave-functions. 

An important test of the nuclear structure calculations required for $\delta_R$
and $\delta_C$ is a comparison of the ${\cal F}t$ values from different decays,
which should agree according to the approximate conserved vector current (CVC)
property of the SM. The ${\cal F}t$ values for nine different super-allowed 
transitions have been measured with a precision of $0.1\%$ or better (for 
recent reviews, see Refs.~\citer{Towner:1995za,Hardy:ci}). 
A fit~\cite{Towner:1998qj} yields ${\cal F}t = 3072.2\pm 0.8$, indicating 
agreement with CVC at the level of a few parts in $10^4$. The largest 
contributors to the overall uncertainty are the estimated uncertainties in 
$\delta _R$ and $\delta_C$.

In order to extract a value of $|V_{ud}|$ from the average ${\cal F}t$ values, 
the calculated correction, $\Delta r_\beta - \Delta r_\mu$, must also be
subtracted, as indicated in Eq.~(\ref{eq:ckm2}). In particular, the correction 
$\Delta r_\beta$ is sensitive to non-perturbative QCD effects, introducing an
additional source of uncertainty beyond those appearing in the average 
${\cal F}t$ value. This uncertainty arises primarily from the box diagram of 
Figure~\ref{diagram} involving the exchange of one $W^\pm$ boson and one 
$\gamma$ that yields the amplitude,
\be
\label{eq:wgbox}
   {\cal M}_{W\gamma} = {G_\mu\over \sqrt{2}}{{\hat\alpha}\over 2\pi} \left[
   \ln\left( {M_W^2\over\Lambda^2} \right) + C_{\gamma W}(\Lambda)\right],
\ee
where the leading logarithmic term can be computed reliably in the SM and 
the constant $C_{\gamma W}(\Lambda)$ parameterizes contributions to the loop 
integral below a scale $\Lambda$. An estimate of $C_{\gamma W}(\Lambda)$ was
given by Sirlin~\cite{Sirlin:1977sv} using nucleon intermediate states in 
the box diagram, and this estimate has been retained by subsequent authors. 
To estimate the corresponding uncertainty, Ref.~\cite{Towner:1995za} used 
the response of the logarithmic term in Eq.~(\ref{eq:wgbox}) when $\Lambda$ is
varied between 400 and 1600~MeV. One obtains~\cite{Hardy:ci},
\be
\label{eq:vud}
   |V_{ud}| = 0.9740 \pm 0.0005,
\ee
where the error is dominated by the estimated uncertainty in $\Delta r_\beta$ 
($\pm 0.0004$). When the results from neutron decay are averaged with 
the result~(\ref{eq:vud}), one obtains the PDG~\cite{Hagiwara:fs} 
value\footnote{The Particle Data Group has augmented the uncertainty 
in~(\ref{eq:vud}) by a factor of two to account for a possible $Z$ dependence
of the super-allowed ${\cal F}t$ values, although there exists no statistically
significant evidence for such an effect.},
\be
\label{eq:vudpdg}
   |V_{ud}| = 0.9734 \pm 0.0008.
\ee
Obtaining a more refined, first principles computation of 
$C_{\gamma W}(\Lambda)$ and the attendant theoretical uncertainty remains 
an open theoretical problem. 

One aim of on-going work in the arena of super-allowed decays is to provide 
further tests of the nuclear corrections $\delta_R$ and $\delta_C$. To that 
end, a comprehensive calculation of these corrections for medium- and 
heavy-mass nuclei has been carried out~\cite{Towner:2002rg}. The computations 
include both the measured cases as well as new transitions whose measurement 
could provide a test of these calculations. A program to determine the $ft$ 
values for a number of these new cases with $18 \leq A\leq 42$ is underway at 
Texas A \& M University and TRIUMF~\cite{Hardy:ci}. Studies of heavier nuclei, 
such as $^{74}$Rb or $^{62}$Gd present new experimental and theoretical 
challenges and may prove problematic for testing the reliability of $\delta_R$
and $\delta_C$ computations for the known cases. These challenges include 
the rapid shape changes of nuclei in the $A\geq 62$ region, the short 
half-lives of the parent nuclei, and the plethora of $1^+$ daughter states
whose branching fractions must be measured accurately in order to extract 
${\cal B}(0^+\to 0^+)$~\cite{Hardy:ci}.

To provide a test of the unitarity requirement in Eq.~(\ref{eq:ckm1}), one must
have in hand reliable values for $|V_{us}|$ and $|V_{ub}|$. The value of 
$|V_{ub}| = 0.0032\pm 0.0009$ is sufficiently small that it may be neglected 
for this purpose. In contrast, the uncertainty in the currently accepted value 
of $|V_{us}|$~\cite{Hagiwara:fs},
\be
\label{eq:vuspdg}
   |V_{us}| =0.2196\pm 0.0026, 
\ee
extracted from kaon leptonic decays ($K_{e3}$) has a similar impact on 
the unitarity test as does the uncertainty in $|V_{ud}|$. Using 
the values~(\ref{eq:vudpdg},\ref{eq:vuspdg}) one obtains for the first row of 
the CKM matrix,
\be
\label{eq:unitaritysum}
   \sum_{j=d,s,b}\ |V_{uj}|^2 = 0.9958 \pm 0.0019,
\ee
indicating a $2.2\sigma$ deviation from unitarity.

Recent experimental and theoretical analyses of $|V_{us}|$ have raised 
questions about both the central value and quoted error in this quantity. A new
analysis of $\sim 70,000$ charged kaon decay events has been carried out by 
the Brookhaven E865 Collaboration~\cite{Sher:2003}, leading to 
$|V_{us}| = 0.2272\pm 0.0023\pm 0.0019$. The first error is the combined 
statistical and systematic error in the partial width $d\Gamma(K^+_{e3})$ while
the second error is associated with the transition form factor, $f_{+}(t)$, 
where $t = (p_K - p_\pi)^2$. Both the value of $f_{+}(0)$ and its slope at 
the photon point, $t=0$, are needed to extract $|V_{us}|$ from the partial 
width,
\be
\label{eq:ke3partial}
   d\Gamma(K^+_{e3}) = C(t) |V_{us}|^2 |f_{+}(0)|^2 \left[ 1 +
   \lambda_{+}{t\over m_\pi^2}\right]^2,
\ee  
where $C(t)$ is a known function and where QED radiative corrections (not shown
here) must also be included in order to determine $|V_{us}|$ to the level of 
precision needed~\cite{Cirigliano:2001mk}. As can be seen, the slope of 
$f_{+}(t)$ at the photon point is characterized by $\lambda_{+}$. 

Independent experimental determinations of $\lambda_+$ have been obtained from
studies of $K_{\ell 3}$ decays~\cite{Hagiwara:fs}. The value of $f_{+}(0)$ 
requires theoretical input. A determination to ${\cal O}(p^4)$ in $\chi$PT was
carried out in Refs.~\cite{Leutwyler:1984je,Gasser:1984ux}, while an estimate 
of the ${\cal O}(p^6)$ terms was obtained using the quark model. 
The uncertainty associated with the latter was estimated to be roughly 1\%. 
More recent computations have explicitly evaluated the ${\cal O}(p^6)$
contributions~\cite{Post:2001si,Bijnens:2003uy}. While the one- and two-loop 
contributions can be evaluated using the known low energy constants (LECs) 
through ${\cal O}(p^4)$, the contributions from the ${\cal O}(p^6)$ LECs cannot
be determined with sufficient precision from existing data. In particular, 
the ${\cal O}(p^6)$ contribution to $f_{+}(0)$ contains~\cite{Bijnens:2003uy},
\be
\label{eq:fplus6}
   f_{+}^{(6)} = 
   -8{(m_\pi^2-m_K^2)^2\over F_\pi^4}\left(C_{12}^r+C_{34}^r\right)+\cdots,
\ee
where the $C_{i}^r$, $i=12,34$, are two of the 94 ${\cal O}(p^6)$ LECs. 

Naive dimensional arguments, as well as those invoking resonance saturation of
relevant pseudoscalar form factors suggest that the impact of the $C_i^r$ on 
the extracted value of $|V_{us}|$ could be substantially larger than the errors
quoted for either the PDG value or the recent Brookhaven result. 
New measurements of pion and kaon form factors, however, could reduce 
the uncertainty in $|V_{us}|$ from the $C_i^r$ to below one percent. As 
demonstrated in Ref.~\cite{Bijnens:2003uy}, a precise measurement of the pion
scalar form factor would allow a sufficiently precise determination of 
$C_{12}^r$, while new determinations of $\lambda_0$, the slope parameter in 
the kaon scalar form factor, using $K_{\mu 3}$ decays, would provide a value 
for $2C_{12}^r + C_{34}^r$. Together, these experimental inputs would yield
the combination, $C_{12}^r + C_{34}^r$, needed for $f_{+}(0)$.

Experimental work in this direction is underway at several facilities. 
Of particular interest are kaon decay branching ratio measurements being
performed by the KLOE experiment~\cite{Moulson:2003zu} at DA$\Phi$NE. 
Preliminary results for the branching ratios of the $K_{S,L}\to \pi^-e^+\nu_e$
and $K_L\to \pi^-\mu^+\nu_\mu$ channels have been reported in 
Ref.~\cite{Moulson:2003zu}. To the extent that isospin is a good symmetry, 
the form factors $f_{+}^{K^0\pi^-}(t)$ and $f_{+}^{K^+\pi^0}(t)$ should be 
identical at $t=0$. The isospin corrections have been computed in $\chi$PT to 
${\cal O}(p^4)$ and found to be $\sim 2\%$~\cite{Gasser:1984ux}. A comparison 
of the product $|V_{us}|f_{+}^{K^0\pi^-}(0)$ for the preliminary KLOE results 
of the three different neutral kaon branching ratios with the previously 
obtained PDG world averages indicates good agreement. The results differ
substantially, however, with the E865 result for $|V_{us}|f_{+}^{K^+\pi^0}(0)$,
which also differs from the previous world average for this quantity. 
The situation may be clarified by future KLOE results for the $K^+_{e3}$ 
branching ratio as well as studies planned by the NA48~\cite{Madigozhin:2003ri}
and KTeV~\cite{Vincenzo:2004} Collaborations. Given the present disagreement,
as well as the theoretical uncertainties associated with the ${\cal O}(p^6)$ 
LECs, it is probably too soon to conclude that the disagreement with CKM 
unitarity has been resolved by a new value of $|V_{us}|$.

Should the present unitarity disagreement persist, then one could draw 
interesting conclusions about NP. In the context of gauge unification, a small 
degree of mixing between left- and right-handed gauge bosons could restore 
the required unitarity. Because $G_F^\beta$ from which $V_{ud}$ is derived 
parameterizes a vector coupling of the $W^\pm$ to quarks, it is insensitive to 
the degree to which the boson is a mixture of $W_L^\pm$ and $W_R^\pm$. In 
contrast, $G_F^\beta$ does depend on the chirality of the outgoing leptons. 
Because the effect of a right-handed component to the neutrino is highly 
suppressed, the coupling of the lightest $W^\pm$ mass eigenstate to 
the left-handed leptons will differ from the $W_L^\pm$ coupling by an amount 
proportional to the left-right mixing angle, $\xi$. Thus, the present unitarity 
disagreement would imply a non-zero value for $\xi$ at the 95\%~CL. Such a result 
would have implications for the interpretation of other precision measurements, 
such as neutrinoless double $\beta$ decay (see Section~\ref{0nu2betadecay}).

First row CKM unitarity has also important implications for SUSY. The results 
of a recent analysis of MSSM radiative corrections to various CC 
observables~\cite{Kurylov:2001zx} indicate that inclusion of these corrections
typically exacerbates the disagreement with the unitarity requirement when 
the SUSY breaking parameters are chosen in 
accord with the most common models of SUSY breaking mediation. The region of 
parameter space favored by weak decays, $(g-2)_\mu$, the current values of 
$M_W$, $m_t$, and the lightest Higgs mass along with theoretical considerations
such as color neutrality of the vacuum implies that the mass of the left-handed
muon superpartner is heavier than that of the first generation scalar quarks, 
in contrast to expectations based on gravity- and gauge-mediated SUSY breaking.
Precision CC data can accommodate SUSY radiative corrections under 
the assumptions of these models if the requirement of $R_P$ conservation is
relaxed~\cite{Kurylov:2003by,Ramsey-Musolf:2000qn,Kurylov:2003zh}. This
would not be attractive from the standpoint of cosmology, since the RPV
interactions that contribute to weak decays would also mediate the decay of 
the $\chi^0$, ruling it out as a candidate for cold dark matter. A test of this
scenario could be performed through a comparison of PV electron scattering 
experiments~\cite{Kurylov:2003zh,Kurylov:2003xa}.

\section{Rare and Forbidden Processes}
\subsection{\it Electric Dipole Moments}
\label{CPandT}
Processes that violate CP and T have been an important arena of study for SM 
physics. Within the SM, the CP violation observed in the decays of neutral 
kaons is accommodated by the presence of the CP violating phase in the CKM 
matrix. Recently, experimental results from the NA48~\cite{Batley:2002gn} and 
KTeV~\cite{Alavi-Harati:2002ye} Collaborations have confirmed the SM prediction
for the ratio of $\epsilon^\prime/\epsilon$ that expresses the relative 
strength of direct CP violation in the decay amplitudes compared to that 
associated with $K^0$-${\bar K}^0$ mixing. With a few possible expceptions 
under study, CP violation in the $b$ quark system is also in agreement with 
the SM. Thus, there appears to be little evidence at present for new sources of
CP violation in these channels.

Searches for CP and T violation in light quark and lepton systems have, to 
date, produced null results. The most powerful probes of this type are searches
for the permanent electric dipole moments (EDMs) of leptons, neutrons, and 
neutral atoms. EDM searches are of interest for several reasons:
\begin{itemize}
\item 
The SM (CKM) predictions for the magnitudes of EDMs are suppressed, falling 
well below the sensitivity of present and prospective measurements. 
Consequently, the observation of a non-zero EDM could signal the presence of 
physics beyond the SM or CP violation in the $SU(3)_C$ sector of the SM. 
The latter arises {\em via\/} a term in the Lagrangian~\cite{Wilczek:pj},
\be
\label{eq:thetacp}
   {\cal L}_{\rm strong\ CP} = \theta_{\rm QCD} \frac{\alpha_s}{8\pi}
   G_{\mu\nu}{\tilde G^{\mu\nu}},
\ee
where $G_{\mu\nu}$ (${\tilde G}_{\mu\nu}$) is the (dual) $SU(3)_C$ field 
strength tensor.
\item
The observed predominance of matter over anti-matter in the universe --- 
the so-called baryon asymmetry of the universe (BAU) --- conflicts with 
expectations based on the SM alone. In particular, the strength of CP violating
effects needed to preserve the matter-antimatter asymmetry during the evolution
of the universe is suppressed in the SM by the Jarlskog 
invariant~\cite{Jarlskog:1985ht}, 
\be
   J = \cos\theta_1\cos\theta_2\cos\theta_3\sin\theta_1^2\sin\theta_2
   \sin\theta_3\sin\delta,
\ee
with the $\theta_i$ and $\delta$ being the angles in the CKM matrix, and by 
light quark masses, rendering a BAU that is far smaller than 
observed\footnote{The SM also does not produce a sufficiently strong first 
order phase transition needed for the BAU.}. On the other hand, candidate 
extensions of the SM that could provide new CP violation of sufficient strength
to accommodate the BAU could also generate EDMs large enough to be seen
experimentally.
\end{itemize}
The literature on EDMs is vast, so we make no attempt to provide an exhaustive
review here (for recent reviews, see Refs.~\citer{Fortson:fi,Dmitriev:2003hs}).
Instead, we highlight the most important experimental developments and 
theoretical issues for the field, and point the reader to other studies for 
more comprehensive reviews. From our standpoint, three aspects of the EDM 
program merit emphasis:
\begin{itemize}
\item
Recent experimental developments have put the field on the verge of 
a revolution. Experimental searches for the electron, muon, neutron, and atomic
EDMs are poised to improve experimental sensitivity by factors of 100 to 10,000
during the next decade. This kind of across-the-board improvement in precision 
by orders of magnitude has never before been seen in the field.
\item 
The lepton, neutron, and atomic EDM searches provide complementary probes of 
new CP violation, as different candidate theories imply different signatures 
for the various moments. For example, the observation of a non-zero neutron or
atomic EDM in conjunction with a null result for the electron EDM at 
a comparable level of sensitivity would point toward the interaction of 
Eq.~(\ref{eq:thetacp}) as the likely source of CP violation. In contrast, 
a non-zero lepton EDM would be a smoking gun for CP violation outside the SM, 
and a comparison with neutron and atomic studies would be essential for 
identifying the particular scenario responsible.
\item 
The precise implications of EDM measurements for the CP violation needed for 
the BAU remains an open theoretical problem. If, for example, CP violation 
arises in the lepton sector {\em via\/} mixing of Majorana neutrinos (see 
Section~\ref{sec:neutrinos}), then it could produce a BAU {\em via\/} $B-L$ 
conserving interactions that both transform $L$ violation into $B$ violation 
and transmit the CP violating effects into the baryon sector. Presumably, such 
processes occur at high scales associated with the see-saw 
mechanism~\citer{Gell-Mann:vs,Yanagida:xy} (see Section~\ref{sec:neutrinos}) 
making them difficult to translate into precise, weak scale computations as 
required for the study of EDMs. Even the more conventional EWB remains subject
to unquantified approximations and theoretical uncertainties, rendering 
the relationship between the BAU and EDMs somewhat opaque.
\end{itemize}
In what follows, we review recent developments relevant to each of these 
points.

\subsubsection{Experimental developments}
Each EDM search relies on the same experimental signature of a non-zero EDM, 
namely, a small shift in the Larmour precession frequency in the presence of 
an applied electric field,
\be
   \hbar\omega = -\mu {\vec J}\cdot {\vec B} - d{\vec J}\cdot{\vec E},
\ee
where $\mu$ and $d$ are the magnetic and electric dipole moments of the system
of interest, ${\vec J}$ is its spin, and ${\vec B}$ and ${\vec E}$ are 
the applied magnetic and electric fields. Under reversal of the direction of 
${\vec E}$, the contribution from the EDM to the precession reverses sign, 
thereby allowing one to isolate the tiny EDM induced shift from the much larger
effect of the magnetic moment. The challenge for experimenters is to apply 
electric fields with as large a magnitude as possible, thereby enhancing 
the sensitivity to $d$, while minimizing various systematic effects, such as 
leakage currents, that can mimic the effect of the $d{\vec J}\cdot {\vec E}$ 
interaction. In many cases, the quest for improved sensitivity is aided by 
various fortuitous enhancement factors that can amplify one's EDM sensitivity.
The present experimental limits on the EDMs of various particles are listed in
Table~\ref{tab:edm}. 
\begin{table}
\begin{center}
\begin{minipage}[t]{16.5 cm}
\caption[]{ Present and prospective EDM limits. Expectations based on SM (CKM)
CP violation are also shown.}
\label{tab:edm}
\vspace*{4pt}
\end{minipage}
\begin{tabular}{|c|r|l|r|c|}
\hline
&&&&\\[-8pt]
System & Present Limit ($e$-cm)& Group & Future Sensitivity & Standard Model (CKM)\\[4pt]
\hline
&&&&\\[-8pt]
$e^-$ & $1.6\times 10^{-27}$  (90\%~CL)  & Berkeley & &  $<10^{-38}$  \\
$e^-$ &  &   Yale & $\sim 10^{-29}$ & \\
$e^-$ &    & LANL & $\sim 10^{-30}$ & \\[4pt]
\hline
&&&&\\[-8pt]
$\mu$ & $ 1.05\times 10^{-18}$ (90\%~CL)  & CERN & &$<10^{-36}$ \\
$\mu$ & & BNL & $\sim 10^{-24}$ & \\[4pt]
\hline
&&&&\\[-8pt]
$n$ & $6.3\times 10^{-26}$ (90\%~CL)  & ILL & $1.5\times 10^{-26}$ & $1.4\times 10^{-33} \to 1.6\times 10^{-31} $ \\
$n$ &  & PSI &  $7\times 10^{-28}$  &\\
$n$ &  & LANL &  $2\times 10^{-28}$ &\\[4pt]
\hline
&&&&\\[-8pt]
$^{199}$Hg & $2.1 \times 10^{-27}$ (95\%~CL) & Seattle & $5\times 10^{-28}$ & $\lsim 10^{-33}$ \\
$^{225}$Ra & & Argonne & $10^{-28}$ & \\
$^{129}$Xe & & Princeton & $10^{-31}$ & $\lsim 10^{-34}$ \\
$D$ & & BNL  & $\sim 10^{-27}$ & \\[-8pt] 
&&&&\\
\hline
\end{tabular}
\end{center}
\end{table}

\subsubsection*{Electron}
The most precise limit on the electron EDM, $d_e$, has been achieved by 
the Berkeley group with a measurement of atomic Thallium~\cite{Regan:ta}. 
The extraction of this limit from the atomic EDM relies on an observation by 
Sandars~\cite{Sandars:1965} that the EDM of a paramagnetic atom, $d_A$, 
induced by an electron EDM can be substantially enhanced. For Tl, 
the enhancement factor is $R = d_e/d_A = -585$~\cite{Liu:1992}. The experiment
employed a pair of atomic beams traversing identical paths but experiencing 
applied electric fields with the same magnitude and opposite sign. One beam 
consisted of sodium atoms that served as a co-magnetometer used to identify and
minimize systematic effects. Interchanging the paths traversed by the Na and Tl
beams would lead to a phase difference $\delta_{\rm EDM}$ for the two Tl paths
due to a non-vanishing $d_e$, but no phase difference for the spin-0 Na beam. 
The absence of any phase difference leads to a limit on $|d_e|$. Corrections 
were applied for residual, motional magnetic field and geometric phase effects,
and conservative upper bounds placed on leakage current, charging current, and 
dielectric absorption. An applied field of strength 410~stat-volts/cm was used.
From the calculated paramagnetic enhancement factor $R$ and the upper limit on 
the phase $\delta_{\rm EDM}$ the Berkeley group obtained the 90\%~CL limit 
given in Table~\ref{tab:edm}.

Further improvements in $|d_e|$ sensitivity are being pursued by two groups. 
The Yale group is employing a PbO molecule, for which near degeneracies of 
opposite parity molecular states enhance the electric field experienced by 
molecular electrons by two or more orders of magnitude relative to 
the corresponding enhancement in atoms~\cite{DeMille:2000,Kawall:2003ga}. 
The possibility for such enhancements in polar molecules was first observed by
Sandars~\cite{Sandars:1967} and subsequently applied to molecules such as 
PbO~\citer{Sushkov:1978,Isaev:2003}. The Yale experiment will rely on 
the $a(1)$ states in PbO, where a degeneracy between states with
electronic angular momentum projection along the internuclear axis, 
${\vec J}_e\cdot {\hat n}=\pm 1$ is lifted by a Coriolis coupling between 
$J_e$ and molecular rotational angular momentum. The splitting for this state 
is $\Delta\Omega_J= 11.2 (2)$~MHz. The presence of a non-zero $d_e$ in 
an applied electric field will lead to a shift in the $M=\pm 1$ projections of
the two $a(1)$ states. The Yale group has recently demonstrated 
the feasibility of this technique by studying Zeeman induced shifts in the $M$
sub-levels and by observing the Stark shift in the $J=1$ doublet in 
the presence of a small ($|{\vec E}| < 2$ V/cm) applied 
field~\cite{Kawall:2003ga}. For a field of this magnitude, the PbO measurement 
would be sensitive to an electron EDM at the scale of the current experimental
limit. The group anticipates that implementation of various experimental
improvements will lead to an EDM sensitivity of $\sim 10^{-29}$~$e$-cm in 
a month of running. 

The Los Alamos group is pursing a $d_e$ measurement using a solid state 
technique~\cite{Liu:2004}. The basic idea, originally developed by 
Shapiro~\cite{Shapiro:68}, relies on changes in the magnetic flux $\Phi_\mu$ in
a macroscopic solid state sample when the orientation of an external ${\vec E}$
is reversed. If the electrons in the solid have a nonzero EDM, then 
the application of ${\vec E}$ will cause the electron spins to align, thereby 
inducing a magnetic flux. Thus, by searching for changes in $\Phi_\mu$ when 
the sign of ${\vec E}$ is reversed, probes for a non-zero $d_e$. 
The Collaboration hopes to achieve a sensitivity of $10^{-30}$~$e$-cm in ten
days of integration using Gadolinium Gallium Garnet polycrystalline material. 

\subsubsection*{Muon} 
Although the present limits on $d_\mu$ fall well below SM expectations, 
a measurement at the level of $10^{-24}$~$e$-cm may be feasible in a storage 
ring experiment~\cite{Semertzidis:2003iq}. Such an experiment would rely on 
a $g-2$ precession type set-up, but with an applied, radial electric field that
would cancel the $g-2$ in plane precession. The precession frequency is given 
by,
\be
\label{eq:muonomega}
   {\vec\omega} = \frac{3}{m_\mu}\left[ a{\vec B} + \Biggl( 
   \frac{1}{\gamma^2-1}-a\Biggr){{\vec\beta}\times{\vec E}\over c} + 
   \frac{\eta}{2}\Biggl({\vec E}{c} + {\vec\beta}\times{\vec B}\Biggr) \right],
\ee
where $a$ is the muon anomalous magnetic moment and $\eta$ is related to 
the muon EDM via,
\be
   d_\mu = \frac{\eta}{2}\frac{e\hbar}{2 m_\mu c} \approx 
   4.7\times 10^{-11}~e\hbox{-cm}.
\ee
For the $g-2$ measurements, the muon energy is tuned to give the \lq\lq magic" 
$\gamma = 29.3$ that eliminates the second term in Eq.~(\ref{eq:muonomega}), 
leaving the precision entirely to $a$. The EDM measurement would use lower 
energy muons and choose the value of the radial electric field ${\vec E}$ so as
to cancel the first two terms. The resulting precession would arise entirely 
from the third term and would cause the spin to tilt out of plane. The degree 
of this tilt would be measured by exploiting the parity violation in the muon 
decay and looking for a vertical asymmetry in the number of decay positrons.

\subsubsection*{Neutron} 
The first searches for the permanent EDM of a quantum system were carried out
with neutrons in the pioneering work of Purcell and 
Ramsey~\cite{Purcell:1950,Smith:ht}. The sensitivity of the original experiment
has been steadily improved upon, culminating in the present limit of 
$6.3 \times 10^{-26}$~$e$-cm achieved at ILL using ultra-cold neutrons
(UCNs)~\cite{Harris:jx}.
The ILL Collaboration achieved a density of 0.6~UCN/cm$^3$. The technique 
involved observing the shift in the Larmour precession when the direction of 
the applied electric field was reversed relative to the static magnetic field.
A Hg co-magnetometer was used to keep track of magnetic field fluctuations. 
The Collaboration expects to improve this sensitivity by roughly a factor of 
four by increasing the number of UCNs collected and stored~\cite{Harris:ap}. 

Future, even more sensitive searches will be carried out by several groups. 
An experiment based at LANSCE~\cite{Mischke:ac} will use a technique involving
super cooling the neutrons by down-scattering in $^4$He and observing 
the neutron spin precession relative to that of a dilute mixture of $^3$He that
will also be present in the cells. The relative alignment of the neutron and 
$^3$He spins will be measured by observing the spin dependent neutron capture 
on $^3$He. The $^3$He will also serve as co-magnetometers. The Collaboration 
hopes to achieve a density of 500~UCN/cm$^3$. The initial goal of 
the experiment is a sensitivity of $9\times 10^{-28}$~$e$-cm at LANL with 
a final target of better than $2\times 10^{-28}$~$e$-cm to be reached at 
the SNS. 

An experiment at PSI is being developed that will use a solid D$_2$ source 
to produce a density of 1000~UCN/cm$^3$~\cite{Aleksandrov:2002}. The technique
involves employing two adjacent UCN cells in which the applied electric fields
have opposite orientation. Neighboring cells of cesium will serve as 
co-magnetometers. The goal of the experiment is a sensitivity of 
$5\times 10^{-28}$~$e$-cm. 

\subsubsection*{Neutral Atoms} 
The most stringent limit on the permanent EDM of any quantum system has been 
achieved for the $^{199}$Hg atom by the Seattle group, 
$|d_A| < 2.1 \times 10^{-27}$~$e$-cm~\cite{Romalis:2000mg}. The experiment 
relied on observing the Zeeman precession frequency of the $^{199}$Hg nuclear 
spin ${\vec I}$ in the presence of magnetic and electric fields. Two adjacent 
$^{199}$Hg cells were used. The orientation of the magnetic fields in the two 
cells was the same, but that of the electric field was opposite. A comparison 
of the precession frequency in the two cells provided a probe for 
the $d_A {\vec I}\cdot {\vec E}$ interaction. An improved version of 
this experiment is underway that uses as magnetometers two additional cells that 
do not experience the applied ${\vec E}$. A 100\%~CO buffer gas will
increase the Hg spin coherence time~\cite{Fortson:2004}. The Collaboration
anticipates a factor of four improvement over the current limit.  

An alternate strategy being employed at Princeton will use liquid 
$^{129}$Xe~\cite{Romalis:2001}. The technique will rely on long range dipole 
interactions that amplify magnetic field gradients. Such gradients would be 
induced by an applied ${\vec E}$ that, in the presence of a non-zero atomic 
EDM, would lead to spatial gradients in the associated shift in the precession 
frequency. Such frequency gradients, in turn, would induce magnetic field 
gradients that get amplified in the $^{129}$Xe liquid, which serves as its own
magnetometer. The Collaboration hopes to probe the $^{129}$Xe EDM with several
orders of magnitude better precision than obtained in the $^{199}$Hg
experiment~\cite{Romalis:2004}. 

A group at Argonne is pursuing a measurement of the $^{225}$Ra EDM using 
optical trapping~\cite{Holt:2004}. As discussed in the next subsection one 
expects the EDM of $^{225}$Ra to be $\sim 400$ times larger than that of 
$^{199}$Hg for a given source of CP violation in the nucleus. The optical 
trapping technique allows one both to reduce systematic effects as well as 
apply large external electric fields. The Argonne group hopes to achieve an EDM
with sensitivity of $\sim 10^{-28}$~$e$-cm. Similar $^{225}$Ra searches are
also being pursued at KVI and TRIUMF. For a brief review of these efforts, see
Ref.~\cite{Behr:2003}

\subsubsection*{Deuteron}
Recently, the possibility of measuring the deuteron EDM using a storage ring 
experiment similar to that for the muon EDM experiment has received 
considerable attention~\cite{Semertzidis:2003iq}. Measuring $d_D$ presents 
several experimental challenges not present in the case of the muon: 
the deuteron does not decay, so the out of plane rotation of its spin must be
detected {\em via\/} the scattering from protons or carbon; its anomalous 
magnetic moment is two orders of magnitude larger than that of the muon; its 
spin coherence time is of the order of 10~s; and the deuteron carries spin one,
so that it can have both vector and tensor polarizations. Efforts are underway
to address these challenges, but initial estimates suggest that a measurement 
with $\sim 10^{-27}$~$e$-cm may be feasible at existing or future storage
rings~\cite{Semertzidis:2003iq}.

\subsubsection{Theoretical implications}
The SM contains two sources of CP violation that can generate an EDM, the phase
$\delta$ in the CKM matrix and the gluonic operator in Eq.~(\ref{eq:thetacp}).
The magnitude of theoretical EDM predictions derived from CKM CP violation --- 
listed in Table~\ref{tab:edm} --- fall well below the present and prospective 
experimental bounds. The CKM induced EDMs of leptons and quarks vanish to 
two-loop order and are thus suppressed~\citer{Shabalin:rs,Bernreuther:1990jx}.
Naive dimensional analysis (NDA) leads to an estimate for the electron EDM,
\be
\label{eq:enda}
   d_e \sim e \frac{\alpha_s (M_W) G_F^2 M_W^2 m_e}{\pi^5} J {m_c\over m_t}
   {m_s\over m_b} \sim \times 10^{-35}~e\hbox{-cm},
\ee
where the presence of $\alpha_s$ is due to gluonic corrections that appear at
three-loop order. Scaling the NDA estimate in Eq.~(\ref{eq:enda}) by  
$m_\mu/m_e$ gives a conservative estimate for $d_\mu$ of 
$\sim 10^{-33}$~$e$-cm. Smaller values, also obtained from dimensional 
arguments, also appear in the literature~\cite{Bernreuther:1990jx} and are used
in Table~\ref{tab:edm}.

The corresponding predictions for the quark EDMs range from $10^{-31}$ to 
$10^{-33}$~$e$-cm~\cite{Shabalin:sg}. The valence quark contribution to 
the neutron EDM is given by,
\be
\label{eq:dnvalence}
   d_n = \frac{1}{3}(4\ d_d - d_u),
\ee
so that one might expect the SM (CKM) contribution to the neutron EDM to be of
the same order of magnitude. Chiral corrections to Eq.~(\ref{eq:dnvalence}) may
lead to order of magnitude enhancements of $d_n$. These effects arise via 
pseudoscalar loops containing $\Delta S = 1$ weak interactions at 
the meson-baryon vertices that are singular in the chiral 
limit~\citer{Gavela:1981sm,He:1989xj}.
A calculation of such contributions contains significant uncertainties, 
however, due to present limitations on carrying out first principles 
computations of the weak baryon-pseudoscalar meson vertices. Typically, they 
rely on approximations of questionable validity, such as factorization and 
vacuum saturation. Moreover, the CKM induced EDM --- which is second order
in the $\Delta S = 1$ weak interaction --- requires inclusion of intermediate 
states containing strange hadrons. The convergence of the $SU(3)$ chiral 
expansion for baryons is slow at best, and higher order operators omitted from
the computations of Refs.~\citer{Gavela:1981sm,He:1989xj} may make significant
contributions. Nevertheless, these studies provide a rough guide to 
the possible magnitude of $d_n$, so for illustrative purposes, we quote 
the range from Ref.~\cite{He:1989xj} in Table~\ref{tab:edm}. 

Within the context of CKM CP violation, the EDMs of nuclei and atoms such as 
$^{199}$Hg are generated primarily by T- and P-odd $NN$ interactions, 
${\hat H}_{NN}^{P,T}$, in the nucleus. Early work in this regard has been 
carried out in Refs.~\cite{Haxton:dq,Flambaum:1984fb}, where it was noted that
a long range, P- and T-odd nuclear force due to CKM CP violation could appear 
in the guise of kaon exchange between the two nucleons. The presence of kaons 
is necessary because the CKM CP violating interaction in light quark systems 
must be second order in the $\Delta S = 1$ weak interaction. In this case, 
the P- and T-odd nuclear force takes on the form,
\be
\label{eq:ptodd}
   {\hat H}_{NN}^{P,T} = \frac{G_F}{\sqrt{2}}\frac{1}{2 m_N}\sum_{ab}
   \Biggl[ \left(\eta_{ab}{\vec\sigma}_a-\eta_{ba}{\vec\sigma}_b \right)
   \cdot{\vec\nabla}_a\delta({\vec r_a}-{\vec r_b}) + \eta^\prime_{ab}
   ({\vec\sigma}_a\times{\vec\sigma}_b)\cdot\left\{({\vec p}_a-{\vec p}_b), 
   \delta({\vec r_a}-{\vec r_b}) \right\} \Biggr],
\ee
where the sum is over the combinations, $pp$, $nn$, and $np$. 
Ref.~\cite{Flambaum:1984fb} estimated the values of the $\eta_{ab}$ to be of 
order $10^{-8}$. However, this estimate did not properly take into account 
the structure of chiral symmetry and likely represents an overestimate of 
an order of magnitude or more~\cite{Donoghue:dd}. 

A complication arises in relating ${\hat H}_{NN}^{P,T}$ to an atomic EDM. 
As first noted by Schiff~\cite{Schiff:1963}, the effect of an external electric
field ${\vec E}$ on a point-like nucleus with an EDM would be entirely screened
out by the field created by distortions induced by ${\vec E}$ in the atomic 
electron cloud. Classically, this effect can be understood by noting that there
can be no net acceleration of a neutral atom in the presence of ${\vec E}$, so 
the acceleration that would be caused by the effect of ${\vec E}$ on 
the charged nucleus must be canceled by a corresponding field created by 
re-arrangements of the atomic electrons. Corrections to this \lq\lq Schiff 
screening" arise from several sources, including finite nuclear size, 
modifications of the {\em magnetic\/} electron-nucleus interaction due to 
${\vec E}$ induced atomic orbital distortion, and higher P- and T-odd nuclear 
moments, such as the magnetic quadrupole moment. In the case of $^{199}$Hg, 
only the first two effects are relevant since the nuclear spin is $I = 1/2$. 
To date, most authors have concentrated on the effects of finite nuclear size.
The corresponding effect on the atomic EDM is driven by the so-called Schiff 
moment, which may be thought of as an $r^2$-weighted moment of a P- and T-odd
component of the nuclear charge density. Ref.~\cite{Flambaum:1984fb} has
provided a simple estimate of the $^{199}$Hg EDM and Schiff moment ($Q$), using
a simplified, schematic model for the nuclear wave-function,
\be
\label{eq:hgschiff}
   d_A(^{199}{\rm Hg}) = -4 \times 10^{-17} \left( {Q\over e~{\rm fm}^3}
   \right)~e\hbox{-cm}, \hspace{50pt} 
   Q(^{199}{\rm Hg}) = -1.4 \times 10^{-8} \eta_{ab}~e~{\rm fm}^3.
\ee
Using the revised estimate, $\eta_{ab}\sim 10^{-9}$, a CKM induced EDM for 
mercury of a few $\times 10^{-34}$~$e$-cm is expected, which is well below 
the current and prospective sensitivities.

Although the effect of CKM CP violation is unlikely to ever be observed in 
$^{199}$Hg, other atoms may offer the possibility of enhanced EDM effects. 
As emphasized in Ref.~\cite{Haxton:dq}, the effect of ${\hat H}_{NN}^{P,T}$ in
certain nuclei can be amplified by accidents in nuclear structure. The effect 
arises because the P- and T-odd nuclear force mixes states of opposite parity,
and for certain nuclei the ground state is part of a nearly-degenerate parity 
doublet. The strength of wrong parity admixture into the ground state is
enhanced by the small energy splitting. Ref.~\cite{Haxton:dq} computed 
the corresponding enhancement factors in the context of the gluonic
CP violation of Eq.~(\ref{eq:thetacp}), but did not consider $^{199}$Hg. 
Moreover, the authors focused exclusively on the enhancements of the nuclear 
EDMs, which are not directly observable in atomic EDM experiments, rather than
on the Schiff moments that are relevant in these cases. Nonetheless, one 
naively expects enhancements of the Schiff moments in cases where the nuclear 
EDMs are also amplified. For non-spherical nuclei the effect can be 
particularly pronounced, leading to enhancements of two to three orders of 
magnitude.

Recently, the nuclear enhancements of the Schiff moment was analyzed for 
$^{225}$Rn~\cite{Engel:2003rz} (see also 
Refs.~\cite{Engel:1999np,Dzuba:2002kg}), an octupole deformed nucleus whose 
atomic EDM measurement is being pursued by the Argonne group. 
The analysis~\cite{Engel:2003rz} reveals an enhancement of $Q(^{225}{\rm Rn})$
relative to $Q(^{199}{\rm Hg})$ by a factor of several hundred, using the work
of Ref.~\cite{Dmitriev:2003sc} on the latter for comparison. While such 
an enhancement is unlikely to put CKM CP violation within reach of atomic EDM 
experiments, it may help to provide added sensitivity to non-SM CP violation.

One advantage of the deuteron EDM experiment~\cite{Semertzidis:2003iq} is that
it would involve deuteron ions rather than neutral atoms and, thus, evade 
the complications due to Schiff screening and provide a direct measurement.
Recently, Ref.~\cite{Khriplovich:1999qr} suggested that $d_D$ may be enhanced 
in the chiral limit. $d_D$ was computed using a $\pi$ exchange model for 
the P- and T-odd $NN$ interaction leading to the result,
\be
\label{eq:deuteronedm}
   d_D=-{eg_A g_1\over 3 m_\pi} \left( {m_N\over\Lambda_\chi} \right) { 1 +
   \xi\over(1+2\xi)^2}\approx 2.5\times 10^{-14} g_1~e\hbox{-cm},
\ee
where $g_1$ gives the strength of the $\Delta I=1$, P- and T-odd pion-nucleon 
coupling,
\be
\label{eq:tvpv1}
   {\cal L}^{\rm P,T}_{\Delta I=1} = g_1({\bar p}p+{\bar n}n)\pi^0,
\ee
and where $\xi=\sqrt{m_N E_B}/m_\pi$ ($E_B$ is the deuteron binding energy).
Thus, a measurement of $d_D$ at the $10^{-27}$~$e$-cm level would be sensitive
to a $g_1$ of ${\cal O} (10^{-13})$. A non-zero value for $g_1$ in the SM --- 
either due to CKM CP violation or ${\cal L}_{\rm strong\ CP} $ --- is 
suppressed, so it can be fairly sensitive to CP violation in models for physics
beyond the SM. 

Given that the CKM induced EDMs of all systems under study are expected to be 
far smaller in magnitude than present or prospective experimental 
sensitivities, the EDM measurements are primarily searches for either strong CP
violation or CP violation that goes beyond the SM. In order to derive 
implications for the latter, one must, however, determine the level at which 
strong CP violation contributes to any EDM. In this respect, a comparison of 
various experiments is essential. In particular, both atomic and neutron EDMs 
are sensitive to $\theta_{\rm QCD}$, whereas the lepton and deuteron EDM 
sensitivity to strong CP violation is highly suppressed. In the case of both 
$d_n$ and $d_A$, the dominant effect of $\theta_{\rm QCD}$ arises {\em via\/} 
an induced, $\Delta I=0$, T- and P-odd $\pi NN$ interaction,
\be
\label{eq:tvpv0}
   {\cal L}^{\rm P,T}_{\Delta I=0} = g_0{\bar N}{\vec\tau}\cdot{\vec \pi} N,
\ee
where the authors of Ref.~\cite{Crewther:1979pi} estimate,
\be
   g_0 = -\theta_{\rm QCD} \left[{(m_\Xi - m_N) m_u m_d\over F_\pi (m_u + m_d)
   (2m_s - m_u - m_d)} \right] \approx - 0.4\ \theta_{\rm QCD}..
\ee
The resulting neutron EDM is singular in the chiral limit,
\be
\label{eq:dnchiral}
   d_n = {g_A g_0\over\Lambda_\chi\pi} \ln\frac{\Lambda_\chi}{m_\pi} +\cdots,
\ee
where the dots indicate subleading contributions. (For other related 
theoretical discussions, see, {\em e.g.}, Ref.~\cite{Kawarabayashi:1980dp}.) 
The corresponding dependence of the atomic EDM on $\theta_{\rm QCD}$ does not 
have a simple analytic expression, since it depends on a complex interplay of 
nuclear and atomic structure computations. Nonetheless, the present limit on 
$d_A(^{199}{\rm Hg})$ leads to the most stringent bound on strong CP violation,
\be
   \theta_{\rm QCD} < 1.5\times 10^{-10} \hspace{20pt} (95\%~{\rm CL}),
\ee
while the present limit from the neutron EDM lead to a bound that is four times
weaker.

Should a future more precise atomic or neutron EDM measurement yield a non-zero
result, one would not be able to determine if the source was $\theta_{\rm QCD}$
or new CP violation without information from the lepton or deuteron 
experiments. Should one of the latter also observe a non-zero EDM, one would 
likely conclude that one had seen evidence for new CP violation. Similarly, 
should future atomic and neutron EDM measurements with comparable sensitivities
to $\theta_{\rm QCD}$ obtain significantly different results that could not be
attributed to hadronic or nuclear structure uncertainties, one would also favor
an explanation in terms of CP violation beyond the SM.

Although free from the effects of ${\cal L}_{\rm strong\ CP} $, the lepton and 
deuteron EDM measurements by themselves would not be sufficient to distinguish
among various scenarios for new CP violation, should any of them find 
a non-zero result. However, the sensitivities of atomic and neutron EDMs to 
these scenarios are sufficiently complementary to those of lepton EDMs that, 
taken together, the full set of EDM studies provides a powerful diagnostic tool
for new CP violation. 

To illustrate, we consider the various EDMs in the context of the MSSM. 
In the case of the electron, for example, one has~\cite{Bernreuther:1990jx},
\be
\label{eq:desusy1}
   d_e = -e\frac{\alpha}{4\pi} M_{\tilde\gamma} \sin 2 \theta_{LR}
   \sin(\phi_A-\phi_{\tilde\gamma}) F(m_{{\tilde e}_i}, M_{\tilde\gamma}),
\ee
when the EDM is dominated by graphs involving selectron (${\tilde e}$)-photino 
(${\tilde\gamma}$) intermediate states. Here, $\theta_{LR}$ is the angle
describing the mixing of chiral electron eigenstates into mass eigenstates with
masses, $m_{{\tilde e}_i}$; $F$ is a calculable function of these masses and 
that of the photino, $M_{\tilde\gamma}$; and $\phi_{A,\tilde\gamma}$ are CP 
violating phases. $\phi_A$ is associated with tri-scalar couplings that arise 
in the soft SUSY breaking Lagrangian, while $\phi_{\tilde\gamma}$ appears in 
Majorana mass terms for the neutral gauginos. Note that $d_e$ depends on
the {\em difference\/} of these phases. For superpartner masses of order 
the weak scale one has,
\be
\label{eq:desusy2}
   d_e\approx -(1\times 10^{-25})\left({M_{\tilde\gamma}\over 100\ {\rm
   GeV}}\right)^{-3} \left({|A_e|\over 100\ {\rm GeV}}\right) 
   \sin(\phi_A-\phi_{\tilde\gamma})~e\hbox{-cm}.
\ee
Thus, for $\sin(\phi_A-\phi_{\tilde\gamma})$ of order unity, one would require
$M_{\tilde\gamma} > 500$ GeV (for $|A_e|\sim 100$ GeV) in order to evade 
the present limit. For significantly lighter gauginos one would need
$\sin(\phi_A-\phi_{\tilde\gamma})$ to be small. Present collider limits on 
$M_{\tilde\gamma}$ are rather weak, so either possibility is currently allowed
by experiment. Even the light gaugino scenario does not imply the individual 
SUSY CP violating phases need be small, but rather that the phase difference be
so. Thus, even within the MSSM, a single EDM measurement may not be sufficient 
to determine an individual CP phase. 

\begin{figure}[t]
\begin{center}
\begin{minipage}[t]{14 cm}
\epsfig{file=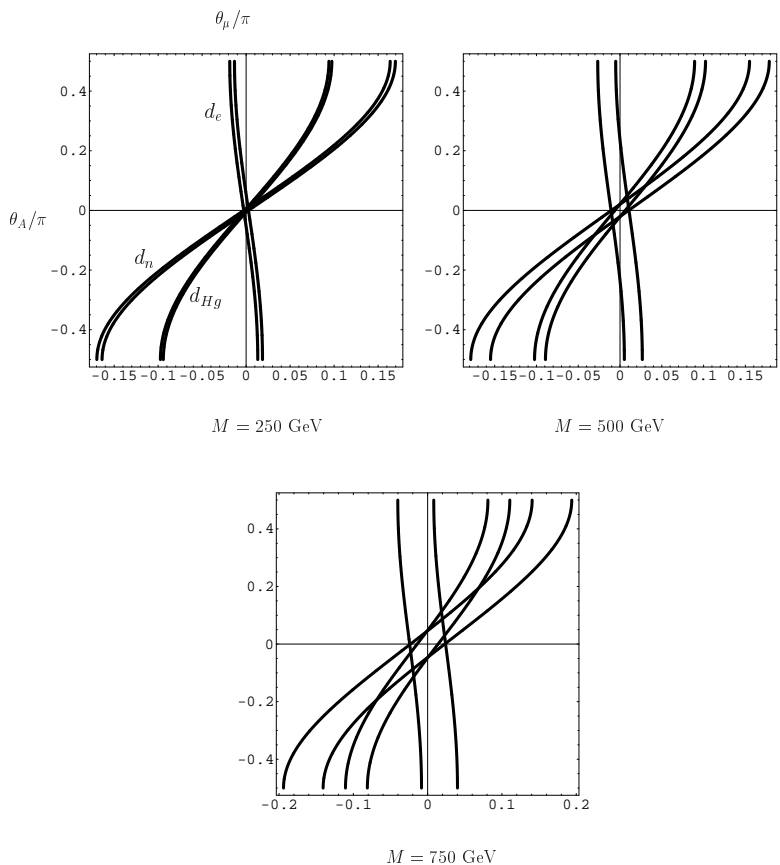,scale=0.58}
\end{minipage}
\vspace*{-20pt}
\begin{minipage}[t]{16.5 cm}
\vspace*{-80pt}
\caption{Constraints on CP violating supersymmetric phases from electron,
neutron, and $^{199}$Hg EDM limits. They are shown assuming three different
values of a common superpartner mass and $\tan\beta=3$. (Figure provided by 
M.~Pospelov.)
\label{EDM}}
\end{minipage}
\end{center}
\end{figure}

Significantly more information may be obtained by comparing EDM experiments. 
As has been noted by several authors, the EDMs of different systems display 
different sensitivities to the CP violating phases in SUSY. A useful 
illustration of this complementarity occurs in the context of gravity mediated
SUSY breaking, in which one has two CP violating phases, $\phi_A$ and 
$\phi_\mu$ associate with the $\mu$-term (see Section~\ref{beyond}). 
This complementary phase dependence, along with the current constraints from 
$d_e$, $d_n$, and $d_A(^{199}{\rm Hg})$ is illustrated in Figure~\ref{EDM} (see
also Ref.~\cite{Falk:1999tm}). When taken together, the EDM experiments imply 
that for light superpartners, $\phi_A$ and $\phi_\mu$ individually --- and not
just $|\phi_A - \phi_\mu|$ --- must be small.  This information provides 
important input for theorists who build models of SUSY breaking mediation, who 
must explain why these SUSY phases are small.

An alternative and more esoteric illustration of the EDM complementarity arises
in the case of new interactions that violate T but conserve P. In the presence
of parity violating SM EW radiative corrections, these time-reversal violating,
parity conserving (TVPC) interactions could give rise to EDMs of various 
particles~\cite{Conti:xn}. Since the only TVPC interactions one can write down
are non-renormalizable, their impact on EDMs in the presence of radiative 
corrections must be treated carefully in the context of an effective field 
theory (EFT)~\cite{Engel:1995vv,Ramsey-Musolf:1999nk}. As noted in 
Ref.~\cite{Ramsey-Musolf:1999nk}, EDMs provide unambiguous constraints on TVPC
interactions only when one assumes that the scale at which P is broken is well 
below that of the breakdown of T. Then the lepton and neutron EDMs are 
considerably more sensitive to these effects than the EDMs of neutral atoms 
such as $^{199}$Hg~\cite{Kurylov:2000ub}. Thus, were one to observe 
non-vanishing EDMs in leptons and the neutron but not in neutral atoms, one 
might favor an explanation in terms of TVPC interactions. This possibility 
contrasts with that of strong CP violation that allows for non-vanishing 
neutron and atomic but not leptonic EDMs. SUSY would naturally imply 
non-vanishing EDMs for all three systems at comparable levels.

\subsubsection{EDMs and the baryon asymmetry of the Universe}

One of the strongest theoretical motivations for postulating the existence of 
CP violation beyond the CKM and strong CP violation of the SM, is the need 
to account for the predominance of matter over antimatter in the Universe. 
As observed by Sakharov in 1967~\cite{Sakharov:dj}, if the initial conditions 
of the Universe were exactly matter-antimatter symmetric, then during 
the subsequent evolution of the Universe three discrete symmetries would have 
to be violated in order to explain the observed baryon asymmetry: baryon number
(B), charge conjugation, and CP. Assuming CPT is an exact symmetry, these other
symmetry violations also need to be accompanied by a departure from thermal 
equilibrium, such as a first order phase transition. 
Many attempts to account for the BAU have focused on the EW phase transition 
and the possibilities for new weak scale CP violation. Presently, a popular 
alternative to such electroweak baryogenesis (EWB) involves CP violation in 
the lepton sector associated with neutrino masses and mixing (leptogenesis). 
Typically, the latter CP violation is associated with very high scales, as in 
see-saw models of neutrino mass~\citer{Gell-Mann:vs,Yanagida:xy} (see 
Section~\ref{sec:neutrinos}). The effects on EDMs of leptonic CP violation are
expected to fall well below those in the SM~\cite{Barr:1988mc}, so we focus 
here on the connections between EDMs and EWB. 

As noted above, the magnitude of CP violation in the SM is far too weak 
to account for the observed BAU. Moreover, numerical studies indicate that 
the strength of the first order SM EW phase transition is too weak to prevent
subsequent wash out of sphaleron-induced baryon number creation. In contrast, 
models that contain new sources of CP violation, such as SUSY or theories with 
a more elaborate Higgs sector, appear to produce a sufficiently strong first 
order phase transition. Within the MSSM, the window of opportunity for 
satisfying this requirement has narrowed, as one needs a fairly light Higgs 
(see {\em e.g.\/} Ref.~\cite{Brignole:1993wv} and references therein). 
However, this requirement can be relaxed by extending the MSSM gauge group. 
Thus, SUSY EWB remains a viable option for producing the BAU. 

Several authors have attempted to compute the BAU in the MSSM using a bubble 
wall picture of the phase transition (for a recent review and references, see
Ref.~\cite{Riotto:1999yt}). In this scenario, a small region of broken EW 
symmetry (the bubble) expands into the unbroken vacuum, and the B-, C-, and
CP violating processes needed for the BAU occur in the plasma at the interface
between the two phases (the bubble wall). Efforts to model this plasma region 
have typically relied on various simplifying assumptions and approximations, 
and not all of the associated uncertainties have been adequately quantified.
Nevertheless, these studies provide some guidance as to the implications of EDM
measurements for EWB. For example, Ref.~\cite{Huet:1995sh} finds for the ratio 
of baryon number density, $\rho_B$, to the entropy, $S$, at photon freeze-out,
\be
\label{eq:huet}
   \frac{\rho_B}{S} = -\gamma_w v_w \left( \frac{\kappa}{\kappa^\prime} \right)
   \Delta\beta\left[ 6.3\sin\phi_\mu+0.23\sin(\phi_A-\phi_\mu)\right ]
   \times 10^{-9},
\ee
where $\gamma_w v_w$ is the velocity of the bubble wall, 
$\kappa/\kappa^\prime$ gives the ratio of weak and strong sphaleron rates, and
$\Delta\beta$ characterizes the change in $\tan\beta$ (see 
Section~\ref{beyond}) across the plasma. The first term in Eq.~(\ref{eq:huet}) 
arises from CP violating interactions between charginos and neutralinos and 
the spacetime dependent Higgs VEVs while the second term is generated by Higgs
VEV-scalar top quark interactions. In arriving at this result, 
Ref.~\cite{Huet:1995sh} assumed fairly light gaugino and squark masses. 
The relative weighting of the two terms in Eq.~(\ref{eq:huet}) will differ 
under other assumptions for the MSSM parameters. From the dominance of 
the first term and the experimental range of $(4-7)\times 10^{-11}$ for 
$\rho_B/s$, one infers~\cite{Huet:1995sh} $|\sin\phi_\mu|\geq 0.025$. 
The constraints from the EDMs shown in Figure~\ref{EDM} imply 
$|\sin\phi_\mu| \lsim 0.03$, a limit that is barely consistent with 
the requirements of the BAU. In order for the interactions involving squarks to
play an appreciable role in the BAU, $|\sin(\phi_A-\phi_\mu)|$ must be 
considerably larger than allowed by the EDM constraints. One might concludes 
that existing EDM measurements have severely constrained the viability of 
supersymmetric EWB --- even with unnaturally small CP violating phases --- and
that future measurements will have an even more substantial impact on 
this scenario.

The extent to which theoretical uncertainties in the analysis leading to 
Eq.~(\ref{eq:huet}) and the EDM limits of Figure~\ref{EDM} allow for the MSSM 
to evade these conclusions is a topic of on-going study. 
Ref.~\cite{Riotto:1998zb}, for example, has argued that memory effects in 
the plasma may significantly enhance the impact of CP violating interactions,
allowing for the production of the observed BAU with considerably smaller CP 
violating phases. If these arguments survive further scrutiny, then the present
EDM constraints would not substantially limit the viability of EWB in the MSSM,
while the impact of future, more precise measurements would have to be analyzed
in more detail.

\subsection{\it Lepton Flavor Violation}
\label{BLviolation}
The observation of neutrino oscillations has inspired renewed interest in 
the more general question of lepton flavor non-conservation. Experimentally, 
the most powerful probes of lepton flavor violation (LFV) involve decays of 
the muon, such as $\mu\to e\gamma$, $\mu+A(Z,N)\to e + A(Z,N)$, or 
$\mu\to 3 e$. The most stringent limit on the process $\mu\to e\gamma$ has been
obtained by the MEGA Collaboration~\cite{Brooks:1999pu},
\be
\label{eq:MEGA}
   {\cal B}(\mu\to e\gamma)\equiv {\Gamma(\mu^+\to e^+\gamma)\over
   \Gamma(\mu^+\to e^+ \nu{\bar\nu})} < 1.2\times 10^{-11}~({\rm 90\%~CL}),
\ee
while the SINDRUM Collaboration~\cite{Wintz:rp} has obtained a limit on the
conversion branching ratio,
\be
\label{eq:SINDRUM}
   R_{\mu\to e}^{A}\equiv {\Gamma [\mu^- + A(N,Z) \to e^- + A(N,Z)] \over\Gamma
   [\mu^- +A(Z,N)\to \nu_\mu A(Z-1,N+1)]} < 6.1\times 10^{-13}~({\rm 90\%~CL}),
\ee
for conversion in Au. The SINDRUM experiment has also constrained 
the $\mu\to 3e$ branching ratio to be less than $1\times 10^{-12}$ 
(90\%~CL)~\cite{Bellgardt:1987du}, and the branching ratio for the lepton 
number violating conversion process in $\mu^-\to e^+$ in Ti is
$< 3.6\times 10^{-11}$~(90\%~CL)~\cite{Kaulard:rb}. A new experiment at PSI 
hopes to improve on the MEGA sensitivity by three orders of 
magnitude~\cite{Yashima:2000qz}, while the MECO experiment at Brookhaven 
expects to achieve a four orders of magnitude improvement~\cite{Molzon:sf}.

The more precise LFV searches are poised to probe NP above the EW scale. 
Although flavor oscillations among light neutrinos will induce LFV decays of 
charged leptons at some level, the corresponding rates are typically suppressed
by factors of $(\Delta m_\nu^2/M_W^2)^2\sim 10^{-50}$ and completely 
negligible. In contrast, flavor violation involving heavy particles that are 
not part of the SM may generate effects that could be seen by experiment. In 
this respect, two broad scenarios have been studied 
theoretically~\citer{Riazuddin:hz,Cirigliano:2004}.

A commonly quoted model involves GUT scale LFV and supersymmetry. GUTs 
naturally provide a link between flavor violation in the quark and lepton 
sectors~\cite{Barbieri:1994pv,Barbieri:1995tw}. In this context, LFV occurs at
the GUT scale {\em via\/} Yukawa couplings involving quark and lepton 
supermultiplets. The assignment of leptons and quarks to the same 
representation of the unification group implies that the large Yukawa coupling
responsible for the top quark mass also appears in the LFV terms, thereby 
leading to unsuppressed LFV couplings. At the weak scale, these can be related 
in a straightforward way to the CKM matrix elements for the top quark, modulo
RG evolution from the GUT scale to the weak scale. To avoid 
GIM~\cite{Glashow:gm} suppression by the light neutrino mass differences or by
inverse powers of the GUT scale, the lepton flavor mixing effects must be 
realized in a sector associated with TeV (or lower) scales. Supersymmetry 
provides a natural mechanism for doing so, as the GUT induced couplings between
quark and lepton supermultiplets can give rise to lepton flavor non-diagonal 
terms in the soft SUSY breaking interactions. Superpartner loops that include 
insertions of the LFV operators can then generate unsuppressed contributions to
$\mu\to e\gamma$, {\em etc.}

For example, in a SUSY $SU(5)$ scenario one has~\cite{Barbieri:1994pv},
\be
   {\cal B}(\mu\to e\gamma) = 2.4\times 10^{-12} \left(
   \frac{|V_{ts}|}{(0.04)}\frac{|V_{td}|}{(0.01)}\right)^2\left(
   \frac{100\ {\rm GeV}}{m_{\tilde\mu}}\right)^4,
\ee
where gaugino mass effects have been neglected for illustrative purposes. 
One expects $R_{\mu\to e}^{\rm Ti}$ to be smaller than 
${\cal B}(\mu\to e\gamma)$ by a factor of about $2.4\ \alpha$, since a virtual 
photon is exchanged between the leptons and nucleus instead of a real photon 
being emitted. The ${\cal O}(\alpha)$ suppression of $R_{\mu\to e}^{A}$ 
relative to ${\cal B}(\mu\to e\gamma)$ is a rather generic feature of GUT scale
LFV. In both cases, the GUT expectation is at least an order of magnitude below
the current experimental bounds.

An alternate scenario involves the generation of LFV at the TeV 
scale~\cite{Riazuddin:hz,Barenboim:1996vu}. This scenario can also be realized 
in the MSSM, though it does not arise naturally from any conventional model of
SUSY breaking mediation. The latter generally entail flavor universality at 
the SUSY breaking scale, so that some alternate mechanism for LFV, such as GUT
scale lepton-quark unification or RPV~\cite{Huitu:1997bi} is needed. In 
contrast, if a GUT breaks down in stages to the SM gauge group~(\ref{su321}), 
then LFV Yukawa interactions can involve particles with intermediate scale 
masses. 

One may consider, {\em e.g.}, a GUT that breaks down to the group 
$SU(3)_C\times SU(2)_L\times SU(2)_R\times U(1)_{B-L}$ of left-right symmetric
models. If the right-handed gauge symmetry is broken at a the TeV scale, 
interactions involving right-handed (Majorana) neutrinos, gauge bosons, and 
Higgs multiplets may give rise to unsuppressed LFV. In contrast to the GUT 
scale LFV scenario, the presence of relatively light Higgs multiplets may lead 
to a substantial enhancement of $R_{\mu\to e}^{A}$, making it comparable in 
magnitude to ${\cal B}(\mu\to e\gamma)$~\cite{Raidal:1997hq,Cirigliano:2004}. 
Specifically, one finds an enhancement by the square of a large logarithm 
involving the $SU(2)_L \times SU(2)_R$ breaking scale and 
$m_\mu$~\cite{Cirigliano:2004}, compensating for the ${\cal O}(\alpha)$ 
suppression of $R_{\mu\to e}^{A}$ relative to ${\cal B}(\mu\to e\gamma)$. One 
therefore expects in this scenario the two LFV ratios to be of comparable 
magnitude. Similar logarithmic enhancements may occur in RPV SUSY 
scenarios~\cite{Huitu:1997bi}.

\section{Lepton Properties}
The observation of neutrino oscillations in atmospheric, solar, and reactor
neutrino experiments has provided the first incontrovertible evidence for
physics beyond the SM. Although a thorough review of neutrino phenomenology 
merits an entire article in itself (see, {\em e.g.}, the recent review by
McKeown and Vogel~\cite{McKeown:2004yq} and references therein), no review of
low energy weak interactions would be complete without some discussion of
neutrinos. Below, we provide a short summary of neutrino properties, with 
an emphasis on experimental results. In brief, the picture that has emerged 
from experiment implies substantial flavor mixing among the three lightest 
neutrinos, in contrast to the situation involving quarks. Labeling 
the neutrino mass eigenstates as $\nu_i$ ($i = 1,2,3$) with masses $m_i$, one 
finds tiny mass-squared differences, $|m_3^2 - m_2^2|\approx 0.002$~eV$^2$ from
atmospheric oscillations and $|m_2^2 - m_1^2|\approx 10^{-5}$~eV$^2$ from solar
neutrinos. The absolute scale of neutrino mass is not yet known, but upper 
bounds exist from various $\beta$-decay studies (see below) as well as 
cosmological observations ($\sum_i m_i < 0.7$~eV for neutrinos light enough
to decouple while still relativistc~\cite{Spergel:2003cb}). A variety of 
neutrino properties remain to be determined by additional experimental work:
\begin{itemize}
\item The charge conjugation properties of the neutrino, {\em i.e.}, whether it
is a Dirac or Majorana particle.
\item The nature of the neutrino mass hierarchy, namely whether it is 
degenerate ($|\Delta m_{ij}^2| \ll m_j^2$ for all $i,j$) or non-degenerate and
whether it is \lq\lq normal" ($m_1^2 < m_2^2 < m_3^2$) or inverted.
\item The size of the mixing angle $\theta_{13}$ and the CP violating Dirac 
phase $\delta$, as well as the two CP violating Majorana phases if neutrinos 
are Majorana particles.
\end{itemize}
In addition, the LSND experiment, which implies a much larger $\Delta m^2$ than
obtained from either the atmospheric or solar oscillation experiments, has 
raised the possibility that there exists either a fourth generation of light 
neutrinos or of CPT violation among the three generations.

The origins of neutrino masses and mixing, as well as the properties of 
neutrinos with respect to various discrete symmetries, have inspired a wealth 
of theoretical proposals for physics beyond the SM. Since a review of these 
ideas goes beyond the scope of this article, we content ourselves with brief 
statements about the simplest ones, along with a compilation of results from 
the various experiments.

\subsection{\it Neutrino masses and mixing}
\label{sec:neutrinos}
In the minimal (without left-handed $\bar\nu$) renormalizable SM as described 
by Eq.~(\ref{Lagrangian}) neutrinos are predicted to be massless. 
In particular, Majorana mass terms~\cite{Majorana:vz} for the left-handed 
neutrinos,
\be
  {\cal L}_{\rm Majorana}= -{m_M\over 2}\bar{\nu} C \bar{\nu}^T,
\label{Majorana}
\ee
where $C$ is the charge-conjugation matrix defined by 
$C \gamma_\mu C^{-1} = - \gamma_\mu^T$, are forbidden by $U(1)_Y$ invariance 
and not produced upon EW symmetry breaking. The simplest way to produce a term
of the form~(\ref{Majorana}) is to add a Higgs triplet~\cite{Gelmini:1980re}
carrying one unit of hypercharge\footnote{A Higgs singlet could also produce
a Majorana mass term, but if it does it necessarily breaks $U(1)_{\rm EM}$.}.
However, unless one has a minuscule Yukawa coupling, one would 
expect\footnote{Triplet VEVs are severely constrained by EW precision data and 
can be at most of order 10~GeV~\cite{Erler:ew}. Values considerably smaller 
than that would require fine-tuning in the Higgs potential order by order in 
perturbation theory~\cite{Langacker:1991nt}.} $m_M = {\cal O} (10~\mbox{GeV})$.
Note that the mass term~(\ref{Majorana}) violates the conservation of any 
additive quantum number that $\nu$ may carry, such as lepton number. This can 
lead to neutrinoless double $\beta$ decay as will be discussed in 
Subsection~\ref{0nu2betadecay}.

Alternatively, the introduction of a left-handed anti-neutrino, $\bar\nu$, 
would allow a Dirac mass term,
\be 
  {\cal L}_{\rm Dirac} = - m_D \bar{\nu} \nu,
\label{Dirac}
\ee
in complete analogy to Eqs.~(\ref{yukawa},\ref{fermionmass},\ref{Lagrangian}).
Thus, $\nu$ and $\bar\nu$ combine to yield a Dirac neutrino with four 
physically distinct degrees of freedom. As before, one generally expects 
$m_D = {\cal O} (G_F^{-1/2})$. If there is more than one Dirac neutrino one
expects CKM mixing analogous to Eq.~(\ref{vckm})~\cite{Maki:mu}. Thus, while 
lepton number is conserved in the Dirac case, lepton flavor number associated 
with a particular lepton generation is not.

In the case of $n$ Majorana neutrinos, $m_M$ becomes in general a (complex) 
symmetric $n \times n$ mass matrix. Each neutrino spinor may absorb 
an unphysical phase, reducing $m_M$ to $n^2$ real observable parameters, namely
$n$ mass eigenvalues, $n (n-1)/2$ mixing angles, and an equal number of CP 
violating phases. A Dirac neutrino corresponds to the special case of a pair of
Majorana neutrinos of equal mass and maximal mixing from their weak interaction
basis. A pure Majorana neutrino corresponds to no mixing. In the most general 
situation, both types of mass terms may be present. Another special case arises
for $n=2$ and $m_M \ll m_D$, {\em i.e.}, the diagonal (Majorana mass) entries 
are much smaller than the off-diagonal (Dirac mass) term. This is called 
a pseudo-Dirac neutrino. On the other hand, if one Majorana mass entry vanishes
and the other one satisfies $m_M \gg m_D$, one obtains eigenvalues of order 
$m_M$ and $m_D^2/m_M$, referred to as the see-saw 
mechanism~\citer{Gell-Mann:vs,Yanagida:xy}. Thus, if the see-saw mechanism is
realized in nature, very small neutrino masses would probe very high mass
scales, $m_M$. More generally a fundamental theory beyond the SM will produce 
non-renormalizable terms in the Lagrangian carrying coefficients of negative 
mass dimensions which are numerically suppressed by the NP scale and could 
therefore naturally produce small neutrino masses. The simplest possibility is
a dimension five term involving two lepton doublets and two Higgs 
doublets~\cite{Barbieri:1979hc}.

Neutrino masses can reveal themselves by a variety of effects. The most 
important classes are kinematic effects in decays, neutrino oscillations, and
possible effects related to the now allowed magnetic and electric dipole 
moments. 

The most precise kinematic limits are obtained from two tritium $\beta$ decay
experiments~\cite{Weinheimer:tn,Lobashev:tp}. Assuming there are no common 
uncertainties the can be combined yielding,
\be
    m^2_{\bar\nu_e} ({\rm eff.}) \equiv \sum\limits_i |U_{ei}|^2 m_{\nu_i}^2 
    = - 2.5 \pm 3.3 \mbox{ eV}^2,
\label{m2nue}
\ee
where $U$ is the neutrino mixing matrix, and the sum is over all mass 
eigenvalues, $m_{\nu_i}$, that cannot be resolved experimentally\footnote{See
Ref.~\cite{Vogel:ee} for a discussion.}. More than 77\% of the range~(\ref{m2nue}) 
is in the unphysical (spontaneously Lorentz invariance
breaking) region\footnote{There are several other tritium $\beta$ decay 
experiments some of which yielding results which are incompatible with
$m^2_{\nu_e} ({\rm eff.}) > 0$. An event excess near the spectrum endpoint 
leading to an apparent $m^2_{\nu_e} ({\rm eff.}) < 0$ is presently not 
understood, but Refs.~\cite{Weinheimer:tn,Lobashev:tp} apply corrections 
to obtain results that can be properly interpreted.} with 
$m^2_{\nu_e} ({\rm eff.}) < 0$. Normalizing the probability conditional on 
$m^2_{\nu_e} ({\rm eff.}) > 0$ yields the 95\%~CL upper limit,
\be
   \sqrt{m^2_{\bar\nu_e} ({\rm eff.})} < 2.3 \mbox{ eV}.
\ee
The KATRIN Collaboration~\cite{Osipowicz:2001sq} at the Forschungszentrum in
Karlsruhe, Germany, will attempt to improve this limit in a next-generation 
tritium $\beta$ decay experiment to the 0.35~eV level. 

Limits analogous to~(\ref{m2nue}) can also be obtained for 
$\nu_\mu$ (from $\pi^\pm$ decays) and for $\nu_\tau$ (from $\tau^\pm$ decays). 
However, these limits are many orders of magnitude weaker and relevant only to
the extent to which the positive results on neutrino mass differences from 
neutrino oscillations are circumvented. For a recent review on absolute
neutrino masses, see Ref.~\cite{Bilenky:2002aw}.

The weak interaction induces one-loop neutrino magnetic moments proportional to
the neutrino masses. However, given the limit~(\ref{m2nue}) the SM 
contribution to the magnetic moment is many orders of magnitude below current
limits. Thus, the search for magnetic and (CP violating) electric dipole 
moments of neutrinos provides tests of physics beyond the SM with massive 
neutrinos. 

If for a given pair of weak eigenstate neutrinos, $\nu_\alpha$ and $\nu_\beta$,
(i) one or both neutrinos are massive, (ii) the mass eigenvalues are not 
identical, $m_{\nu_\alpha} \neq m_{\nu_\beta}$, and (iii) the mass eigenstates 
differ from the weak interaction eigenstates by a mixing angle, $\theta$, 
$\nu_\alpha = \cos\theta \nu_1 + \sin\theta \nu_2$ (where $\cos\theta\neq 1$),
one predicts the phenomenon of neutrino 
oscillations~\citer{Pontecorvo:cp,Bilenky:1998dt}. For highly relativistic, 
$m_i\ll E_i$, and monochromatic neutrinos one finds the probability for
oscillations in vacuum,
\be
   P(\nu_\alpha\rightarrow\nu_\beta) = 1 - |<\nu_\alpha(0)|\nu_\alpha(t)>|^2 =
   \sin^2 2 \theta \sin^2\left[ {(m_1^2 - m_2^2) t\over 4 E} \right] \approx
   \sin^2 2 \theta \sin^2\left[ 1.267\Delta m^2 {L\over E} \right],
\label{oscprob}
\ee
where the second form assumes that the masses are given in eV, the distance, 
$L \approx t$, traveled from production to detection in km, and the neutrino 
energies, $E \approx E_1 \approx E_2$, in GeV. It is assumed that both $L$ and 
$E$ are fixed and known with negligible uncertainty. Otherwise, one has to take
appropriate averages. For example, one may have to correct for finite source or
detector effects or non-monochromaticity. In the limit of very large 
oscillations, $P(\nu_\alpha\rightarrow\nu_\beta)\rightarrow\sin^2(2\theta)/2$.
Quantities of the form $P(\nu_\alpha\rightarrow\nu_\beta)$ are measured in what
are called $\nu_\beta$ appearance experiments, while 
$P(\nu_\alpha\rightarrow\nu_\alpha)$ is the object of interest in $\nu_\alpha$ 
disappearance experiments. 

This formalism is easily extended to multi-flavor oscillations, where the three
standard neutrinos (even if their masses are of Dirac type) will in general 
produce CP violation in the lepton sector. These flavor oscillations are also
called {\em first class}. If one wishes to introduce a fourth neutrino, this 
would almost certainly be sterile (see Section~\ref{SM}), as the number of 
active neutrinos, $N_\nu = 2.986\pm 0.007$, with $m_\nu \ll M_Z/2$ is strongly 
constrained by LEP (see Section~\ref{SMstatus}). Allowing $m_\nu \gsim M_Z/2$ 
to avoid this would either require the introduction of a whole set 
of fermions (such as a fourth generation) to cancel gauge anomalies which in 
turn (due to chiral non-decoupling effects) would clash with EW precision data;
or would require a vector-like pair of lepton doublets. In any case, 
the experimental hints and signals to be discussed in the Subsections below, 
all involve very small $\Delta m^2$, so that the LEP bound on active neutrinos
holds. Oscillations of ordinary into sterile neutrinos~\cite{Bilenky:1976yj}, 
$\nu_s$, are called {\em second class}.

To simplify the discussion and to facilitate the comparison between different
experiments we will always assume simple two-neutrino oscillation scenarios.
We primarily summarize the experiments yielding the most stringent 
limits~\cite{Murayama:eb}.

\subsection{\it Solar Neutrinos}
Detectors observing solar neutrinos fall into two categories\footnote{For 
a recent overview of solar neutrino experiments, see 
Refs.~\cite{Nakamura:dz,Goswami:2003bh}.}. Radiochemical detectors (Cl or Ga) 
are blind to any spectral or directional information, and time variation studies
are limited by the extraction periods of typically several weeks. But they have
the great advantage of relatively low neutrino energy thresholds set by the CC 
reactions,
\be
   \nu_e + {}^{37}{\rm Cl} \rightarrow e^- + {}^{37}{\rm Ar}, \hspace{50pt}
   \nu_e + {}^{71}{\rm Ga} \rightarrow e^- + {}^{71}{\rm Ge}.
\label{radiochemical}
\ee
$\check{\rm C}$herenkov detectors (light or heavy water) are only sensitive to
the highest energy neutrinos, but are real time experiments. In addition,
energy and direction of the incident neutrino can be estimated, and there is
a NC component in the basic electron scattering (ES) process,
\be
   \nu + e^- \rightarrow \nu + e^- .
\label{escattering}
\ee

\subsubsection{Rates}
\label{sec:rates}
The experimental results of the capture rates in the radiochemical experiments
and of the neutrino fluxes in the $\check{\rm C}$herenkov detectors are
compared to predictions based on the Standard Solar Model (SSM). We use here 
the latest available results~\cite{Bahcall:2001cb} (updated from 
Ref.~\cite{Bahcall:2000nu}) which include the recent 
measurement~\cite{Junghans:2001ee} of the low energy Beryllium-proton fusion 
cross section\footnote{The cross section quoted in Ref.~\cite{Junghans:2001ee}
is larger (but consistent) and more precise than the one quoted in 
Ref.~\cite{Adelberger:1998qm} which was employed in Ref.~\cite{Bahcall:2000nu}.
This drives the SSM predictions for all fluxes and capture rates up. The most 
recent result by the ISOLDE Collaboration~\cite{Baby:2002hj} is slightly lower,
of the same precision, and consistent within 1.1~$\sigma$ when compared to 
the one in Ref.~\cite{Junghans:2001ee} (which is, however, currently under 
revision~\cite{Baby:2002hj}).} Experiments using $\check{\rm C}$herenkov 
detectors quote results assuming no neutrino oscillations and an undistorted 
neutrino spectrum, where the latter has been verified experimentally. 

The Cl detector in the Homestake mine in South Dakota represented the first 
solar neutrino experiment and also the one with the longest running time 
(1970-1994). The combined result of 108 extractions of solar induced 
${}^{37}{\rm Ar}$ is~\cite{Cleveland:nv} (1 SNU $= 1$ interaction per second 
and $10^{36}$ target atoms),
\be
   2.56 \pm 0.16 \mbox{ (stat.)} \pm 0.16 \mbox{ (syst.) SNU}
   \hspace{50pt} [\mbox{SSM: } 8.59^{+1.1}_{-1.2} \mbox{ SNU}].
\ee
Thus, the ratio of the observed to expected capture rates,
\be
   R\ (^{37}{\rm Cl}) = 0.298^{+0.049}_{-0.046},
\label{clratio}
\ee
corresponds to a deficit\footnote{The uncertainties are mainly fractional, so 
that it is more appropriate to assume an (asymmetric) log-normal rather than 
a normal distribution. Note also that Gaussian error distributions are not 
well defined for positive-definite quantities like capture rates or ratios 
thereof.} of 7.3~$\sigma$. The SSM also estimates that almost 79\% of 
the neutrinos with energies above the Cl threshold of 0.814~MeV originate from
${}^8$B disintegration, and more than 13\% from ${}^7$Be electron capture.

The deep underground SAGE experiment at the Baksan Neutrino Observatory in 
Russia was the first Ga experiment to take data and is still on-going. 
The advantage of Ga over Cl is the lower neutrino threshold of only 0.233~MeV
allowing the detection of the dominant proton-proton fusion neutrinos (pp), 
which contribute about 54\% of the flux, with ${}^7$Be (26\%) and ${}^8$B 
(11\%) subleading. Recently, the SAGE Collaboration presented 
results~\cite{Abdurashitov:2002nt} of their first 12 years (92 runs) of 
observation from January 1990 through December 2001,
\be
      70.8^{+5.3}_{-5.2} \mbox{ (stat.)}^{+3.7}_{-3.2} \mbox{ (syst.) SNU}. 
\label{sage}
\ee
A similar experiment with the GALLEX detector at the Gran Sasso Underground 
Laboratories collected a total of 65 runs from May~1991 through January~1997.
The GALLEX Collaboration found a capture rate~\cite{Hampel:1998xg},
\be
      77.5 \pm 6.2 \mbox{ (stat.)}^{+4.3}_{-4.7} \mbox{ (syst.) SNU},
\label{gallex}
\ee
in agreement with SAGE. Subsequently, GALLEX was succeeded by GNO, an upgraded
and improved Ga experiment. The first published capture rate by the GNO 
Collaboration~\cite{Altmann:2000ft},
\be
      65.8^{+10.2}_{-\;\;9.6} \mbox{ (stat.)}^{+3.4}_{-3.6}\mbox{ (syst.) SNU},
\label{gno}
\ee
is based on 19 runs between May~1998 and January~2000 and consistent with both
SAGE and GALLEX. The combined result of GALLEX and the first GNO period 
is~\cite{Altmann:2000ft},
\be
      74.1 \pm 5.4 \mbox{ (stat.)}^{+4.0}_{-4.2} \mbox{ (syst.) SNU}.
\label{gransasso}
\ee
Finally, we combine the results~(\ref{sage}) and (\ref{gransasso}) by
(conservatively) assuming the smaller systematic uncertainty (SAGE) as common 
to SAGE and the Gran Sasso experiments. We obtain a total observed Ga capture 
rate,
\be
      72.2^{+5.3}_{-5.1} \mbox{ SNU} \hspace{50pt} 
      [\mbox{SSM: } 130^{+9}_{-7} \mbox{ SNU}].
\ee
and for the ratio of the observed to expected capture rates,
\be
   R\ (^{71}{\rm Ga}) = 0.556^{+0.051}_{-0.055},
\label{garatio}
\ee
corresponding to a deficit\addtocounter{footnote}{-1}\footnotemark\ of 
6.4~$\sigma$. Thus, both types of radiochemical experiments see very large and
statistically significant deficits. Moreover, taking the ratio of 
the ratios~(\ref{clratio}) and (\ref{garatio}) shows at the 3.2~$\sigma$ level (or 
more, considering that the uncertainties from the SSM partly cancel in 
the double ratio) that the deficits do not correspond to an overall reduction.

The Kamiokande water $\check{\rm C}$herenkov detector in the Kamioka mine in
Japan studied solar neutrinos {\em via\/} electron 
scattering~(\ref{escattering}) from January~1987 to February~1995. The recoil 
electron energy threshold was chosen relatively high to avoid large backgrounds
and varied between 7.0 and 9.3~MeV. This implies sensitivity almost exclusively
to the ${}^8$B neutrino flux, $\Phi (^8{\rm B})$, including a few per mille 
neutrino contribution from Helium-proton (hep) fusion\footnote{We will ignore 
hep neutrinos in the following as they have not been treated identically in 
the various papers.}. The Kamiokande Collaboration found 
the flux~\cite{Fukuda:1996sz}, 
\be
   \Phi (^8{\rm B}) = 2.80 \pm 0.19 \mbox{ (stat.)} \pm 0.33 \mbox{ (syst.)}
   \times 10^6 {\rm cm}^{-2} {\rm s}^{-1}.
\label{kamiokande}
\ee
Subsequently, the Kamiokande detector was succeeded by the Super-Kamiokande 
upgrade including an increase in fiducial volume by a factor of~33. The first 
phase (Super-Kamiokande-I~\cite{Fukuda:2002pe}) had an electron energy 
threshold of 5~MeV and took place from May~1996 through July~2001. It resulted 
in a flux measurement~\cite{Smy:2002rz},
\be
   \Phi (^8{\rm B}) = 2.35 \pm 0.02 \mbox{ (stat.)} \pm 0.08 \mbox{ (syst.)} 
   \times 10^6 {\rm cm}^{-2} {\rm s}^{-1}
\ee
which is consistent with the flux~(\ref{kamiokande}). 

SNO is a heavy water $\check{\rm C}$herenkov detector in the Creighton mine in
Ontario, Canada. During the first phase of the experiment, from November~1999 
to May~2001, only D$_2$O was present in the detector. In the final analysis of
the completed first phase an electron energy threshold identical to that of 
Super-Kamiokande was chosen. The SNO Collaboration reported~\cite{Ahmad:2002jz}
a $^8$B flux based on electron scattering~(\ref{escattering}),
\be
   \Phi (^8{\rm B}) = 2.39^{+0.24}_{-0.23}\mbox{ (stat.)} \pm 0.12
   \mbox{ (syst.)} \times 10^6 {\rm cm}^{-2} {\rm s}^{-1},
\ee
which is in good agreement with the Kamioka experiments. The different energy 
threshold\footnote{The different electron energy threshold may also have 
a small effect (which we ignore) on the NC contribution to the electron 
scattering cross section.} at Kamiokande and the different liquid at SNO 
should imply negligible common systematics among the three experiments. 
Assuming this, we find the combined $^8$B flux from electron scattering,
\be
   \Phi (^8{\rm B;ES}) = 2.37 \pm 0.08 \times 10^6 {\rm cm}^{-2} {\rm s}^{-1} 
   \hspace{50pt} 
   [\mbox{SSM: } 5.93^{+0.83}_{-0.89} \times 10^6 {\rm cm}^{-2} {\rm s}^{-1}],
\label{ESflux}
\ee
and for the ratio of the observed to expected $^8$B fluxes,
\be
   R\ (^8{\rm B;ES}) = 0.400_{-0.057}^{+0.061},
\label{ESratio}
\ee
again showing a 6.0~$\sigma$ deficit\addtocounter{footnote}{-3}\footnotemark. 
Unique to SNO is the ability to study CC and NC deuterium breakup,
\be
   \nu_e + d \rightarrow e^- + p + p, \hspace{50pt}
   \nu   + d \rightarrow \nu + p + n,
\ee
which in its first phase resulted, respectively, in flux measurements,
\bea
   &&\Phi (^8{\rm B;CC}) = 1.76^{+0.06}_{-0.05}\mbox{ (stat.)} \pm 0.09 
   \mbox{ (syst.)} \times 10^6 {\rm cm}^{-2} {\rm s}^{-1},  \\ \nonumber
   &&\Phi (^8{\rm B;NC}) = 5.09^{+0.44}_{-0.43}\mbox{ (stat.)}^{+0.46}_{-0.43} 
   \mbox{ (syst.)} \times 10^6 {\rm cm}^{-2} {\rm s}^{-1}, 
\label{CCandNCfluxes}
\eea
and in ratios of observed to expected $^8$B fluxes,
\be
   R\ (^8{\rm B;CC}) = 0.297_{-0.045}^{+0.048}, \hspace{50pt}
   R\ (^8{\rm B;NC}) = 0.86_{-0.16}^{+0.17}.
\label{CCandNCratios}
\ee
The CC ratio which shows another 7.5~$\sigma$ deficit is virtually identical to
$R\ (^{37}{\rm Cl})$ implying that $^7$Be and $^8$B neutrino depletions are 
approximately equal. In contrast, the NC ratio is consistent with SSM 
expectations. After completion of the first phase of SNO, NaCl was added to 
the detector which enhances the capture efficiency of the NC neutrons 
drastically. The third phase (scheduled for summer of~2003) will be 
characterized by the addition of $^3$He proportional counter tubes as 
an alternative way to increase the NC efficiency.

Thus, there are not only statistically convincing neutrino deficits in all 
three CC channels ($^{37}$Cl, $^{71}$Ga, and $^8$B), but there is also strong 
evidence (at the $\gsim 4~\sigma$~CL) that the low energy (mainly pp) neutrinos
are suppressed less than $^7$Be or $^8$B neutrinos. On the other hand, 
$R\ (^8{\rm B;NC})$ is consistent with SSM expectations implying that 
the solar neutrino flux is dominated by active neutrinos. There is redundancy
in the data since $\Phi (^8{\rm B;ES})$ can be expressed in terms of 
$\Phi (^8{\rm B;CC})$ and $\Phi (^8{\rm B;NC})$,
\be
   \Phi (^8{\rm B;ES}) = (1 - \epsilon) \Phi (^8{\rm B;CC}) + 
                              \epsilon  \Phi (^8{\rm B;NC}),
\label{ESCCNC}
\ee
where $1/\epsilon =6.48$ is the ratio of theoretical cross sections for $\nu_e$
relative to the sum of $\nu_\mu$ and $\nu_\tau$. The ratios~(\ref{CCandNCfluxes}) 
then imply, $\Phi (^8{\rm B;ES}) = 2.27^{+0.13}_{-0.12}\times 10^6 
{\rm cm}^{-2} {\rm s}^{-1}$, in perfect agreement with the measurement~(\ref{ESflux}). 
Thus, flux measurements of solar neutrinos provide very strong evidence for 
$(\nu_e \rightarrow \nu_\mu, \nu_\tau)$ flavor oscillations with any component
of oscillations into $\nu_s$ strongly constrained. The Super-Kamiokande 
Collaboration also sets limits~\cite{Gando:2002ub} on conversions into 
$\bar\nu_e$.

Note that with the crucial SNO result, 
$\Phi (^8{\rm B;NC}) \gg \Phi (^8{\rm B;CC})$, relation~(\ref{ESCCNC}) fully 
explains the observation, $R\ (^8{\rm B;ES}) > R\ (^{37}{\rm Cl})$. Due to 
(i) a downward shift in $R\ (^8{\rm B;ES})$ (Super-Kamiokande and SNO relative
to Kamiokande); (ii) an upward shift in $R\ (^{37}{\rm Cl})$ (final relative to
preliminary Homestake results); and (iii) an upward shift in 
the Beryllium-proton cross section\addtocounter{footnote}{-4}\footnotemark\ 
(Ref.~\cite{Junghans:2001ee} relative to Ref.~\cite{Adelberger:1998qm}), 
the current data do not favor strongest depletion for $^7$Be neutrinos (or even
a vanishing $^7$Be flux at the best fit~\citer{Bludman:1993tk,Krastev:1994nx})
as used to be the case in the past. A new experiment, 
BOREXINO~\cite{Alimonti:2000xc}, with the capability to settle the question of
the $^7$Be flux, is currently under construction at Gran Sasso. It is 
a (real time) liquid scintillator detector and with a threshold energy of only
250~keV, it combines the advantages of the radiochemical and 
$\check{\rm C}$herenkov type detectors. Completion of the construction phase is
expected in 2003 (a counting test facility is already in operation).

\subsubsection{Time variations and spectral distortions}
Signal variations on various time scales can give additional hints as to what 
the origin of the solar neutrino deficit might be. The Homestake, Kamiokande,
and SAGE experiments took data over time spans of the order of solar cycles.
With the exception of a small and controversial hint at an anti-correlation of 
the capture rate in some of Homestake's preliminary results, there is no 
evidence for a solar neutrino flux correlation with sun spot activity.

Super-Kamiokande finds a seasonal rate variation at the 2.5~$\sigma$~CL, 
consistent with the expectation from the distance variation between the sun and
the Earth. 

Neither Kamiokande nor Super-Kamiokande reported a significant day-night 
electron scattering cross section asymmetry. For example, Super-Kamiokande 
quotes,
\be
   A_{\rm DN} ({\rm ES})\equiv 2 {\sigma_D - \sigma_N\over \sigma_D + \sigma_N}
   = - 0.021 \pm 0.020 \mbox{ (stat.)}^{+0.013}_{-0.012} \mbox{ (syst.)}.
\label{ADNSuperK}
\ee
On the other hand, SNO finds (at the 2.2~$\sigma$~CL) a stronger night time CC 
cross section~\cite{Ahmad:2002ka},
\be
\ba{l}
   A_{\rm DN} ({\rm ES}) =   0.174 \pm 0.195 \mbox{ (stat.)}^{+0.022}_{-0.024}
                                             \mbox{ (syst.)},\\[4pt]
   A_{\rm DN} ({\rm CC}) = - 0.140 \pm 0.063 \mbox{ (stat.)}^{+0.014}_{-0.015}
                                             \mbox{ (syst.)},\\[4pt]
   A_{\rm DN} ({\rm NC}) =   0.204 \pm 0.169 \mbox{ (stat.)}^{+0.025}_{-0.024}
                                             \mbox{ (syst.)}.
\label{ADNSNO}
\ea
\ee
As is the case for the rates, there is also redundancy in the asymmetry 
measurements,
\be
  A_{\rm DN} ({\rm ES}) = {A_{\rm DN} ({\rm CC})\over 1 + {\epsilon
  R\ (^8{\rm B;NC})\over (1 - \epsilon) R\ (^8{\rm B;CC})}} + 
  {A_{\rm DN} ({\rm NC})\over 1 + {(1 - \epsilon) R\ (^8{\rm B;CC})\over 
  \epsilon R\ (^8{\rm B;NC})}}.
\label{AESCCNC}
\ee
From the measurements~(\ref{CCandNCfluxes}) we find independently of the solar model,
$R\ (^8{\rm B;CC})/R\ (^8{\rm B;NC}) = 0.346^{+0.046}_{-0.048}$, and with that
from Eq.~(\ref{AESCCNC}) the central value, $A_{\rm DN} ({\rm ES}) = - 0.021$,
in agreement with the result~(\ref{ADNSuperK}) and the first result~(\ref{ADNSNO}), but 
the uncertainties are large and strongly anti-correlated~\cite{Ahmad:2002ka} 
between $A_{\rm DN} ({\rm CC})$ and $A_{\rm DN} ({\rm NC})$. If one assumes that
flavor oscillations into active neutrinos saturate the solar flux, 
$R\ (^8{\rm B;NC}) = 1$ and $A_{\rm DN} ({\rm NC}) = 0$, then there is only one
independent asymmetry and with these constraints SNO finds~\cite{Ahmad:2002ka},
\be
   A_{\rm DN} ({\rm CC}) = - 0.070 \pm 0.049 \mbox{ (stat.)}^{+0.012}_{-0.013}
                                             \mbox{ (syst.)}.
\ee
This agrees with the Super-Kamiokande result obtained from~(\ref{ADNSuperK}) 
and Eq.~(\ref{AESCCNC}), 
\be
   A_{\rm DN} ({\rm CC}) = - 0.032 \pm 0.031 \mbox{ (stat.)}^{+0.020}_{-0.018} 
                                             \mbox{ (syst.)},
\ee
and can be combined with the SNO result to give,
\be
   A_{\rm DN} ({\rm CC}) = - 0.045^{+0.030}_{-0.029}, 
\ee
which is non-vanishing at the 1.5~$\sigma$~CL.

Super-Kamiokande and SNO have enough statistics to divide their data 
{\em simultaneously\/} into energy and zenith angle (day-night in the case of
SNO) bins. The observed energy dependence is in both cases consistent with SSM
expectations.

\subsubsection{The large mixing angle MSW solution to the solar neutrino 
problem}
\begin{figure}[t]
\begin{minipage}[t]{10 cm}
\hspace{60pt}\epsfig{file=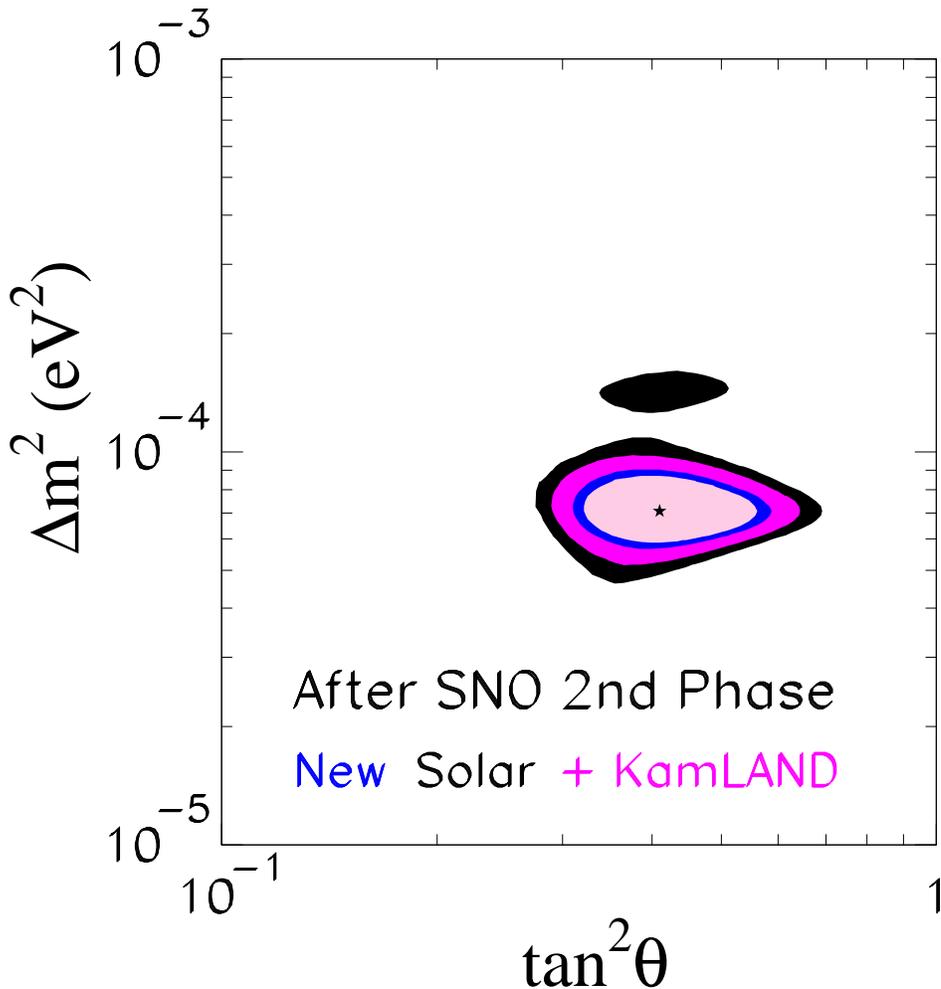,scale=0.95}
\end{minipage}
\begin{center}
\begin{minipage}[t]{16.5 cm}
\caption{The allowed parameter space (from Ref.~\cite{Bahcall:2003ce}) at 90\%,
95\%, 99\%, and 99.73\%~CL in the $\tan^2 \theta_\odot$-$\Delta m^2_\odot$ 
plane. Note that without loss of generality one can take $\Delta m^2_\odot$ 
non-negative. $\theta_\odot$ is then defined in the interval $[0,\pi/2]$. 
$\theta_\odot > \pi/4$ is allowed by the KamLAND reactor neutrinos since 
Eq.~(\ref{oscprob}) (valid in vacuum) is symmetric under 
$\theta \rightarrow \pi/2 - \theta$, but $\theta_\odot < \pi/4$ is selected by
the requirement to have an MSW resonance effect.
\label{bahcall}}
\end{minipage}
\end{center}
\end{figure}

In the past, there have been several solutions of the solar neutrino deficit
problem in terms of neutrino oscillations depending on the mass splitting,
$\Delta m^2_\odot$, and the mixing angle, $\theta_\odot$. The quality of 
the best fit within each solution has varied significantly over time. All but
one of these solutions are by now highly disfavored. For example, the absence
of spectral distortions as mentioned in the previous paragraph rules out any 
solution at small mixing~\cite{Fukuda:2002pe} independently of solar model flux
predictions. The same conclusion applies~\cite{Fukuda:2002pe} to the vacuum 
oscillation solution characterized by very small mass splittings. Some other
parameter regions predict large day-night variations (by matter effects in 
the Earth) contradicting observations still independent of any solar model. 
The predictions of the SSM enter only at the level of rate comparisons which 
strongly favor a solution proposed by Mikheev, Smirnov~\cite{Mikheev:gs}, and 
Wolfenstein~\cite{Wolfenstein:1977ue} (MSW). MSW type solutions invoke 
a resonant enhancement (for $\theta_\odot > \pi/4$ one predicts a reduction) of
flavor oscillations by matter effects in the sun and can occur for a variety of
parameters. But only the MSW solution with a large mixing angle and relatively
large $\Delta m^2_\odot$ gives a good description of all data. 
Moreover, this solution has been confirmed by the KamLAND reactor experiment as
discussed in Section~\ref{reactornus}. Combining the KamLAND and all solar
neutrino experiments~\citer{Barger:2002at,Bahcall:2003ce} splits the large 
angle MSW solution region into two parts\footnote{After this Section had been
completed, the NaCl phase data of SNO~\cite{Ahmed:2003kj} (see 
Section~\ref{sec:rates}) and some new results by the SAGE and 
GNO~Collaborations appeared. These are included in Figure~\ref{bahcall}, and 
have the effect of almost eliminating the upper part of the allowed region 
which now appears only at the 99.73\%~CL.} as shown in Figure~\ref{bahcall}.

Since the MSW effect is based on neutrino interactions in the sun it can also
(in parts of parameter space) distinguish between oscillations into active 
{\em vs.\/} sterile neutrinos. The large mixing angle solution fits the data 
only for oscillations dominantly into active neutrinos, in perfect agreement 
with the NC cross section seen by SNO.

\subsection{\it Atmospheric Neutrinos}
Cosmic rays hitting the upper atmosphere of the Earth produce a rich source of
neutrinos with variation in flavor (primarily $\nu_e$, $\bar\nu_e$, $\nu_\mu$, 
and $\bar\nu_\mu$), energy (experiments probe the range from 
${\cal O} (100~\mbox{MeV})$ to ${\cal O} (100~\mbox{GeV})$, and distance 
(depending on the zenith angle $L$ ranges from ${\cal O} (10~\mbox{km})$ to 
the Earth radius of about 13,000 km). Neutrinos with energies of no more than 
a few GeV produce events that are fully contained (FC) in the detectors. Muons
produced by neutrinos outside the detector can also be studied if they are 
energetic enough to enter the detector and if they are going up so that they 
can be distinguished from cosmic ray muons. If such muons have energies of 
${\cal O} (10~\mbox{GeV})$ they are likely to stop within the detector and are
referred to as partially contained (PC). If they have energies even larger
than this they go through the detector.

\subsubsection{Rates}
The Kamiokande-II Collaboration was the first to observe~\cite{Hirata:1992ku}
that the ratio\footnote{There are also several experiments quoting observed to 
expected $\nu_\mu$ flux ratios, but the uncertainty from the flux prediction is
much larger than in $R_{\mu e}$. Nevertheless, most of these experiments are
consistent with the atmospheric neutrino deficit.}, $R_{\mu e}$, of muon 
neutrinos, $\nu_\mu$ or $\bar\nu_\mu$, to electron neutrinos, $\nu_e$ or 
$\bar\nu_e$, is smaller than the expected value of $R_{\mu e} \approx 2$. 
Kamiokande determines~\cite{Fukuda:1994mc} the double ratio of observed 
to expected events (which from now on we will take as the definition of 
$R_{\mu e}$),
\be
   R_{\mu e} = 0.60^{+0.06}_{-0.05} \mbox{ (stat.)} \pm 0.05 \mbox{ (syst.)},
   \hspace{50pt}
   R_{\mu e} = 0.57^{+0.08}_{-0.07} \mbox{ (stat.)} \pm 0.07 \mbox{ (syst.)},
\label{kamiokandeatm}
\ee
where the first (second) value is for $\nu$s with an average energy in 
the sub-GeV (multi-GeV) region. These correspond to 3.9~$\sigma$ and 
2.3~$\sigma$ deficits, respectively. A similar deficit was 
seen~\cite{Becker-Szendy:hq} with the IMB-3 $\check{\rm C}$herenkov detector in
the Morton salt mine in Ohio, but it was not observed~\cite{Daum:bf} with iron 
calorimeters such as the detector in the Frejus Underground Laboratory close
to the French-Italian border. On the other hand, the 3.3 times larger 
Super-Kamiokande detector confirmed~\cite{Fukuda:1998tw,Fukuda:1998ub} 
the results~(\ref{kamiokandeatm}),
\be
   R_{\mu e} = 0.61 \pm 0.03 \mbox{ (stat.)} \pm 0.05 \mbox{ (syst.)},
   \hspace{50pt}
   R_{\mu e} = 0.66 \pm 0.06 \mbox{ (stat.)} \pm 0.08 \mbox{ (syst.)}.
\label{superkatm}
\ee
Subsequently, the Soudan-2 Collaboration in the Tower-Soudan Iron Mine in 
Minnesota reported~\cite{Allison:1999ms} the precise measurement for sub-GeV
neutrinos,
\be
   R_{\mu e} = 0.64 \pm 0.11 \mbox{ (stat.)} \pm 0.06 \mbox{ (syst.)}.
\label{soudanatm}
\ee
This is only a 2.3~$\sigma$ effect but it is in perfect agreement 
with~(\ref{kamiokandeatm}) and (\ref{superkatm}) and the first time that 
an iron calorimeter confirmed the atmospheric neutrino deficit seen with water
$\check{\rm C}$herenkov detectors.

We combine the results~(\ref{kamiokandeatm}), (\ref{superkatm}), and (\ref{soudanatm})
assuming that the smaller systematic uncertainty between Kamiokande and 
Super-Kamiokande is common to both, and that Soudan-2 is uncorrelated with 
the water $\check{\rm C}$herenkov results. The results,
\be
   R_{\mu e} = 0.614^{+0.052}_{-0.051}, \hspace{50pt} R_{\mu e} = 0.63\pm 0.09,
\label{allatm}
\ee
correspond to 5.8~$\sigma$ and 3.2~$\sigma$ deficits, respectively.

\subsubsection{Zenith angle and spectral distributions}
Super-Kamiokande~\cite{Fukuda:1998ub} also formed the ratio of the number of 
upward to downward $\mu$-like events,
\be
   R^{\uparrow\downarrow} 
   = 0.52^{+0.07}_{-0.06}\mbox{ (stat.)} \pm 0.01\mbox{ (syst.)}\hspace{50pt}
   [\mbox{expected: } 0.98 \pm 0.03 \mbox{ (stat.)} \pm 0.02\mbox{ (syst.)}].
\ee
This means a 4.5~$\sigma$ effect in a ratio that is independent of the overall
flux prediction, while the analogous ratio for $e$-like events was consistent 
with unity. The significance in both ratios, $R_{\mu e}$ and 
$R^{\uparrow\downarrow}$, increased further and in Ref.~\cite{Fukuda:1998mi}
Super-Kamiokande provided detailed distributions of the event rates {\em vs.\/}
zenith angle and $L/E_\nu$, demonstrating an excellent fit for two-flavor 
oscillations into $\nu_\tau$ or $\nu_s$ with $\sin^2 2\theta > 0.82$ and 
0.0005~eV$^2 < \Delta m^2 < 0.006$~eV$^2$ at 90\%~CL. Agreement of the $e$-like
data with expectations disfavored oscillations into $\nu_e$ in agreement with 
the exclusion regions established by reactor experiments as discussed in 
Section~\ref{reactornus}. Furthermore, a study of upward through-going muons of
minimum energy 1.6~GeV by Super-Kamiokande found that the absolute flux is in 
agreement with the prediction~\cite{Fukuda:1998ah} while the zenith angle 
dependence does not agree with no-oscillation predictions. Rather, the observed
distortion suggested $\nu_\mu\rightarrow\nu_\tau$ or $\nu_\mu\rightarrow\nu_s$
oscillations and so did the double ratio of observed to expected PC neutrinos 
relative to through-going ones~\cite{Fukuda:1999pp},
\be
   0.22 \pm 0.02 \mbox{ (stat.)} \pm 0.01 \mbox{ (syst.)} \hspace{50pt}
   [\mbox{expected: } 0.37^{+0.05}_{-0.04} \mbox{ (theo.)}],
\ee
which revealed another 3.5~$\sigma$ deficit. A simultaneous fit to zenith angle
distributions of upward stopping and through-going muons yielded,
$\sin^2 2\theta > 0.7$ and 0.0015~eV$^2 < \Delta m^2 < 0.015$~eV$^2$ at 90\%~CL
in agreement with the FC sample. Finally, the difference in zenith angle 
distribution due to NCs and matter effects rejected the hypothesis of 
oscillations into sterile neutrinos at the 99\%~CL, while 
$\nu_\mu\rightarrow\nu_\tau$ oscillations are sufficient to explain all 
results~\cite{Fukuda:2000np}.

The MACRO detector at Gran Sasso confirmed~\cite{Ambrosio:2001je} 
the preference for $\nu_\mu\rightarrow\nu_\tau$ oscillations of atmospheric 
neutrinos using their through-going muon sample. They also 
studied~\cite{Ambrosio:2003yz} the spectral and angular dependence of 
through-going muons, and provide a measurement of the ratio of vertical 
($\cos\theta \leq - 0.7$) to horizontal ($\cos\theta \leq - 0.7$) events,
\be
   1.48 \pm 0.13 \hspace{50pt} [\mbox{expected: } 2.20 \pm 0.17]
\ee
which differs by 3.4~$\sigma$ from the no-oscillation hypothesis while it is in
reasonable agreement with the oscillation prediction, $1.70\pm 0.14$.

The latest piece of evidence for atmospheric neutrino oscillations came from 
the K2K accelerator neutrino experiments as discussed in 
Section~\ref{acceleratornus}. Ref.~\cite{Fogli:2003th} presents a simultaneous
analysis of the Super-Kamiokande and K2K results and quote,
\be
   \sin^2 2\theta = 1.00^{+0.00}_{-0.05}, \hspace{50pt} 
   \Delta m^2 = 0.0026 \pm 0.0004 \mbox{ eV}^2.
\ee

\subsection{\it Accelerator Neutrinos}
\label{acceleratornus}
As can be seen from Eq.~(\ref{oscprob}), an ideal experiment searching for
two-neutrino oscillations is characterized by only two parameters, 
$P(\nu_\alpha\rightarrow\nu_\beta)$ and $L/E$. The oscillation probability for
a realistic experiment in which there is some uncertainty or variation in $P$,
$L$, and/or $E$ can be much more complicated (and may be based on 
an event-by-event likelihood), but the resulting limits are often still 
characterized by only two parameters although these may not describe every 
detail of an exclusion region~\cite{Groom:ec}. This is frequently the case
for accelerator and reactor neutrino experiments with known neutrino fluxes,
and with a single detector whose size is small compared to $L$. In such
cases $P(\nu_\alpha\rightarrow\nu_\beta)$ and $L/E$ (or effective definitions
thereof) translate into a lower limit on $\sin^2 2\theta$ at large $\Delta m^2$
independent of $L/E$, and a lower limit on $\Delta m^2$ at maximal mixing,
$\sin^2 2\theta = 1$. 

The KARMEN Collaboration searched for neutrino oscillations at the ISIS Neutron
Spallation Source at the Rutherford Appleton Laboratory in England studying 
$\mu^+$ decays at rest. Their oscillation length was $L = 17.6$~m, and 
the neutrino energy at their kinematic endpoint was $E = 52.8$~MeV. 
In the $\bar\nu_\mu \rightarrow \bar\nu_e$ appearance mode, studied in inverse
$\beta$ decay, $\bar\nu_e p \rightarrow e^+ n$, 15~candidate reactions were
seen with the KARMEN~2 detector (operating from February~1997 through
March~2000) which is consistent with $15.8 \pm 0.5$ expected background events.
The Collaboration sets the 90\%~CL limits~\cite{Armbruster:2002mp}, 
\be
\ba{lcl}
   \sin^2 2\theta < 0.0017 &\hspace{50pt}& (\Delta m^2 > 100 \mbox{ eV}^2),\\
   \Delta m^2 < 0.055 \mbox{ eV}^2 &\hspace{50pt}& (\sin^2 2\theta = 1).
\ea
\label{karmen}
\ee
This excludes most of the parameter space to which the LSND Collaboration at 
the Los Alamos Neutron Science Center had access to in a similar experiment,
but with $L$ dominantly at 110~m and 135~m (1993-1995) and at $L = 30$~m 
(1996-1998). In contrast to KARMEN, LSND reports~\cite{Aguilar:2001ty}
a signal of $87.9 \pm 22.4 \pm 6.0$ inverse $\beta$ decay events above 
the expected background. This excess corresponds to an oscillation probability 
of $P = 0.264 \pm 0.067 \pm 0.045$ (a 3.3~$\sigma$ effect), and a best fit 
occurring at,
\be
   \sin^2 2\theta = 0.003, \hspace{50pt} \Delta m^2 = 1.2 \mbox{ eV}^2.
\ee
This is excluded by the limits~(\ref{karmen}), but its larger values of $L$ extends
the sensitivity of LSND to slightly smaller values of $\Delta m^2$ so that 
the findings of the two groups do not necessarily conflict. The BooNE
Collaboration~\cite{Church:1997jc} at FNAL is able to test the LSND result in 
both the $\nu_\mu \rightarrow \nu_e$ appearance and the $\nu_\mu$ disappearance
modes. Data taking of the single detector version of the experiment, MiniBooNE,
commenced in August~2002. If a positive signal is seen, the experiment will be
upgraded by a second detector at a distance optimized for a precision 
determination of oscillation parameters and searches for CP and CPT 
violation\footnote{If the LSND signal can be confirmed, the most likely 
interpretation of it will not be in terms of a forth neutrino but rather in 
terms of a more fundamental type of NP such as CPT 
violation~\cite{Barenboim:2001ac} since any additional neutrino is strongly 
constrained~\cite{Barger:2003zg} by big bang nucleosynthesis. This is true even
for sterile neutrinos as long as there is appreciable mixing with active 
neutrinos as would apply to LSND.} in the lepton sector.

The results of the reactor disappearance experiments discussed in 
Subsection~\ref{reactornus} can be used as constraints on 
$\nu_e \rightarrow \nu_\tau$ oscillations. In addition, there were two 
$\nu_\tau$ appearance experiments, CHORUS~\cite{Eskut:2000de} and 
NOMAD~\cite{Astier:2001yj}, at the SPS accelerator at CERN operating from 1994
through 1997. The 450~GeV protons of the SPS were aimed at a Beryllium target 
located 850~m (940~m) from the CHORUS (NOMAD) detector. The average values of 
$L$ for the two experiments were a few hundred meters shorter than this. 
The average neutrino energy of the almost pure $\nu_\mu$ wide-band beam was 
$E \approx 25$~GeV. Due to the relatively small $L/E$, these experiments were 
only sensitive to larger $\Delta m^2$, but the small $\nu_e$ component of 
the beam was sufficient to achieve stronger limits on $\nu_e - \nu_\tau$ mixing
at large mass splittings than the reactor experiments discussed in 
Subsection~\ref{reactornus},
\be
\ba{lcl}
\sin^2 2\theta < 0.052 &\hspace{50pt}& (\mbox{CHORUS at large } \Delta m^2), \\
\sin^2 2\theta < 0.015 &\hspace{50pt}& (\mbox{NOMAD  at large } \Delta m^2).
\ea
\label{chorusandnomad}
\ee
The primary physics goal of CHORUS and NOMAD, however, were studies of
$\nu_\mu \rightarrow \nu_\tau$ oscillations. CHORUS set limits,
\be
\ba{lcl}
   \sin^2 2\theta < 0.00068 &\hspace{50pt}& (\mbox{large } \Delta m^2), \\
   \Delta m^2 < 0.6 \mbox{ eV}^2 &\hspace{50pt}& (\sin^2 2\theta = 1),
\ea
\label{chorus}
\ee
while NOMAD finds,
\be
\ba{lcl}
   \sin^2 2\theta < 0.00033 &\hspace{50pt}& (\mbox{large } \Delta m^2), \\
   \Delta m^2 < 0.7 \mbox{ eV}^2 &\hspace{50pt}& (\sin^2 2\theta = 1).
\ea
\label{nomad}
\ee
These limits are complemented at large mixing by an older (1983) $\nu_\mu$
disappearance experiment~\cite{Dydak:1983zq} at the PS accelerator at CERN.
The Collaboration (which continued later as CDHSW Collaboration) took data at 
distances, $L = 130$~m and 885~m from the target and excluded,
\be
   0.23 \mbox{ eV}^2 < \Delta m^2 < 100 \mbox{ eV}^2 \hspace{50pt} 
   (\sin^2 2\theta = 1).
\label{cdhsw}
\ee  
Thus, the lower limit was not superseded by the corresponding ones from CHORUS 
and NOMAD. It is a limit of considerable interest since the atmospheric 
neutrino analysis of the Super-Kamiokande Collaboration implies large 
$\nu_\mu - \nu_\tau$ mixing. However, even the limit~(\ref{cdhsw}) is 
still about two orders of magnitude above the $\Delta m^2$ values preferred
by Super-Kamiokande. 

An accelerator neutrino disappearance experiment, K2K, with a long enough 
($L = 250$~km) baseline to test atmospheric neutrino oscillations has been 
designed using a 12~GeV proton beam from the KEK proton synchrotron. 
The produced neutrino beam consists to 98\% of $\nu_\mu$ with an average energy
of 1.3 GeV. The $\nu_\mu$ beam is monitored by a near detector 300~m from 
the target. Located at the far end is the Super-Kamiokande detector. 
A reduction of the $\nu_\mu$ flux (56~events were observed and 
$80.1^{+6.2}_{-5.4}$ expected) together with a distortion of the energy 
spectrum indicate neutrino oscillations~\cite{Ahn:2002up} at the 99\%~CL. 
The oscillation parameters are in perfect agreement with the results implied by
the atmospheric neutrinos.

A similar $\nu_\mu$ disappearance experiment, MINOS~\cite{Saakian:sd}, is 
planned using the NuMI beam currently under construction at FNAL. NuMI will be
produced with the 120~GeV proton beam of the Main Injector striking a movable 
carbon target which will allow a variation in the average neutrino energy 
between about 3 and 15~GeV. The 731~km baseline will pass through a near 
detector at FNAL and a similar detector in the Soudan mine. MINOS will provide
large statistics and controlled systematics enabling the Collaboration 
to address the atmospheric neutrino oscillation hypothesis in both CC and NC
channels. Limits (or measurements) on $\nu_\mu \rightarrow \nu_e$ and 
$\nu_\mu \rightarrow \nu_s$ oscillations are also physics goals. 

Two experiments at Gran Sasso~\cite{Duchesneau:2002yq}, Opera and Icarus,
plan to prove the $\nu_\mu \rightarrow \nu_\tau$ oscillation hypothesis by 
directly detecting the $\nu_\tau$. They will have virtually identical baselines
compared to MINOS ($L = 732$~km) and will be using the future CERN high energy 
$\nu_\mu$ beam, CNGS. In addition, they will also be able to test a possible 
subleading $\nu_\mu \rightarrow \nu_e$ transition in $\nu_e$ appearance mode. 
Completion of CNGS and begin of data taking is scheduled for~2006.

\subsection{\it Reactor Neutrinos}
\label{reactornus}
Neutrinos produced at nuclear power plants have been studied in $\bar\nu_e$
disappearance experiments. The reaction in which $\bar\nu_e$ oscillate into 
$\bar\nu_\mu$ is the CP conjugate relative to the KARMEN and LSND accelerator 
experiments discussed in Subsection~\ref{acceleratornus} and are therefore 
complementary. The reactor experiments in France and Arizona are short-baseline
experiments, while KamLAND with two orders of magnitude longer baselines is
motivated by and sensitive to the oscillation parameters suggested by the LMA 
solution to the solar neutrino problem.

A high statistics measurement~\cite{Declais:1994su} of neutrino energy spectra
was carried out at $L = 15$, 40 and 95~m from the Bugey reactor in France. 
No oscillations were reported, and the limits,
\be
\ba{lcl}
   \sin^2 2\theta < 0.02 &\hspace{50pt}& (\mbox{large } \Delta m^2),\\
   \Delta m^2 < 0.01 \mbox{ eV}^2 &\hspace{50pt}& (\sin^2 2\theta = 1),
\ea
\label{bugey}
\ee
could be set, limiting the unexcluded LSND parameter space to small mixing.

In comparison to the Bugey experiment, the Chooz and Palo Verde experiments had
sensitivity to smaller $\Delta m^2$ but not as small mixing. The experiment 
near Chooz in France studied neutrinos with $E \sim 3$~MeV that originated from
two reactors about 1~km away. The 200~day exposure included substantial periods
with only one or the other reactor operating, in addition to about 143~days of
background studies with both reactors turned off.
The results~\cite{Apollonio:2002gd},
\be
\ba{lcl}
   \sin^2 2\theta < 0.10 &\hspace{50pt}& (\mbox{large } \Delta m^2), \\
   \Delta m^2 < 0.0007 \mbox{ eV}^2 &\hspace{50pt}& (\sin^2 2\theta = 1),
\ea
\label{chooz}
\ee
excluded $\Delta m^2$ values at large mixing almost down to the LMA solution of
the solar neutrino problem. The Collaboration also presented results of lower 
sensitivity (due to statistical limitations) but independent of the absolute
normalization of the $\bar\nu_e$ flux, the cross section, the number of target
protons, and detector efficiencies. They were based only on the comparison of 
the positron spectra from the two nuclear reactors located at different 
distances, $\Delta L = 116.7$~m.

An experiment~\cite{Boehm:2001ik} with similar reach took data between 
September~1998 and July~2000 near the Palo Verde Nuclear Generating Station in
Arizona. A detector was exposed to the reactor for 350~days at distances 
$L = 750$ and 890~m. This included 108~days with one of the three reactors 
turned off, which could be used for additional background studies. 
Non-observation of neutrino oscillations in the $\bar\nu_e$ disappearance mode
resulted in the 90\%~CL limits,
\be
\ba{lcl}
   \sin^2 2\theta < 0.17 &\hspace{50pt}& (\mbox{large } \Delta m^2),\\
   \Delta m^2 < 0.0011 \mbox{ eV}^2 &\hspace{50pt}& (\sin^2 2\theta = 1).
\ea
\label{paloverde}
\ee

None of the limits of these three reactor experiments probes values of 
$\Delta m^2$ small enough to address the solar neutrino problem. For this one 
needs to extend the baseline, $L$, by about two orders of magnitude. This is 
the purpose of KamLAND, the largest low energy anti-neutrino detector, located 
in the Kamioka mine. The $\bar\nu_e$ flux contains energies up to about 3.5~MeV
and originates from a variety of reactors with an average $L \approx 180$~km. 
The dominant fraction of 79\% of the flux is produced by reactors with 
distances between 138 and 214~km from KamLAND. 6.7\% of the flux is due to 
a reactor at $L = 88$~km, with the rest traveling $L \geq 295$~km. 
Very recently, the KamLAND Collaboration presented results of their first 
145.1~days of data taking. The number of observed neutrinos (reduced by 
the expected background) divided by the number of expected signal events 
assuming no oscillations is quoted as~\cite{Eguchi:2002dm},
\be
   {N_{\rm obs}\over N_{\rm expected}} = 0.611 \pm 0.085 \mbox{ (stat.)} \pm 
                                                   0.041 \mbox{ (syst.)}.
\ee
This is a 4.1~$\sigma$ deficit, strongly suggestive of $\bar\nu_e$ 
disappearance, and therefore the first reactor indication for neutrino
oscillations. Moreover, the KamLAND Collaboration used spectral shape 
information to perform a fit to two-neutrino oscillation parameters, yielding
$\sin^2 2\theta = 1.01$ and $\Delta m^2 = 6.9 \times 10^{-5}$~eV$^2$ at 
the minimum of the $\chi^2$ function. These values are in good agreement with 
the LMA solution of the solar neutrino problem.

\subsection{\it Neutrinoless Double $\beta$ Decay}
\label{0nu2betadecay}
A fundamental question left open by the neutrino oscillation experiments is 
whether the neutrino is a Dirac or Majorana particle. As discussed above, in
most scenarios that generate a massive neutrino, such as the see-saw
mechanism~\citer{Gell-Mann:vs,Yanagida:xy}, the neutrino has a Majorana mass 
term. Perhaps the most direct test of this possibility is carried out through 
the search for a neutrinoless final state in the double $\beta$ decay of heavy
$0^+$ nuclei,
\be
\label{eq:0nu1}
   A(Z,N) \rightarrow A(Z \mp 2,N \pm 2) + 2 e^\pm,
\ee
Since these modes entail a change of total lepton number (L) by two units, 
a non-zero result would imply that the neutrino is its own antiparticle.
Experimentally, the signature for the $0\nu$ mode is characterized by a sharp 
peak at the endpoint of the $2\nu\beta\beta$ decay spectrum. Until recently, no
statistically significant $0\nu$ peak had been observed, and this absence led 
to upper bounds on the rate for neutrinoless decay.

A summary of representative, present limits on $T_{1/2}^{0\nu}$ from a variety
of experiments is given in Table~\ref{tab:doublebeta}. The most stringent 
\begin{table}
\begin{center}
\begin{minipage}[t]{16.5 cm}
\caption[]{Representative present (upper part) and prospective (lower part) 
$0\nu\beta\beta$ limits, compiled from Refs.~\cite{McKeown:2004yq,Elliott:2002xe}. 
The forth column shows a range of theoretical expectations for $T_{1/2}^{0\nu\beta\beta}$
for $\langle m_\nu\rangle = 50$~eV given various nuclear matrix element computations.
The limits on $\langle m_\nu\rangle$ in the last colmun of the upper part of the Table
are those quoted in the experimental papers; the ranges in the lower part indicate 
proposed sensitivities.}
\label{tab:doublebeta}
\vspace*{4pt}
\end{minipage}
\begin{tabular}{|c|l|r|r|r|}
\hline
&&&&\\[-8pt]
Isotope & Experiment & $T_{1/2}^{0\nu}(y)$ [90\%~CL] & $T_{1/2}^{0\nu}(y)$~[Theory] & $\langle m_\nu \rangle$~[eV] \\[4pt]
\hline 
&&&&\\[-8pt]
 $^{76}$Ge & HM~\cite{Baudis:1999xd,Klapdor-Kleingrothaus:2000sn} & $1.9\times 10^{25}$ & $6.8-70.8\times 10^{26}$ & $<0.35$      \\
 $^{76}$Ge & IGEX~\cite{Aalseth:ji,Aalseth:ud}                    & $1.6\times 10^{25}$ & $6.8-70.8\times 10^{26}$ & $<0.33-1.25$ \\ 
$^{136}$Xe & Gotthard~\cite{Luscher:sd}                           & $4.4\times 10^{23}$ & $7.2-48.4\times 10^{26}$ & $<1.8-5.2$   \\
$^{130}$Te & MIBETA~\cite{Alessandrello:kt}                       & $1.4\times 10^{23}$ & $0.6-23.2\times 10^{26}$ & $<1.1-2.6$   \\
$^{100}$Mo & ELEGANTS~\cite{Ejiri:2001fx}                         & $5.5\times 10^{22}$ & $1.2-15.6\times 10^{26}$ & $<2.1$       \\[4pt]
\hline 
&&&&\\[-8pt]
 $^{76}$Ge & GENIUS~\cite{Klapdor-Kleingrothaus:2000ue}           & $1  \times 10^{28}$ & $6.8-70.8\times 10^{26}$ & $0.013-0.042$\\
 $^{76}$Ge & Majorana~\cite{Aalseth:2002sy}                       & $3  \times 10^{27}$ & $6.8-70.8\times 10^{26}$ & $0.024-0.077$\\
$^{136}$Xe & EXO~\cite{Danevich:2000cf}                           & $8  \times 10^{26}$ & $7.2-48.4\times 10^{26}$ & $0.050-0.120$\\
$^{130}$Te & CUORE~\cite{Arnaboldi:hb}                            & $2  \times 10^{26}$ & $0.6-23.2\times 10^{26}$ & $0.050-0.170$\\
$^{100}$Mo & MOON~\cite{Ejiri:1999rk}                             & $1  \times 10^{27}$ & $1.2-15.6\times 10^{26}$ & $0.017-0.060$\\[-8pt]
 &&&&\\
\hline
\end{tabular}
\end{center}
\end{table}
limits had been obtained from two experiments using Ge detectors, 
the Heidelberg-Moscow (HM)~\cite{Baudis:1999xd,Klapdor-Kleingrothaus:2000sn}
and IGEX experiments~\cite{Aalseth:ji,Aalseth:ud}. The two experiments employed
comparable quantities of 86\% enriched $^{76}$Ge and quoted similar background
levels of 0.20~counts/(keV$\cdot$kg$\cdot$yr). However, a different limiting 
factor in background suppression was identified in the two experiments. 
The HM Collaboration saw the presence of copper in the cryostat as the dominant
limitation, whereas in the IGEX experiment, the presence of radon and activated
$^{68}$Ge were considered the most serious backgrounds. The $T_{1/2}^{0\nu}$ 
limits for the two experiments are comparable. 

Recently, a subset of the HM Collaboration has reported a signal for 
neutrinoless decay. While the initial 
report~\cite{Klapdor-Kleingrothaus:2001ke} generated 
controversy~\cite{Aalseth:2002dt}, an analysis of additional data indicates 
a 4~$\sigma$ effect, with $T_{1/2}^{0\nu} = (0.69\to 4.18)\times 10^{25}$~y at 
99.73\% CL~\cite{Klapdor-Kleingrothaus:2004na}. Using an estimate for 
the uncertainty in nuclear matrix elements (see below), the authors derive 
a range for the effective mass of the exchanged Majorana neutrino ---  
$\langle m_\nu\rangle = (0.1\to 0.9)$ eV --- that would indicate a highly 
degenerate neutrino spectrum. Given the potential implications of this result,
independent experimental confirmation ({\em e.g.}, using a different nucleus 
and experimental method) is clearly warranted.

A number of proposals have been developed for improving the rate sensitivity by
several orders of magnitude. These proposals are discussed extensively in 
the recent review by Elliot and Vogel~\cite{Elliott:2002xe}, and we do not 
repeat their discussion here. However, in Table~\ref{tab:doublebeta}, we 
summarize the goals of the five most widely discussed proposals: 
GENIUS~\cite{Klapdor-Kleingrothaus:2000ue} (GErmanium NItrogen Underground 
Setup) that would consist of an array of 2.5~kg Ge crystals; 
Majorana~\cite{Aalseth:2002sy} that would build a 500~kg detector from 210 
segmented Ge crystals; EXO~\cite{Danevich:2000cf} (Enriched Xenon Observatory)
that proposes to employ 10~tons of 60-80\% enriched $^{136}$Xe in the Waste 
Isolation Pilot Plant (WIPP); CUORE~\cite{Arnaboldi:hb} (Cryogenic Underground
Observatory for Rare Events) at Gran Sasso that would use 1000~TeO$_2$ 
crystals; and MOON (MOlybdenum Observatory of Neutrinos) that would use 34~tons
of natural Mo. The designs of the two Ge experiments follow from 
the differently identified limiting backgrounds in the HM and IGEX experiments,
respectively. The segmentation of the crystals in the Majorana detector would 
permit veto of $^{68}$Ge induced events that were highlighted in the IGEX 
experiment. In contrast, the Ge crystals in the GENIUS experiment would be 
isolated in a liquid nitrogen bath from the external sources of radioactivity 
considered most problematic by the HM Collaboration. 

From a theoretical standpoint, the reliable extraction of limits on neutrino 
properties from the $T_{1/2}^{0\nu}$ lower bounds have proved as challenging as
the experiments themselves. The most common theoretical analyses assume that 
neutrinoless decay is mediated by the exchange of a light Majorana neutrino. 
In this case, the rate is proportional to the square of the effective mass,
\be
\label{eq:0nu2}
   \langle m_\nu\rangle=\left|\sum_{j} |U_{ej}|^2 e^{i\alpha_j} m_j\right|,
\ee
where the sum runs over all light neutrino generations, the $U_{ej}$ connects
left handed electron neutrino weak eigenstate to the Majorana mass eigenstates 
labeled by the index, $j$, and the $\alpha_j$ are the $N-1$ Majorana phases in 
the general neutrino mass matrix. Note that the presence of these phases allows
for the possibility of cancellations in the sum over light mass eigenstates. 

The remaining theoretical input to the rate involves nuclear matrix elements. 
For the $0^+\to 0^+$ decays of interest here, one has,
\be
\label{eq:0nu3}
   {1\over T_{1/2}^{0\nu}}=G^{0\nu}(E_0,Z) \left| M_{GT}^{0\nu} - 
   {g_V^2 \over g_A^2}M_{F}^{0\nu}\right|^2 \langle m_\nu\rangle^2,
\ee
where $M_{F}^{0\nu}$ and $M_{GT}^{0\nu}$ are the Fermi and Gamow-Teller matrix
elements for the $\Delta Q=2$ nuclear transition, respectively, $g_V$ ($g_A$) 
are the (axial) vector coupling of the $W$ boson to quarks, and 
$G^{0\nu}(E_0,Z)$ is a calculable phase space factor that depends on 
the energy, $E_0$, released to the two electrons. 

The matrix elements $M_{F,GT}^{0\nu}$ depend on a joint neutrino-nucleus 
Green's function that involves a sum over all possible intermediate nuclear 
states. Nuclear model studies indicate that the sum is dominated by states 
within a fairly narrow range of energies and that it may be rather well 
approximated by using closure with an appropriate choice for the average 
excitation energy. Matrix elements of the resulting two-body operator must then
be computed between the initial and final ground state wave-functions. For 
the kinematics of the neutrinoless decay, the operator samples relatively 
high momentum components of the wave-functions that are not well constrained by
existing measurements of nuclear properties. Different model calculations have
led to a fairly broad spread in the values of the $M_{F,GT}^{0\nu}$.

Two types of approaches have been taken~\cite{Elliott:2002xe}, namely nuclear 
shell model computations and those involving the quasi-particle random phase 
approximation (QRPA). In the former case, the huge size of the single particle 
basis requires truncation of the model space to perform tractable computations.
In order to account for omitted states, one generally employs effective 
operators that act within the truncated model space to compensate for 
the truncation. Since no one has, as yet, derived these effective operators by
rigorously integrating out the omitted states, phenomenological tests of 
the effective operators must be employed to evaluate their degree of 
effectiveness. While such tests have proved successful in analyzing nuclear 
systematics such as magnetic moments or single particle Gamow-Teller 
transitions, there exist few, if any, independent tests of effective operators
relevant to $M_{F,GT}^{0\nu}$.  

The QRPA approach entails performing a complete sum over particle-hole or 
particle-particle \lq\lq bubbles". The strength of the particle-particle
interaction, $g_{pp}$  can be adjusted to yield agreement with measured 
$2\nu\beta\beta$ decay Gamow-Teller strengths and then used in calculations of
$M_{F,GT}^{0\nu}$. However, the value of $g_{pp}$ needed to account for 
$M_{GT}^{2\nu}$ is close to the limit of validity for the QRPA equations, 
indicating the importance of more complicated configurations than the two 
particle or particle-hole configurations included in the QRPA framework. Were 
it not for the truncation problem, the shell model would provide a more 
realistic framework, since diagonalization of the shell model space includes 
the effect of complex configurations. As it stands, however, neither approach 
has proved completely satisfactory, and this situation is reflected in the wide
variation in QPRA and shell model predictions for the $M_{F,GT}^{0\nu}$. One 
rough guide to the theoretical uncertainty associated with these approximations
is provided by a comparison of different calculations. In the case of 
$^{76}$Ge, for example, the spread in rates among various calculations is 
roughly an order of magnitude, leading to a factor of three spread in 
the values of $\langle m_\nu\rangle$. One should bear in mind, however, that 
such comparisons may overlook systematic shortcomings pertinent to all 
published calculations (such as the approximations employed in a given method)
and, therefore, may underestimate the degree of theoretical uncertainty. 
Clearly, new ideas are needed for providing more robust computations of
the $M_{F,GT}^{0\nu}$ for the light Majorana neutrino scenario. These caveats 
notwithstanding, the proposed future $0\nu\beta\beta$ decay experiments
summarized above would improve the $\langle m_\nu\rangle$ sensitivity to 
the 50~meV range. 

The exchange of a light Majorana neutrino is not, however, the only mechanism 
by which the neutrinoless mode may occur. It is also possible that the exchange
of a heavy particle drives the decay. In the case of the see-saw 
model~\citer{Gell-Mann:vs,Yanagida:xy}, for example, cancellations in 
the sum~(\ref{eq:0nu2}) may suppress the effect of the light neutrinos, thereby
exposing the effect of exchanged heavier Majorana neutrinos. Alternatively, 
the exchange of some other Majorana particle, such as the neutralino ($\chi^0$)
in SUSY scenarios may be dominant~\cite{Faessler:1996ph,Paes:2000vn}. Recently,
the framework for treating such heavy particle exchange in the nucleus has been
placed on a firmer theoretical footing through the use of  
an EFT~\cite{Prezeau:2003xn}. The EFT approach involves integrating out 
the heavy particle, and leaving dimension nine operators containing four quark
and two lepton fields. In the most general case one has~\cite{Prezeau:2003xn},
\be
\label{eq:0nu4}
   {\cal L}_{0\nu\beta\beta} = {G_F^2\over \Lambda_{\beta\beta}}
   \sum_{j=1}^{14} C_j(\mu) {\hat O}_j {\bar e} \Gamma_j e^c,
\ee
where $\Lambda_{\beta\beta}$ is a mass scale associated with the heavy lepton 
number violating physics ({\em e.g.}, the mass of the $\chi^0$), 
the ${\hat O}_j$ are four-quark operators, the $\Gamma_j$ are Dirac matrices, 
and the $C_j(\mu)$ are constants determined by a particular particle physics 
model. In order to compute the effects of ${\cal L}_{0\nu\beta\beta}$ in 
$\Delta Q=2$ $0^+\to 0^+$ transitions, one can construct the appropriate set of
hadronic operators containing nucleon and pion fields that reflect 
the spacetime and chiral transformation properties of the various ${\cal O}_j$.
The hadronic operators can be expanded systematically in powers of 
$Q/\Lambda_{\rm had}$, where $Q$ is some momentum or mass relevant to nuclear
processes and $\Lambda_{\rm had}$ is the hadronic scale at which one matches 
the ${\cal O}_j$ onto the hadronic basis. One may then systematically construct
an appropriate set of nuclear operators using this tower of effective hadronic 
interactions, and use them to compute ground state-to-ground state transition
matrix elements. 

The genesis for this development appeared in the work of 
Ref.~\cite{Faessler:1996ph}, where the particular case of RPV SUSY was studied
and where it was pointed out that long range, pion exchange nuclear operators
could be generated by the relevant four-quark operators. A more comprehensive 
and systematic development of this framework using the ideas of EFTs was 
subsequently given in Ref.~\cite{Prezeau:2003xn}. In that work, additional
scenarios for $0\nu\beta\beta$ decay were discussed, and the importance of 
considering the spacetime and chiral properties of the relevant ${\cal O}_j$
illustrated. Under certain scenarios, these symmetries imply an enhancement of 
the nuclear matrix element via the existence of lower order operators that are
otherwise forbidden. Additional theoretical work is needed, however, in order 
to fully implement these ideas. In particular, a refined treatment of QCD
effects --- such as the RG evolution of the $C_j(\mu)$ from the scale
$\Lambda_{\beta\beta}$ down to $\Lambda_{\rm had}$ --- remains to be carried 
out. In addition, the degree to which the nuclear matrix elements reflect the
systematic power counting in $Q/\Lambda_{\rm had}$ of the hadronic and nuclear
operators is an open question, as is the appropriate treatment of short range 
correlations in taking nuclear matrix elements. In this respect, new studies of
parity violation in hadronic systems may provide useful guidance.

\subsection{\it Anomalous Magnetic Moment of the Muon}
The magnetic moment of the electron, $g_e$, provides the best determination of 
the fine structure constant, but is currently not measured precisely enough 
to give a sensitive probe of EW physics which is suppressed by a factor 
$(\alpha/\pi) m_e^2/M_W^2 \sim {\cal O}(10^{-13})$. Meaningful (but still weak)
bounds on NP contributions can be set only if they enter without loop
suppression.  On the other hand, EW contributions to the anomalous magnetic 
moment of the muon\footnote{$a_\tau$ has not yet been observed 
experimentally.}~\cite{Kinoshita:1990}, $a_\mu = (g_\mu - 2)/2$, are enhanced 
by a factor $m_\mu^2/m_e^2 \sim 4\times10^4$, which renders them sizable enough
to be detectable at the E821 experiment at the AGS at BNL.

Recently, the E821 Collaboration~\cite{Bennett:2004pv} published its final
result, leading to the new world average,
\be
   a_{\mu}({\rm exp.}) - {\alpha\over 2\pi} = (4511.07 \pm 0.63)\times 10^{-9}.
\label{gminus2}
\ee
Individual results on positively and negatively charged muons are in good 
agreement with each other and serve as a test of the CPT theorem. 
The value~(\ref{gminus2}) is precise enough to provide a good laboratory 
to test the SM and probe theories beyond it~\cite{Czarnecki:2001pv}. 
For example, scenarios of low energy supersymmetry with large $\tan\beta$ and 
moderately light superparticle masses can give large contributions to 
$a_\mu$~\cite{Lopez:1993vi}.

Unfortunately, the interpretation of $a_\mu$ is compromised by large 
theoretical uncertainties introduced by hadronic effects. Two- and three-loop 
vacuum diagrams containing light quark loops\footnote{In contrast, 
the contributions from heavy quarks and from light quarks in the perturbative
regime are known in analytic form~\cite{Erler:2000nx}.} cannot be calculated 
reliably in perturbative QCD (PQCD). Instead they are obtained by computing 
dispersion integrals over measured (at low energies) and theoretical (at higher
energies) hadronic cross sections. At the two-loop level~\cite{Gourdin:dm},
\be
   a_\mu (\mbox{had; 2-loop}) = \left(\frac{\alpha\; m_\mu}{3\pi}\right)^2
   \int_{4m_\pi^2}^\infty \frac{ds}{s^2} \hat{K}\left(s\right) R(s),
\label{eq9}
\ee
where $R(s)$ is the cross section of $e^+e^- \to \mbox{ hadrons}$, normalized
to the tree level cross section of $e^+e^- \to \mu^+\mu^-$, and where
\be
   \hat{K}(s) = \int_0^1 dx \frac{3 x^2(1-x)} {1-x + x^2 m_\mu^2/s}.
\label{ks}
\ee
The uncertainty introduced by this procedure is comparable to the experimental 
one in~(\ref{gminus2}). An analogous uncertainty occurs in the QED coupling
constant, $\hat\alpha(\mu)$, preventing its precise theoretical computation 
from the fine structure constant, $\alpha$, for $\mu \gsim 2 m_{\pi^0}$. 
Note that knowledge of $\hat\alpha (M_Z)$ is indispensable for the extraction
of $M_H$ from EW precision data (see Section~\ref{SMstatus}). As a result, 
these hadronic uncertainties are strongly correlated with each 
other~\cite{Erler:2000nx}, with the renormalization group evolution of 
the weak mixing angle, and also with other fundamental SM parameters, such as 
the strong coupling constant and the heavy quark masses. In addition to $R(s)$, 
one can obtain additional experimental information using $\tau$ spectral 
functions and isospin symmetry when isospin breaking effects are properly taken 
into account. A $\tau$-based analysis yields~\cite{Davier:2003pw},
\be
   a_\mu - {\alpha\over 2\pi} = (4509.83 \pm 0.58) \times 10^{-9},
\label{amutau}
\ee
which is marginally (within 1.4~$\sigma$) consistent with the result~(\ref{gminus2}).
On the other hand, using only $R(s)$ including the most recent data from 
the CMD-2 Collaboration~\cite{Akhmetshin:2001ig} yields~\cite{Davier:2003pw},
\be
   a_\mu - {\alpha\over 2\pi} = (4508.36 \pm 0.72) \times 10^{-9},
\label{amurs}
\ee
{\em i.e.}, a 2.8~$\sigma$ deviation. The $R(s)$ driven value is also 
consistent with the findings of Refs.~\cite{Hagiwara:2003da,Ghozzi:2003yn} 
which defined a higher cutoff for the transition from experimental data to 
PQCD. The difference between the evaluations~(\ref{amutau}) and (\ref{amurs}) can be 
traced to a discrepancy between the 2$\pi$ spectral functions obtained from 
the two methods. For example, if one uses the $e^+ e^-$ data and CVC to predict
the branching ratio for $\tau^- \rightarrow \nu_\tau \pi^- \pi^0$ decays one 
obtains $24.52\pm 0.32$\%~\cite{Davier:2003pw} while the average of 
the measured branching ratios by DELPHI~\cite{Matorras:2003}, ALEPH, CLEO, L3, 
and OPAL~\cite{Davier:2003pw} yields $25.43 \pm 0.09$\%, which is 2.8~$\sigma$ 
higher. It is important to understand the origin of this difference and 
to obtain additional experimental information ({\it e.g.}, from the radiative
return method~\cite{Binner:1999bt}). Fortunately, this problem is less 
pronounced as far as $a_\mu^{\rm had}$ is concerned: due to the suppression at 
large $s$ (from where the conflict originates) the difference is only 
1.7~$\sigma$ (or 1.9~$\sigma$ if one adds the 4~$\pi$ channel which by itself 
is consistent between the two methods). Note also that a part of 
this difference is due to the older $e^+e^-$ data~\cite{Davier:2003pw}, and 
the direct conflict between $\tau$ decay data and CMD~2 is less significant.  
Isospin violating corrections have been estimated in 
Ref.~\cite{Cirigliano:2002pv} and found to be under control. The largest effect
is due to higher order EW corrections~\cite{Marciano:vm} but 
introduces a negligible uncertainty~\cite{Erler:2002mv}. If we view 
the 1.7~$\sigma$ difference as a fluctuation and use all available information
as constraints in a SM fit we find, 
\be
  a_\mu ({\rm SM}) - {\alpha\over 2\pi} = (4509.15 \pm 0.49) \times 10^{-9},
\label{amusm}
\ee
which corresponds to a 2.4~$\sigma$ deviations. Thus, regardless of whether one
trusts the $\tau$-based analysis or follows Ref.~\cite{Ghozzi:2003yn} which
argues that CVC breaking effects ({\em e.g.}, through a relatively large mass 
difference between the $\rho^\pm$ and $\rho^0$ vector mesons) may be larger 
than expected\footnote{Large CVC breaking effects would also be relevant in 
the context of the NuTeV discrepancy discussed in 
Section~\ref{nun}~\cite{Ghozzi:2003yn}.}, one concludes in both cases that 
there is a deviation at the level of two and a half standard deviations. 

The error in the prediction~(\ref{amusm}) is from the hadronic uncertainties excluding 
parametric ones such as from $\alpha_s$ and the heavy quark masses. We estimate
its correlation with $\Delta \alpha (M_Z)$ to $-24$\%. The uncertainty in~(\ref{amusm}) 
includes a contribution from the hadronic three-loop 
light-by-light scattering amplitude~\cite{Knecht:2001qf}, 
$a_\mu({\rm LBLS}) = (+0.83\pm 0.19)\times 10^{-9}$, which has been estimated 
within a form factor approach. Its sign is opposite relative to earlier work,
and has subsequently been confirmed by two other 
groups~\cite{Hayakawa:2001bb,Bijnens:2001cq}. A more rigorous calculation 
based on $\chi$PT confirmed the value of $a_\mu({\rm LBLS})$ but could not
exclude the possibility that its uncertainty might have been 
underestimated~\cite{Ramsey-Musolf:2002cy}. In this context, 
$a_\mu({\rm LBLS})$ depends on several {\em a priori\/} unknown constants, not 
all of which can be determined in a model-independent way from other 
measurements. The uncertainty in $a_\mu({\rm LBLS})$ associated with 
the unknown constants could be substantially larger than what is quoted for 
the model calculation of Ref.~\cite{Knecht:2001qf}, potentially reducing 
the significance of the deviation from the SM value~(\ref{amusm}). 
Other hadronic effects at three-loop order~\cite{Krause:1996rf} contribute 
$a_\mu^{\rm had} (\alpha^3) = ( - 1.00 \pm 0.06) \times 10^{-9}$.

{\em Note added:} After the completion of this Section, 
an update~\cite{Kinoshita:2004wi} appeared of the four-loop QED contribution 
reducing the discrepancy of $a_\mu$ with the SM by 0.22~$\sigma$. There was
also a new estimate~\cite{Melnikov:2003xd} of
$a_\mu({\rm LBLS}) = (+1.36\pm 0.25)\times 10^{-9}$, which would reduce
the discrepancy further by $2/3$ of a standard deviation.

\section{Weak Probes of the Strong Interaction}
\subsection{\it Hadronic Parity Violation}
\label{sec:HPV}
Despite decades of experimental and theoretical study, achieving a solid 
understanding of the weak interaction between quarks remains an elusive goal. 
In contrast to the analysis of purely leptonic and semileptonic weak 
interactions, one cannot disentangle the effects of non-perturbative strong 
interactions between quarks from those generated by the weak interaction.
This difficulty is reflected by a number of on-going puzzles in 
the $\Delta S=1$ sector, the best known being the $\Delta I=1/2$ rule. 
While the latter has been well established empirically, its dynamical origins 
remain mysterious. Indeed, even the most general considerations of symmetries 
in QCD do not provide much insight in this case. Similarly, the weak decays of
strange baryons do not appear to respect the symmetries of QCD. The long 
standing $S$-wave/$P$-wave discrepancy, wherein a simultaneous accounting for
the parity conserving and parity violating amplitudes evades the grasp of 
chiral dynamics, points to a breakdown of QCD symmetries in the hadronic weak 
interaction (HWI). Similar issues also arise in weak radiative decays of 
polarized hyperons, which display PV asymmetries that should vanish in 
the limit of exact $SU(3)$ symmetry and that are unnaturally large from 
the standpoint of naive $SU(3)$ breaking expectations. 

These issues pertaining to $\Delta S=1$ HWIs have been studied and reviewed
extensively elsewhere, so we do not treat them in detail here (see, {\em e.g.},
Refs.~\cite{Donoghue:dd1,Borasoy:1998ku} and references therein). We note, 
however, that they pose a more general question: are the apparent failures of 
QCD symmetries in strangeness changing HWIs due to the presence of a strange 
quark with mass close to the QCD scale, or do they reflect poorly understood 
dynamics at a more fundamental level? In order to address this question, one 
would ideally study the $\Delta S=0$ HWI --- for which one expects strange 
quark effects to be sub-dominant --- with the same level of intensity as one 
has scrutinized the $\Delta S=1$ sector. In practice, doing so has presented 
challenges of its own. As we outline below, however, recent experimental and 
theoretical developments may have put this field on a new and more promising 
footing.

Because the parity conserving $\Delta S=0$ HWI is masked at low energies by
much larger strong and electromagnetic effects, one must isolate the former by
studying PV processes. Historically, such studies have been carried out largely
in nuclei, where various features of nuclear structure may amplify tiny PV
effects. One generically expects the size of low energy, hadronic PV 
observables to be of order $G_F F_\pi^2\sim 10^{-7}$, thereby making
their observation an enormously challenging task. Indeed, only two PV 
observables of this magnitude in the more pristine few-body system have been 
measured: $A_L^{pp}$, the longitudinal analyzing power in polarized 
proton-proton scattering~\citer{Potter:1974gu,Berdoz:2002sn}, and 
the corresponding analyzing power in elastic ${\vec p}\alpha$ scattering,
$A_L^{p\alpha}$~\cite{Lang:jv,Lang:nw}. 

The remaining hadronic PV observables have involved nuclei that enhance the
effects by several orders of magnitude. The largest effects have been observed
in the scattering of polarized, epithermal neutrons from $^{139}$La, where 
a $\sim 10\%$ asymmetry was seen~\citer{Alfimenkov:1983,Yuan:1991},
\be
   A_z = (9.55\pm 0.35) \times 10^{-2}.
\ee
Unfortunately, the theoretical interpretation of measurements involving heavy 
nuclei is difficult, since one cannot perform {\em ab initio\/} nuclear 
structure computations for such systems. A somewhat more tractable situation 
arises in the case of the lighter $S$--$D$ shell nuclei, such as $^{18}$F. 
Here, fortuitous near degeneracies of opposite parity states can enhance 
the magnitude of parity mixing in the nuclear wave-function by factors of ten 
or more. When probed electromagnetically, these large parity mixing effects can
be additionally amplified since the parity mixing allows contributions from
large, parity forbidden multipole matrix elements to the transition amplitudes
of interest. The resulting observables can be as large as a few 
$\times 10^{-4}$. For example, one has for the directional asymmetry in 
the $\gamma$ decay of polarized $^{19}$F from its first excited state,
\be
  A_\gamma(^{19}{\rm F}) = -(8.5\pm 2.6)\times 10^{-5}~\cite{Adelberger:1983},
  \hspace{50pt} A_\gamma(^{19}{\rm F}) = 
  -(6.8\pm 1.8)\times 10^{-5}~\cite{Elsener:vp,Elsener:sx},
\ee
for the circular polarization in the $\gamma$ decay of excited $^{21}$Ne,
\be
  P_\gamma(^{21}{\rm Ne}) = (24\pm 24)\times 10^{-4}~\cite{Snover:1978},
  \hspace{50pt} P_\gamma(^{21}{\rm Ne}) = 
  (3\pm 16)\times 10^{-4}~\cite{Earle:ji},
\ee
or for the circular polarization in the $\gamma$ decay of $^{18}$F (in units of
$10^{-4}$),
\be
   P_\gamma(^{18}{\rm F}) = - 7 \pm 20~\cite{Barnes:sq}, \hspace{7pt} 
   P_\gamma(^{18}{\rm F}) =   3 \pm  6~\cite{Bini:1985}, \hspace{7pt} 
   P_\gamma(^{18}{\rm F}) = -10 \pm 18~\cite{Ahrens:1982}, \hspace{7pt} 
   P_\gamma(^{18}{\rm F}) =   2 \pm 6~\cite{Page:ak}.
\ee
Note that the $P_\gamma$ results are consistent with zero. However, owing to 
the nuclear enhancements expected in $^{21}$Ne and $^{18}$F, measurements at
the $10^{-4}$ level can probe ${\cal O}(10^{-7})$ strangeness conserving HWIs. 

Given the short range ($\sim 0.002$ fm) of the HWI, it is improbable that 
direct exchange of $W$ and $Z$ bosons between nucleons are responsible for 
the PV effects observed in these nuclei. Indeed, the repulsive hard core of 
the strong $NN$ interaction that becomes dominant below 0.4~fm highly 
suppresses direct $W$ and $Z$ exchange. Consequently, theorists have 
conventionally relied upon a meson exchange framework for describing the PV 
$NN$ interaction. In this picture, the PV HWI resides within one of the two 
meson-nucleon vertices appearing in the meson exchange amplitude (the second 
vertex involves a purely strong, parity conserving interaction). Moreover, 
the Compton wavelength of the lightest pseudoscalar and vector mesons is 
sufficient to overcome the short range repulsion of the $NN$ interaction. Thus,
one has a physically natural mechanism for understanding the observed PV 
effects in nuclei, and one that is consistent with the long standing framework 
for understanding the strong $NN$ interaction. 

Unfortunately, the application of this picture to nuclear PV entailed 
the reliance on a number of model approximations --- not all of which are well
controlled --- and resulted in a less than successful phenomenology. The first 
approximation has been to severely truncate the spectrum of exchanged mesons, 
retaining only the $\pi$, $\rho$, and $\omega$. The resulting model for the PV
$NN$ potential then depends on seven, {\em a priori\/} unknown PV meson-nucleon
couplings, $h_M^I$, that appear in the PV $NN$ potential,
\begin{eqnarray}
\label{eq:ddh}
   {\cal H}_{\rm PV}&=&{\hpinn\over\sqrt{2}} \bar{N}(\tau\times\pi)_3N -
   \bar{N} \left (h_\rho^0\tau\cdot\rho^\mu + h_\rho^1\rho_3^\mu +
   {h_\rho^2\over 2\sqrt{6}}(3\tau_3\rho_3^\mu - \tau\cdot\rho^\mu) \right)
   \gamma_\mu\gamma_5N\nonumber\\
   & & -\bar{N} \left( h_\omega^0\omega^\mu + h_\omega^1\tau_3\omega^\mu\right)
   \gamma_\mu\gamma_5N + h_\rho^{'1} \bar{N} (\tau\times\rho^\mu)_3
   {\sigma_{\mu\nu}k^\nu\over 2m_N} \gamma_5 N.
\end{eqnarray}
Here, the superscripts in the $h_M^I$ denote the isospin ($I=0,1,2$) of 
the corresponding meson-nucleon interaction. The longest range effect that 
arises from the exchange of charged pions is characterized by a single PV 
Yukawa coupling, $h_\pi^1$.

The goal of experiment has been to provide a self-consistent determination of 
these couplings. In doing so, one has relied upon many-body nuclear 
calculations in order to relate the PV $NN$ model potential to the PV 
observables. This portion of the analysis thereby entails a second level of
approximation associated with the nuclear many-body problem. In some cases, 
such as $P_\gamma(^{18}{\rm F})$, one may attempt to constrain the nuclear
theory uncertainties by calibrating the parity mixing matrix elements to those 
of the two-body axial charge operator that contribute to the analog forbidden
$\beta$ decays. This semi-empirical approach, however, is not feasible in all 
cases. A recent example involves the anapole moment of $^{133}$Cs (see 
Section~\ref{APV}), where the state of the art involves a shell model 
computation using a truncated model space. In fact, a comparison of 
the $^{133}$Cs analysis with that of $^{18}$F presents a particularly puzzling 
phenomenological inconsistency. While the analysis of 
the $P_\gamma(^{18}{\rm F})$ measurements imply a value for $h_\pi^1$ 
consistent with zero (assuming the DDH ranges for $h_\rho^1$ and $h_\omega^1$ 
discussed below)~\cite{Haxton:2001mi},
\be
   -0.6 \lsim \hpinn-0.11 h_\rho^1-0.19 h_\omega^1 \lsim 1.2,
\ee
a combination of $A_L^{pp}$ results and the $^{133}$Cs anapole moment suggests
that $h_\pi^1$ is considerably larger. Whether the origin of this discrepancy 
lies with experiment, the nuclear many-body theory, or the adoption of 
an inadequate model for the PV $NN$ potential is unknown at present. One 
expects that a precise measurement of $A_\gamma({\vec n}p\to d\gamma)$, 
the photon asymmetry in the capture of polarized neutrons on hydrogen that is 
underway at LANSCE~\cite{Wilburn:1998xq,Mitchell:2004fn}, will shed new light 
on this question.

Even if one were able to extract a self-consistent set of values for 
the $h_M^I$ from experiment, their interpretation in terms of the underlying
weak quark-quark interaction would remain problematic. It is not entirely 
clear, for example, that the $h_M^I$ represent the couplings of the physical 
mesons to the nucleon rather than some effective parameters that include 
the contributions from a tower of heavier meson exchanges. Under the assumption
that the $h_M^I$ do represent physical meson-nucleon couplings, Desplanques,
Donoghue, and Holstein (DDH) used symmetry arguments and quark model 
computations to predict values for the PV couplings from the underlying 
$\Delta S=0$ PV quark-quark interactions~\cite{Desplanques:1979hn}.
The so-called DDH \lq\lq reasonable ranges" and 
their updates~\cite{Dubovik:pj,Feldman:tj} (see Table~\ref{tab:ddh}) have 
\begin{table}
\begin{center}
\begin{minipage}[t]{16.5 cm}
\caption[]{Weak, PV meson-nucleon couplings as calculated in 
Refs.~\citer{Desplanques:1979hn,Feldman:tj}. All numbers are quoted in units of
the ``sum rule" value $g_\pi = 3.8\times 10^{-8}$~\cite{Desplanques:1979hn}.}
\label{tab:ddh}
\vspace*{4pt}
\end{minipage}
\begin{tabular}{|c|c|c|c|c|}
\hline
&&&&\\[-8pt]
Coupling & DDH Reasonable Range~\cite{Desplanques:1979hn} & DDH Best Value~\cite{Desplanques:1979hn} 
& DZ~\cite{Dubovik:pj} & FCDH~\cite{Feldman:tj}\\[4pt]
\hline
&&&&\\[-8pt]
$h_\pi^1$   & $~~~~0\rightarrow +30$& +12   & +~3 & +~7    \\
$h_\rho^0$  & $+30\rightarrow -81$&$-30$  &$-22$&$-10$     \\
$h_\rho^1$  & $-~~1\rightarrow ~~~~0$&$-0.5$ & +~~1 &$-~~1$\\
$h_\rho^2$  & $-20\rightarrow -29$&$-25$  &$-18$&$-18$     \\
$h_\omega^0$& $+15\rightarrow -27$&$-~5$  &$-10$&$-13$     \\
$h_\omega^1$& $-~5\rightarrow -~2$&$-~3$  &$-~6$&$-~6$     \\[-8pt] 
&&&&\\
\hline
\end{tabular}
\end{center}
\end{table}
become the benchmark for comparison with experiment, and their latitude reflect
the theoretical strong interaction uncertainties entering the treatment of 
HWIs. Even the symmetry arguments used by DDH may not be entirely well founded.
For example, the effects of chiral symmetry breaking on the value of $h_\pi^1$ 
--- not included in the DDH treatment --- may be anomalously 
large~\cite{Zhu:2000gn}.

These theoretical and phenomenological issues triggered a reformulation of 
the hadronic PV problem~\citer{Holstein:2004,Holstein:2004a}. The basic thrust 
of that work was to recast the PV $NN$ interaction into the framework of 
an EFT, thereby eliminating the dependence on the model dependent meson 
exchange picture. The use of EFT allows one to expand ${\cal H}_{\rm PV}$ in 
the most general set of PV operators, and systematically classifying them in 
powers of $Q/\Lambda_\chi$, where $Q \ll 1$ GeV is either the pion mass or 
a typical momentum transfer associated with the $NN$ interaction. Doing so 
allows one to truncate the expansion at a given order in $Q/\Lambda_\chi$ with
a reasonable expectation for the magnitude of the truncation error. 
The coefficients of the operators are left as {\em a priori\/} unknown 
\lq\lq low energy constants" (LECs) that are to be determined from experiment.
This approach is model independent, since, to a given order in the expansion, 
one includes all operators consistent with the symmetries of the underlying EW 
and strong interactions. In essence, the meson exchange model dictates certain
relations among these operators --- relations that may not, in fact, be 
consistent with experiment. 

For very low energies, one may treat the pion as heavy and integrate it out 
from the EFT, leaving only short range, four-nucleon operators in 
${\cal H}_{\rm PV}$. Through ${\cal O}(Q/\Lambda_\chi)$, there exist only five 
independent operators and, thus, only five independent LECs that must be taken
from experiment,
\be
\label{eq:pvlec}
   \lambda_s^{pp}, \hspace{10pt} 
   \lambda_s^{pn}, \hspace{10pt} 
   \lambda_s^{nn}, \hspace{10pt}
   \lambda_t,      \hspace{10pt} {\rm and} \hspace{10pt}
   \rho_t.
\ee 
For higher energy processes, for which $E\gsim m_\pi$, the pion must be treated
as a dynamical degree of freedom and additional operators appear in the EFT. 
In this case, the leading contribution in ${\cal H}_{\rm PV}$ arises at 
${\cal O}(\Lambda_\chi/Q)$ and is the same long range pion exchange operator
containing $\hpinn$ in Eq.~(\ref{eq:ddh}). 

The future program for hadronic PV that is laid out in 
Refs.~\citer{Holstein:2004,Holstein:2004a} involves performing a set of 
few-body measurements to determine the coefficients of the operators appearing
in the EFT. In the case of the EFT without pions, 
the constants~(\ref{eq:pvlec}) could be obtained by combining the ${\vec p} p$ 
and ${\vec p}\alpha$ results with three additional observables: the circular 
polarization, $P_\gamma$, and photon asymmetry, $A_\gamma$, in radiative $np$ 
capture and the rotation of neutron spin as it passes through a gas of $^4$He,
$d\phi^{n\alpha}/ dz$. Using somewhat idealized few-body wave-functions, 
Refs.~\citer{Holstein:2004,Holstein:2004a} obtained approximate relations 
between the PV constants and these observables, 
\begin{eqnarray}
   m_N \lambda_s^{pp} &=& -1.22 A_L^{pp}(45\mbox{ MeV}),           \nonumber \\
   m_N \rho_t         &=& -9.35 A_\gamma (np\rightarrow   d\gamma),\nonumber \\
   m_N \lambda_s^{pn} &=&  1.6  A_L^{pp}(45\mbox{ MeV})   -  3.7 A_L^{p\alpha}
   (46\mbox{ MeV}) + 37 A_\gamma (np\rightarrow d\gamma)  - 
   2 P_\gamma (np\rightarrow d\gamma), \label{eq:holstein} \\
   m_N \lambda_t      &=&  0.4  A_L^{pp}(45\mbox{ MeV})   -  0.7 A_L^{p\alpha}
   (46\mbox{ MeV}) +  7 A_\gamma (np\rightarrow d\gamma)  + 
     P_\gamma (np\rightarrow d\gamma), \nonumber \\
   m_N \lambda_s^{nn} &=&  0.83 {d\phi^{n\alpha}\over dz} - 33.3 A_\gamma
   (np\rightarrow d\gamma) - 0.69 A_L^{pp}(45\mbox{ MeV}) + 1.18 A_L^{p\alpha}
   (46\mbox{ MeV}) - 1.08 P_\gamma (np\rightarrow d\gamma).        \nonumber
\end{eqnarray}
Although a more precise version of these relations will require the use of more
sophisticated few-body methods, Eqs.~(\ref{eq:holstein}) provide a rough guide
for a future experimental program. In the case of the EFT with explicit pions,
additional experiments, such as radiative $nd$ capture or $np$ spin rotation, 
will be needed in order to determine all of the PV parameters.

Once such a program is completed, the LECs, $\hpinn$, $\lambda_s^{pp}$, 
{\em etc.}, would provide model independent input for theoretical analyses. One
direction would involve first principles studies that would address 
the fundamental question of QCD symmetries and the HWI outlined above. For 
example, are the magnitudes of the various LECs consistent with what one 
expects based on these symmetry considerations? Or, does one find surprising
enhancements as in the $\Delta S=1$ sector? Alternately, the PV $NN$ 
interaction, ${\cal H}_{\rm PV}$, could be used to re-analyze the PV 
observables in many-body systems. To the extent that one is able to understand 
the latter using the more \lq\lq primordial" PV $NN$ interaction derived from
EFT and few body experiments, one would have some confidence of 
the applicability of the EFT to complex nuclei --- a situation that might open 
the way to applying EFT methods to other weak processes, such as 
$0\nu\beta\beta$ decay discussed in Section~\ref{0nu2betadecay}.

\subsection{\it Probes of Nucleon Strangeness}
\label{strangeness}
As noted in the foregoing discussion, the strange quark --- with its mass of 
order $\Lambda_{\rm QCD}$ --- presents a variety of theoretical challenges from
the standpoint of HWIs. It is a similarly problematic object for those seeking
to understand the structure of hadrons. Its mass is not sufficiently heavy 
to apply heavy quark effective theory, and it may not be sufficiently light 
to apply chiral dynamics with a high degree of confidence. In the simplest 
quark model description of the lowest lying non-strange baryons, the strange
quark does not appear as a constituent degree of freedom. On the other hand, 
there have been indications that it contributes substantially to the nucleon 
mass and possibly to the nucleon spin. For these reasons, there has been 
considerable interest within the hadron structure community over the past 
decade to probe other aspects of strange quarks in the nucleon, in the hope of
gaining new insights into the non-perturbative strong interaction.

In this regard, the weak NC interaction between leptons and quarks has provided
a new tool to study strange quarks. The basic idea, first proposed by Kaplan 
and Manohar~\cite{Kaplan:1988ku}, is that the hadronic vector weak NC contains
a different linear combination of the various quark currents, 
${\bar q}\gamma_\mu q$ ($q=u,d,s$) than the electromagnetic current. Similarly,
the hadronic axial vector NC contains a term involving 
${\bar q}\gamma_\mu\gamma_5 q$ that is not simply related {\em via\/} 
an $SU(3)$ flavor rotation to the charge changing weak currents relevant 
to weak decays or CC lepton-nucleon scattering. By performing measurements of 
the hadronic weak NC form factors, one obtains additional information that 
allows one to experimentally separate the up-, down-, and strange quark vector 
and axial vector current matrix elements in the nucleon. Then one can in 
principle determine the role played by the strange quark in the nucleon's 
electromagnetic and spin structure. For example, in the case of the vector 
currents, one has for the proton~\cite{Musolf:1993tb,Ramsey-Musolf:2003dd},
\be
\label{eq:flavorsep}
   G_{E,M}^{NC} (Q^2) = 2 Q_W(p) G_{E,M}^{p}(Q^2) + 2 Q_W(n) G_{E,M}^{I=0}(Q^2)
   - G_{E,M}^{s}(Q^2),
\ee
where $Q_W(p) = 1/2 - 2\sin^2\theta_W$ and $Q_W(n) = - 1/2$ are the tree level 
proton and neutron weak charges (see Section~\ref{escatt}) and the $G_{E,M}^a$ 
are the Sachs electric ($E$) and magnetic ($M$) proton ($a=p$) and neutron
($a=n$) electromagnetic and strange quark ($a=s$) form factors. Since one has 
experimental information on $G_{E,M}^{p,n}(Q^2)$ from parity conserving 
electron scattering, measurements of $G_{E,M}^{NC}(Q^2)$ with parity violating
electron scattering (PVES) effectively provide determinations of 
$G_{E,M}^s(Q^2)$. Note that at very low momentum transfer $G_E^n(Q^2)$ and 
$G_E^s(Q^2)$ vanish linearly with $Q^2$, so that a measurement of 
$G_E^{NC}(Q^2)$ at sufficiently low $Q^2$ provides a determination of 
the proton's weak charge. 

A substantial program involving PVES has been generated by these ideas. Since 
this program has been extensively reviewed in Refs.~\cite{Musolf:1993tb} and
\citer{Beck:2001dz,Spayde:2003nr}, we give only a brief summary here. The PVES
program involves measuring the PV helicity asymmetry for scattering 
longitudinally polarized electrons from a hadronic target. For elastic 
scattering one has,
\be
\label{eq:pvasym}
   A_{\rm LR} = {d\sigma_+ - d\sigma_- \over d\sigma_+ + d\sigma_-} = 
   {G_F Q^2\over 4\sqrt{2}\pi\alpha} \left[ Q_W + F(Q^2,\theta) \right],
\ee
where $d\sigma_{+(-)}$ is the cross section for scattering electrons with 
positive (negative) helicity; $F(Q^2,\theta)$ is a term that depends on 
the target's electromagnetic, axial vector, and strange quark form factors; 
$Q^2=|{\vec q}|^2-q_0^2$; and $\theta$ is the laboratory scattering angle.
The dependence on $Q^2$ and $\theta$ can be exploited to isolate the dependence
on various form factor components ({\em e.g.}, electric, magnetic, and axial). 

A variety of measurements of the vector NC form factors are being carried out 
at MIT-Bates, the MAMI facility in Mainz, and at Jefferson Lab. A summary of 
the present and future strange quark form factor determinations from 
these experiments is given in Table~\ref{tab:strange}.
\begin{table}[t]
\begin{center}
\begin{minipage}[t]{16.5 cm}
\caption[]{Present (upper part) and future (lower part) strange quark form factor 
(FF) determinations with PVES and neutrino scattering. Statistical and systematic 
errors have been combined in quadrature.}
\label{tab:strange} 
\vspace*{4pt}
\end{minipage}
\begin{tabular}{|c|c|c|c|c|c|} 
\hline
&&&&&\\[-8pt]
Experiment & Lepton & Target & FF Combination & $|Q^2|$ (GeV)$^2$ & Result \\[4pt]
\hline
&&&&&\\[-8pt]
SAMPLE~\cite{Spayde:2003nr,Ito:2003mr}  & $e$                       & $p$, $^2$H & $G_M^s$             & $0.1$      & $0.37 \pm 0.34$ \\
HAPPEX~\cite{Aniol:2004hp}              & $e$                       & $p$        & $G_E^s+0.392 G_M^s$ & $0.477$    & $0.014\pm 0.022$\\
A4~\cite{Maas:2004ta}                   & $e$                       & $p$        & $G_E^s+0.225G_M^s$  & $0.230$    & $0.039\pm 0.034$\\
BNL E734~\cite{Ahrens:xe,Garvey:1992cg} & $\nu_\mu$,${\bar\nu}_\mu$ & $^{12}$C   & $0.2\to 1.2$        & $G_A^s(0)$ & $-0.21\pm 0.10$ \\
                                        & $\nu_\mu$,${\bar\nu}_\mu$ & $^{12}$C   & $0.2\to 1.2$        & $G_M^s(0)$ & $-0.40\pm 0.72$ \\
                           & $\nu_\mu$,${\bar\nu}_\mu$ & $^{12}$C & $0.2\to 1.2$ & $ m_N^2\ d G_E^s(0)/d Q^2$ & $-0.57\pm 0.62$   \\[4pt]
\hline
&&&&&\\[-8pt]
HAPPEX~\cite{Cates:1999,Kumar:2004}     & $e$                       & $p$        & $ G_E^s+0.08 G_M^s$ & $0.01$     & $\pm 0.01$  \\
HAPPEX~\cite{Kumar:2004,Armstrong:2000} & $e$                       & $^4$He     & $G_E^s$             & $0.01$     & $\pm 0.015$ \\
G\O\ \cite{Beck:2000,Pitt:2004}         & $e$                       & $p$        & $G_E^s$             & $0.300$    & $\pm 0.112$ \\
 & $e$ & $p$ &  $G_E^s$   & $0.500$ &$\pm 0.073$ \\
 & $e$ & $p$ &  $G_E^s$   & $0.800$ &$\pm 0.051$ \\
 & $e$ & $p$ &  $G_M^s$   & $0.300$ &$\pm 0.031$ \\
 & $e$ & $p$ &  $G_M^s$   & $0.500$ &$\pm 0.033$ \\
 & $e$ & $p$ &  $G_M^s$   & $0.800$ &$\pm 0.034$ \\
FINeSSE~\cite{Bugel:2004yk}             & $\nu_\mu$                 & $^{12}$C   & $0.5\to 1.0$        & $G_A^s$    & $\pm0.04$   \\[-8pt]
&&&&&\\
\hline
\end{tabular}
\end{center}
\end{table} 

Theoretical uncertainties in the EW radiative corrections preclude the use of 
PVES for a determination of $G_A^s$~\cite{Musolf:ts}. The cross section for 
neutrino-nucleon scattering is free from such uncertainties, however, and thus 
provides a theoretically clean means of obtaining this form factor. Several 
neutrino scattering measurements have been used. The analysis of 
these measurements is complicated by the kinematics of neutrino NC scattering,
for which one detects recoil nucleons rather than the outgoing lepton. 
The experimental cross sections are, thus, effective integrals over $Q^2$, so 
only the form factor for an average $<Q^2>$ or fits to form factor moments can
be quoted. The values shown in Table~\ref{tab:strange} for the BNL E734 
experiment~\cite{Ahrens:xe} are taken from a reanalysis by Garvey 
\etal~\cite{Garvey:1992cg} in which the form factors and their slopes at 
$Q^2 = 0$ were fit to the data (for other reanalyses, see 
Refs.~\cite{Alberico:1998qw,Pate:2003rk}). Note that the experiments performed 
to date have entailed the use of QE neutrino-nucleus scattering, which 
introduces some small degree of nuclear model dependence. A future measurement
at Fermilab (FINeSSE) has been proposed, but it is yet to be approved for 
running~\cite{Bugel:2004yk}. For a review of the theoretical hadron structure 
implications of the current experimental results, see {\em e.g.\/} 
Ref.~\cite{Spayde:2003nr}. 

In addition to the use of PVES to probe strangeness, it may also provide 
a probe of various aspects of nucleon structure. In particular, the PAVEX 
experiment at Jefferson Lab will use elastic PVES from $^{208}$Pb to measure 
the Fourier transform of the neutron distribution in that 
nucleus~\cite{Michaels:1999}. In contrast
to the photon, the $Z$ couples preferentially to neutrons rather than protons 
at low momentum transfer, making it a particularly powerful probe of neutron 
properties in the nuclear medium. In the past, strongly interacting probes, 
such as pions and protons, have been used in attempts to determine such neutron
properties. In contrast, the purely leptonic probe of PVES provides 
a theoretically cleaner means of studying the neutron distribution.

\section{Summary and Outlook} 
The next decade promises to be a time of discovery in the field of EW physics.
With the operation of the Large Hadron Collider at CERN, one expects 
to discover at least one Higgs boson and to find evidence for other new 
particles, such as the superpartners of SUSY or additional light gauge bosons 
of grand unified theories. As outlined in this article, progress on the high 
energy collider front is being matched by new strides in the study of low 
energy weak interactions involving light quarks and leptons. The precision of
experiments in this arena are making them sensitive to NP at the TeV scale and
beyond in ways that complement the reach of colliders. From our perspective, 
several aspects of the future program of low energy studies merit emphasizing:

\begin{itemize}
\item {\em Neutrino properties}. 
Now that neutrino oscillations have provided the clearest experimental 
\lq\lq smoking gun" for physics beyond the SM, it is essential to map out 
the characteristics of neutrinos and their interactions in as detailed a manner
as possible. In particular, one would like to know the scale of neutrino mass,
whether there exist more than three generations of light neutrinos, whether 
the neutrino is a Majorana particle, and whether the neutrino mixing angle
$\theta_{13}$ is sufficiently large to lead to significant CP violating effects
in the lepton sector. From this standpoint, more precise measurements of 
tritium $\beta$ decay and neutrinoless double $\beta$ decay, along with 1-2~km
reactor-based studies and long baseline experiments should provide important
insights. 

\item {\em Electric Dipole Moments}. 
The prospects that one or more EDM searches will find a non-zero result during
the next decade are strong. Breakthroughs in experimental techniques for 
probing the EDMs of the electron, neutron, and neutral atoms are pushing 
the sensitivity to a range expected from models of EWB. Complementary efforts 
are also underway involving the muon and deuteron. 

\item {\em Weak mixing}. 
The running of $\sin^2\theta_W$ below the $Z$ pole remains a largely untested
feature of the SM. The PV M\o ller experiment at SLAC will test this running 
precisely by measuring $\sin^2\theta_W$ at a specific low energy value of 
the momentum transfer, and the PV $ep$ experiment at Jefferson Lab will provide
a complementary test at the same scale. Together, these experiments, along with
the results from APV and neutrino-nucleus deep inelastic scattering, begin
to map out the scale dependence of $\sin^2\theta_W$ and either provide 
indications of NP or lead to stringent constraints on deviations from the SM. 
Additional tests of low scale weak mixing, however, would be ideal.

\item {\em Weak decays of light quarks}. 
The weak decays of light quarks are being studied with increasingly high 
precision on a number of fronts. An issue that clearly calls for resolution is
that of the unitarity of the CKM matrix. In this respect, new experiments in 
neutron $\beta$ decay and new studies of kaon leptonic decays are underway, as
are measurements of nuclear and pion $\beta$ decay. Open theoretical questions
involving the effects of strong interactions in EW radiative corrections and 
the momentum dependence of kaon form factors also require further scrutiny. 
Should the long standing deviation from unitarity persist, light quark decays 
would provide important constraints on scenarios for NP.

\item {\em Muon physics}. 
In the near future, the TWIST experiment will produce a substantially more 
precise determination of the Michel parameters, by up to factors of $\sim 60$ 
in some cases. Similarly, the MECO experiment at Brookhaven will improve 
the sensitivity to $\mu\to e$ conversion by a factor of $10^{4}$, providing 
a new window to lepton flavor violation at the GUT scale in SUSY models or 
the TeV scale in other scenarios. Measurements of the muon anomalous magnetic 
moment and electric dipole moment provide complementary probes of physics at 
the TeV scale. In particular, the recent determination of $(g-2)_\mu$ has 
provided a tantalizing hint at SUSY, and a more precise measurement would 
complement future collider searches for superpartners.
\end{itemize}

Throughout our discussion of these studies, an important theme has been 
the complementarity of low energy and high energy probes. Just as in the case 
of the top quark, where the comparison of indirect constraints from precision 
measurements and direct measurements at the Tevatron provided an important test
of the self-consistency of the SM, similar comparisons will be essential in 
the future for testing whatever the \lq\lq new Standard Model" is to be. 
In some cases, such as $0\nu\beta\beta$ decay, electric dipole moments, and 
$\mu\to e(\gamma)$, low energy studies can provide a more direct and more
powerful probe of certain aspects of NP than one may hope to achieve with 
colliders during the next decade. 

At the same time, many of these low energy probes require having in hand 
a sufficiently clear theoretical picture. In the case of quantities such as
$(g-2)_\mu$, for which the SM predicts a non-zero result, obtaining 
a sufficiently reliable treatment of strong interaction contributions is 
essential if one is going to make strong inferences about NP from any report of
a deviation. In this respect, there exists considerable room for future 
interplay between EW and QCD theory. More generally, it is important 
to emphasize the role of theory in assembling the information obtained from 
the broad range of studies --- both low and high energy --- and identifying 
which of the many candidates for NP provides the most coherent account. 

In summary, our view is that the field of low energy tests of the weak 
interaction will continue to be an exciting one during the next decade, 
providing for a rich and fruitful interplay between particle, nuclear, and 
atomic physicists --- in both experiment and theory.

\section*{Acknowledgments}

We gratefully acknowledge the receipt of information from, and useful 
discussions with,
V.~Cirigliano,
D.~DeMille,
B.~Filippone,
N.~Fortson,
S.~Gardner,
J.~Hardy,
P.~Herczeg,
D.~Hertzog,
B.~Holstein, 
R.~Holt,
K.~Kumar,
A.~Kurylov,
S.~Lamoreaux,
P.~Langacker,
F.~Maas,
W.~Marciano,
J.~Martin,
L.~Orozco,
C.~Pe\~na-Garay,
M.~Pitt,
M.~Pospelov,
M.~Romalis,
J.~Stalnaker,
S.~Su,
R.~Tribble,
and
P.~Vogel.  We also thank the Institute for Nuclear Theory at the University of
Washington, where part of this work was carried out. MR-M thanks the Department
of Theoretical Physics at the Physics Institute at UNAM for hospitality during
the completion of this review. This work was supported in part by U.S.\
Department of Energy contracts DE--FG02--00ER41146, DE--FG03--02ER41215, and
DE--FG03--00ER41132, by the National Science Foundation Award PHY00--71856, by 
CONACYT (M\'exico) contract 42026--F, and by DGAPA--UNAM contract PAPIIT 
IN112902.


\begin{thebibliography}{999}
\itemsep -2pt 

\bibitem{Weinberg:tq}
S.~Weinberg, \Journal{\PRL}{19}{1264}{1967}

\bibitem{Salam:rm} 
A.~Salam, p.~367 of Ref.~\cite{Svartholm:1969}

\bibitem{Svartholm:1969}
{\it Elementary Particle Theory}, N.~Svartholm (ed.), (Almquist and Wiksells, 
Stockholm, 1969)

\bibitem{Wilczek:pj}
F.~Wilczek, \Journal{\PRL}{40}{279}{1978}

\bibitem{Nir:2001ge}
Y.~Nir, preprint hep-ph/0109090, lectures at {\sl 55th Scottish 
Universities Summer School in Physics: Heavy Flavor Physics}, 7--23 Aug 2001,
St. Andrews, Scotland, C.~Davies and S.~M.~Playfer (eds.)

\bibitem{Fleischer:2002ys}
R.~Fleischer, \Journal{\PREP}{370}{537}{2002}

\bibitem{Nir:2002gu}
Y.~Nir, \Journal{\NPPS}{117}{111}{2003}

\bibitem{Hurth:2003vb}
T.~Hurth, \Journal{\RMP}{75}{1159}{2003}

\bibitem{Battaglia:2003in}
M.~Battaglia \etal, preprint hep-ph/0304132

\bibitem{Isidori:2004rd}
G.~Isidori, preprint hep-ph/0401079, talk presented at Beauty~2003

\bibitem{Marciano:pd}
W.~J.~Marciano and A.~Sirlin, \Journal{\PRD}{22}{2695}{1980}

\bibitem{Czarnecki:2000ic}
A.~Czarnecki and W.~J.~Marciano, \Journal{\INT}{15}{2365}{2000}

\bibitem{Wood:zq}
C.~S.~Wood \etal, \Journal{\SCI}{275}{1759}{1997}

\bibitem{Bennett:1999pd}
S.~C.~Bennett and C.~E.~Wieman, \Journal{\PRL}{82}{2484}{1999}

\bibitem{Zeller:2001hh}
NuTeV Collaboration: G.~P.~Zeller \etal, \Journal{\PRL}{88}{091802}{2002}

\bibitem{Anthony:2003ub}
SLAC-E-158 Collaboration: P.~L.~Anthony \etal, preprint hep-ex/0312035

\bibitem{Armstrong:2001}
Qweak Collaboration: D.~Armstrong \etal, proposal JLAB--E--02--020, \\
http://www.jlab.org/exp\_prog/proposals/02/PR02-020.pdf

\bibitem{Close:2004da}
F.~E.~Close and Q.~Zhao, preprint hep-ph/0403159

\bibitem{Cabibbo:yz}
N.~Cabibbo, \Journal{\PRL}{10}{531}{1963}

\bibitem{Kobayashi:fv}
M.~Kobayashi and T.~Maskawa, \Journal{\PRO}{49}{652}{1973}

\bibitem{Abe:1995hr}
CDF Collaboration: F.~Abe \etal, \Journal{\PLB}{74}{2626}{1995}

\bibitem{Kodama:2000mp}
DONUT Collaboration: K.~Kodama \etal, \Journal{\PLB}{504}{218}{2001}

\bibitem{Erler:1999ug}
J.~Erler, preprint hep-ph/0005084, in {\sl Physics At Run II: QCD And Weak 
Boson Physics Workshop: 2nd General Meeting}, 3--4 Jun 1999, Batavia, IL,
U. Baur \etal\ (eds.)

\bibitem{Abbaneo:2002}
ALEPH, DELPHI, L3, and OPAL Collaborations, LEP Electroweak Working Group and 
SLD Heavy Flavor Group: D. Abbaneo \etal, preprint hep-ex/0212036

\bibitem{Ackerstaff:1997rc}
OPAL Collaboration: K.~Ackerstaff \etal, \Journal{\ZPC}{76}{387}{1997}

\bibitem{Abreu:1999cj}
DELPHI Collaboration: P.~Abreu \etal, \Journal{\EPJ}{14}{613}{2000}

\bibitem{Abe:2000uc}
SLD Collaboration: K.~Abe \etal, \Journal{\PRL}{85}{5059}{2000}

\bibitem{Abe:2000dq}
SLD Collaboration: K.~Abe \etal, \Journal{\PRL}{84}{5945}{2000}

\bibitem{Abe:2000hk}
SLD Collaboration: K.~Abe \etal, \Journal{\PRL}{86}{1162}{2001}

\bibitem{Holzner:2002ft}
A.~G.~Holzner, preprint hep-ex/0208045, talk presented at BEACH 2002

\bibitem{Abe:1996ef}
SLD Collaboration: K.~Abe \etal, \Journal{\PRL}{78}{17}{1997}

\bibitem{Abbott:1998dc}
D\O\ Collaboration: B.~Abbott \etal, \Journal{\PRD}{58}{052001}{1998}

\bibitem{Abe:1998bf}
CDF Collaboration: F.~Abe \etal, \Journal{\PRL}{82}{271}{1999}

\bibitem{Alitti:1991dk}
UA2 Collaboration: J.~Alitti \etal, \Journal{\PLB}{276}{354}{1992}

\bibitem{Abbott:1999ns}
D\O\ Collaboration: B.~Abbott \etal, \Journal{\PRL}{84}{222}{2000}

\bibitem{Affolder:2000bp}
CDF Collaboration: T.~Affolder \etal, \Journal{\PRD}{64}{052001}{2001}

\bibitem{Smith:1996xz}
M.~C.~Smith and S.~S.~Willenbrock, \Journal{\PRL}{79}{3825}{1997}

\bibitem{Erler:1998sy}
J.~Erler, \Journal{\PRD}{59}{054008}{1999}

\bibitem{Erler:2002bu}
J.~Erler and M.~Luo, \Journal{\PLB}{558}{125}{2003}

\bibitem{Erler:2000cr}
J.~Erler, \Journal{\PRD}{63}{071301}{2001}

\bibitem{Abazov:2004cs}
D\O\ Collaboration: V.M.~Abazov \etal, \Journal{\NATURE}{429}{638}{2004}

\bibitem{Erler:2004nh}
J.~Erler and P.~Langacker, preprint hep-ph/0407097, in Ref.~\cite{PDG:2004}

\bibitem{PDG:2004}
Particle Data Group: S.~Eidelman \etal, \Journal{\PLB}{592}{1}{2004}

\bibitem{Primack:2002th}
J.~R.~Primack, {\em Nucl. Phys. Proc. Suppl.} 124 (2003) 3

\bibitem{'tHooft:bx}
G.~'t Hooft and M.~J.~Veltman, {\em Ann. Poincare Phys. Theor.} A 20 (1974) 69

\bibitem{Wess:1973kz}
J.~Wess and B.~Zumino, \Journal{\PLB}{49}{1974}{52},
\Journal{\NPB}{70}{39}{1974} and 78 (1974) 1

\bibitem{Dimopoulos:1981zb}
S.~Dimopoulos and H.~Georgi, \Journal{\NPB}{193}{150}{1981}

\bibitem{Sakai:1981gr}
N.~Sakai, \Journal{\ZPC}{11}{153}{1981}

\bibitem{Grisaru:1979wc}
M.~T.~Grisaru, W.~Siegel and M.~Rocek, \Journal{\NPB}{159}{429}{1979}

\bibitem{Fischler:1981zk}
W.~Fischler, H.~P.~Nilles, J.~Polchinski, S.~Raby and L.~Susskind,
\Journal{\PRL}{47}{757}{1981}

\bibitem{Witten:nf}
E.~Witten, \Journal{\NPB}{188}{513}{1981}

\bibitem{Dimopoulos:1981yj}
S.~Dimopoulos, S.~Raby and F.~Wilczek, \Journal{\PRD}{24}{1681}{1981}

\bibitem{Ibanez:yh}
L.~E.~Iba\~nez and G.~G.~Ross, \Journal{\PLB}{105}{439}{1981}

\bibitem{Amaldi:1987fu}
U.~Amaldi \etal, \Journal{\PRD}{36}{1385}{1987}

\bibitem{Carena:1995wu}
M.~Carena, M.~Quiros and C.~E.~Wagner, \Journal{\NPB}{461}{407}{1996}

\bibitem{Haber:1996fp}
H.~E.~Haber, R.~Hempfling and A.~H.~Hoang, \Journal{\ZPC}{75}{539}{1997}

\bibitem{Heinemeyer:1998np}
S.~Heinemeyer, W.~Hollik and G.~Weiglein, \Journal{\EPJ}{9}{343}{1999}

\bibitem{Espinosa:2000df}
J.~R.~Espinosa and R.~J.~Zhang, \Journal{\NPB}{586}{3}{2000}

\bibitem{Brignole:2001jy}
A.~Brignole, G.~Degrassi, P.~Slavich and F.~Zwirner, {\NPB}{631}{195}{2002}

\bibitem{Degrassi:2002fi}
G.~Degrassi \etal, \Journal{\EPJ}{28}{133}{2003}

\bibitem{Girardello:1981wz}
L.~Girardello and M.~T.~Grisaru, \Journal{\NPB}{194}{65}{1982}

\bibitem{Kim:1983dt}
J.~E.~Kim and H.~P.~Nilles, \Journal{\PLB}{138}{150}{1984}

\bibitem{Salam:1974xa}
A.~Salam and J.~Strathdee, \Journal{\NPB}{87}{85}{1975}

\bibitem{Fayet:1974pd}
P.~Fayet, \Journal{\NPB}{90}{104}{1975}

\bibitem{Fayet:1976cr}
P.~Fayet and S.~Ferrara, \Journal{\PREP}{32}{249}{1977}

\bibitem{Sohnius:qm}
M.~F.~Sohnius, \Journal{\PREP}{128}{39}{1985}

\bibitem{Nilles:1983ge}
H.~P.~Nilles, \Journal{\PREP}{110}{1}{1984}

\bibitem{Haber:1984rc}
H.~E.~Haber and G.~L.~Kane, \Journal{\PREP}{117}{75}{1985}

\bibitem{Farhi:1980xs}
E.~Farhi and L.~Susskind, \Journal{\PREP}{74}{277}{1981}

\bibitem{Langacker:1980js}
P.~Langacker, \Journal{\PREP}{72}{185}{1981}

\bibitem{Duff:hr}
M.~J.~Duff, B.~E.~Nilsson and C.~N.~Pope, \Journal{\PREP}{130}{1}{1986}

\bibitem{Green:sp}
M.~B.~Green, J.~H.~Schwarz and E.~Witten, {\it Superstring Theory}, 
Vol. 1 and 2 (Cambridge Monographs On Mathematical Physics, 1987)

\bibitem{Hewett:1988xc}
J.~L.~Hewett and T.~G.~Rizzo, \Journal{\PREP}{183}{193}{1989}

\bibitem{Cvetic:1995rj}
M.~Cveti$\check{\rm c}$ and P.~Langacker, \Journal{\PRD}{54}{3570}{1996}

\bibitem{Erler:2000wu}
J.~Erler, \Journal{\NPB}{586}{73}{2000}

\bibitem{Bellucci:1981bs}
S.~Bellucci, M.~Lusignoli and L.~Maiani, \Journal{\NPB}{189}{329}{1981}

\bibitem{Bardin:1980fe}
D.~Y.~Bardin, P.~K.~Khristova and O.~M.~Fedorenko, 
\Journal{\NPB}{175}{435}{1980} and 197 (1982)~1

\bibitem{Hasert:ff}
Gargamelle Neutrino Collaboration: F.~J.~Hasert \etal, 
\Journal{\PLB}{46}{138}{1973}

\bibitem{LlewellynSmith:ie}
C.~H.~Llewellyn Smith, \Journal{\NPB}{228}{205}{1983}

\bibitem{Paschos:1972kj}
E.~A.~Paschos and L.~Wolfenstein, \Journal{\PRD}{7}{91}{1973}

\bibitem{Barnett:1976kh}
R.~M.~Barnett, \Journal{\PRD}{14}{70}{1976}

\bibitem{Georgi:1976ve}
H.~Georgi and H.~D.~Politzer, \Journal{\PRD}{14}{1829}{1976}

\bibitem{Blondel:1989ev}
CDHS Collaboration: A.~Blondel \etal, \Journal{\ZPC}{45}{361}{1990}

\bibitem{Allaby:1987vr}
CHARM Collaboration: J.~V.~Allaby \etal, \Journal{\ZPC}{36}{611}{1987}

\bibitem{Perrier:qg}
F.~Perrier, p.~385 of Ref.~\cite{Langacker:qb}

\bibitem{Langacker:qb}
{\it Precision Tests of the Standard Electroweak Model}, P.~Langacker (ed.),
Advanced series on directions in high energy physics: 14 
(World Scientific, Singapore, 1995)

\bibitem{McFarland:1997wx}
CCFR Collaboration: K.~S.~McFarland \etal, \Journal{\EPJ}{1}{509}{1998}

\bibitem{Miller:2002xh}
G.~A.~Miller and A.~W.~Thomas, preprint hep-ex/0204007

\bibitem{Zeller:2002et}
NuTeV Collaboration: G.~P.~Zeller \etal, preprint hep-ex/0207052

\bibitem{Kovalenko:2002xe}
S.~Kovalenko, I.~Schmidt and J.~J.~Yang, \Journal{\PLB}{546}{68}{2002}

\bibitem{Davidson:2001ji}
S.~Davidson, S.~Forte, P.~Gambino, N.~Rius and A.~Strumia, 
\Journal{\JHEP}{0202}{037}{2002}

\bibitem{Goncharov:2001qe}
NuTeV Collaboration: M.~Goncharov \etal, \Journal{\PRD}{64}{112006}{2001}, as
quoted in Ref.~\cite{Davidson:2002fb}

\bibitem{Davidson:2002fb}
S.~Davidson, preprint hep-ph/0209316, talk presented at the NuFact02 Summer 
Institute

\bibitem{Zeller:2002du}
NuTeV Collaboration: G.~P.~Zeller \etal, \Journal{\PRD}{65}{111103}{2002}

\bibitem{Bodek:1999bb}
A.~Bodek \etal, \Journal{\PRL}{83}{2892}{1999}

\bibitem{Dobrescu:2003ta}
B.~A.~Dobrescu and R.~K.~Ellis, preprint hep-ph/0310154

\bibitem{Kretzer:2003wy}
S.~Kretzer \etal, preprint hep-ph/0312322

\bibitem{Kurylov:2003by}
A.~Kurylov, M.~J.~Ramsey-Musolf and S.~Su, \Journal{\NPB}{667}{321}{2003}

\bibitem{Langacker:2000ju}
P.~Langacker and M.~Pl\"umacher, \Journal{\PRD}{62}{013006}{2000}

\bibitem{Erler:1999nx}
J.~Erler and P.~Langacker, \Journal{\PRL}{84}{212}{2000}

\bibitem{Babu:2002en}
K.~S.~Babu and J.~C.~Pati, preprint hep-ph/0203029

\bibitem{Loinaz:2002ep}
W.~Loinaz, N.~Okamura, T.~Takeuchi and L.~C.~Wijewardhana, 
\Journal{\PRD}{67}{073012}{2003}

\bibitem{Conrad:2004gw}
J.~M.~Conrad, J.~M.~Link and M.~H.~Shaevitz, preprint hep-ex/0403048.

\bibitem{Mann:qh}
A.~K.~Mann, p.~491 of Ref.~\cite{Langacker:qb}

\bibitem{Ahrens:xe}
BNL-E-734 Collaboration: L.~A.~Ahrens \etal, \Journal{\PRD}{35}{785}{1987}

\bibitem{Panman:rg}
J.~Panman, p.~504 of Ref.~\cite{Langacker:qb}

\bibitem{Sarantakos:1982bp}
S.~Sarantakos, A.~Sirlin and W.~J.~Marciano, \Journal{\NPB}{217}{84}{1983}

\bibitem{Bardin:1983yb}
D.~Y.~Bardin and V.~A.~Dokuchaeva, \Journal{\NPB}{246}{221}{1984}

\bibitem{Dorenbosch:1988is}
CHARM Collaboration: J.~Dorenbosch \etal, Journal{\ZPC}{41}{567}{1989}

\bibitem{Vilain:1994qy}
CHARM-II Collaboration: P.~Vilain \etal, \Journal{\PLB}{335}{246}{1994}

\bibitem{Ahrens:fp}
BNL-E-734 Collaboration: L.~A.~Ahrens \etal, \Journal{\PRD}{41}{3297}{1990}

\bibitem{Erler:ew}
J.~Erler and P.~Langacker, p.~98 of Ref.~\cite{Hagiwara:fs}

\bibitem{Hagiwara:fs}
Particle Data Group: K.~Hagiwara \etal, \Journal{\PRD}{66}{010001}{2002}

\bibitem{Allen:qe}
R.~C.~Allen \etal, \Journal{\PRD}{47}{11}{1993}

\bibitem{Auerbach:2001wg}
LSND Collaboration: L.~B.~Auerbach \etal, \Journal{\PRD}{63}{112001}{2001}

\bibitem{Reines:pv}
F.~Reines, H.~S.~Gurr and H.~W.~Sobel, \Journal{\PRL}{37}{315}{1976}

\bibitem{Hung:yg}
P.~Q.~Hung and J.~J.~Sakurai, \Journal{\PLB}{63}{295}{1976}

\bibitem{Wheater:ym}
J.~F.~Wheater, \Journal{\PLB}{105}{483}{1981}

\bibitem{Marciano:1982mm}
W.~J.~Marciano and A.~Sirlin, \Journal{\PRD}{27}{552}{1983} and 29 (1984) 75

\bibitem{Marciano:1993ep}
W.~J.~Marciano, p.~170 of Ref.~\cite{Langacker:qb}

\bibitem{Derman:sp}
E.~Derman, \Journal{\PRD}{7}{2755}{1973}

\bibitem{Berman:1973pt}
S.~M.~Berman and J.~R.~Primack, \Journal{\PRD}{9}{2171}{1974}

\bibitem{Prescott:1978tm}
SLAC-E-122 Collaboration: C.~Y.~Prescott \etal, \Journal{\PLB}{77}{347}{1978}

\bibitem{Prescott:1979dh}
SLAC-E-122 Collaboration: C.~Y.~Prescott \etal, \Journal{\PLB}{84}{524}{1979}

\bibitem{Cahn:1977uu}
R.~N.~Cahn and F.~J.~Gilman, \Journal{\PRD}{17}{1313}{1978}

\bibitem{Bjorken:1978ry}
J.~D.~Bjorken, \Journal{\PRD}{18}{3239}{1978}

\bibitem{Wolfenstein:1978rr}
L.~Wolfenstein, \Journal{\NPB}{146}{477}{1978}

\bibitem{Fritzsch:1978ku}
H.~Fritzsch, \Journal{\ZPC}{1}{321}{1979}

\bibitem{Kim:sa}
J.~E.~Kim, P.~Langacker, M.~Levine and H.~H.~Williams, 
\Journal{\RMP}{53}{211}{1981}

\bibitem{Hung:1981nv}
P.~Q.~Hung and J.~J.~Sakurai, \Journal{\ARNPS}{31}{375}{1981}

\bibitem{Argento:1982tq}
CERN-NA-004 Collaboration: A.~Argento \etal, \Journal{\PLB}{120}{245}{1983}

\bibitem{Heil:dz}
W.~Heil \etal, \Journal{\NPB}{327}{1}{1989}

\bibitem{Hoffmann:1978xy}
E.~Hoffmann and E.~Reya, \Journal{\PRD}{18}{3230}{1978}

\bibitem{Feinberg:cg}
G.~Feinberg, \Journal{\PRD}{12}{3575}{1975}

\bibitem{Walecka:1977us}
J.~D.~Walecka, \Journal{\NPA}{285}{349}{1977}

\bibitem{Musolf:1993tb}
M.~J.~Musolf \etal, \Journal{\PREP}{239}{1}{1994}

\bibitem{Souder:ia}
P.~A.~Souder \etal, \Journal{\PRL}{65}{694}{1990}

\bibitem{Hasty:2001ep}
SAMPLE Collaboration: R.~Hasty \etal, \Journal{\SCI}{290}{2117}{2000}

\bibitem{Musolf:ts}
M.~J.~Musolf and B.~R.~Holstein, \Journal{\PLB}{242}{461}{1990}

\bibitem{Zhu:2000gn}
S.~L.~Zhu, S.~J.~Puglia, B.~R.~Holstein and M.~J.~Ramsey-Musolf,
\Journal{\PRD}{62}{033008}{2000}

\bibitem{Souder:qk}
P.~A.~Souder, p.~599 of Ref.~\cite{Langacker:qb}

\bibitem{Erler:2003yk}
J.~Erler, A.~Kurylov and M.~J.~Ramsey-Musolf, \Journal{\PRD}{68}{016006}{2003}

\bibitem{Bosted:2003} 
P.~E.~Bosted \etal, proposal at SLAC in preparation

\bibitem{Ramsey-Musolf:1999qk}
M.~J.~Ramsey-Musolf, \Journal{\PRC}{60}{015501}{1999}

\bibitem{Masterson:qi}
B.~P.~Masterson and C.~E.~Wieman, p.~545 of Ref.~\cite{Langacker:qb}

\bibitem{Emmons:vn}
T.~P.~Emmons, J.~M.~Reeves and E.~N.~Fortson, \Journal{\PRL}{51}{2089}{1983}

\bibitem{Hollister:1981}
J.~H.~Hollister \etal, \Journal{\PRL}{46}{642}{1981}

\bibitem{MacPherson:1991} 
M.~J.~D.~MacPherson \etal, \Journal{\PRL}{67}{2784}{1991}

\bibitem{Meekhof:1993}
D.~M.~Meekhof \etal, \Journal{\PRL}{71}{3442}{1993}

\bibitem{Wolfeden:1991} 
T.~Wolfeden, P.~Baird and P.~Sandars, \Journal{\EP}{15}{731}{1991}

\bibitem{Drell:mx} 
P.~S.~Drell and E.~D.~Commins, \Journal{\PRL}{53}{968}{1984}

\bibitem{Edwards:1995} 
N.~H.~Edwards \etal, \Journal{\PRL}{74}{2654}{1995}

\bibitem{Vetter:vf}
P.~A.~Vetter \etal, \Journal{\PRL}{74}{2658}{1995}

\bibitem{Gilbert:ki}
S.~L.~Gilbert, M.~C.~Noecker, R.~N.~Watts and C.~E.~Wieman,
\Journal{\PRL}{55}{2680}{1985}

\bibitem{Bouchiat:1986} 
M.~A.~Bouchiat \etal, {\em J. Phys.} (Paris) 47 (1986) 1709

\bibitem{Noecker:ys}
M.~C.~Noecker, B.~P.~Masterson and C.~E.~Wieman, \Journal{\PRL}{61}{310}{1988}

\bibitem{DeMille:1995}
D.~DeMille, \Journal{\PRL}{74}{4165}{1995}

\bibitem{Kimball:2001}
D.~F.~Kimball, \Journal{\PRA}{63}{052113}{2001}

\bibitem{Stalnaker:2004} 
J.~Stalnaker, private communication

\bibitem{Fortson:2004} 
N.~Fortson, private communication

\bibitem{Orozco:2004} 
L. Orozco, private communication

\bibitem{Haxton:2001ay}
W.~C.~Haxton and C.~E.~Wieman, \Journal{\ARNPS}{51}{261}{2001}

\bibitem{Pollock:1999ec}
S.~J.~Pollock and M.~C.~Welliver, \Journal{\PLB}{464}{177}{1999}

\bibitem{Blundell:1990ji}
S.~A.~Blundell, W.~R.~Johnson and J.~Sapirstein, \Journal{\PRL}{65}{1411}{1990}

\bibitem{Blundell:ss}
S.~A.~Blundell, J.~R.~Sapirstein and W.~R.~Johnson, 
\Journal{\PRD}{45}{1602}{1992}

\bibitem{Dzuba:yu}
V.~A.~Dzuba, V.~V.~Flambaum and O.~P.~Sushkov, \Journal{\PLA}{141}{147}{1989}

\bibitem{Rosner:1999cy}
J.~L.~Rosner, \Journal{\PRD}{61}{016006}{2000}

\bibitem{Casalbuoni:1999mw}
R.~Casalbuoni \etal, preprint hep-ph/0001215, talk given at 2nd ECFA/DESY Study
1998-2001

\bibitem{Barger:2000gv}
V.~D.~Barger and K.~M.~Cheung, \Journal{\PLB}{480}{149}{2000}

\bibitem{Ramsey-Musolf:2000qn}
M.~J.~Ramsey-Musolf, \Journal{\PRD}{62}{056009}{2000}

\bibitem{Derevianko:2000dt}
A.~Derevianko, \Journal{\PRL}{85}{1618}{2000}

\bibitem{Dzuba:2001}
V.~A.~Dzuba, C.~Harabati, W.~R.~Johnson, and M.~S.~Safronova, 
\Journal{\PRA}{63}{044103}{2001}

\bibitem{Johnson:2001nk}
W.~R.~Johnson, I.~Bednyakov and G.~Soff, \Journal{\PRL}{87}{233001}{2001}

\bibitem{Milstein:2002ai}
A.~I.~Milstein, O.~P.~Sushkov and I.~S.~Terekhov, 
\Journal{\PRL}{89}{283003}{2002}

\bibitem{Dzuba:2002kx}
V.~A.~Dzuba, V.~V.~Flambaum and J.~S.~Ginges, \Journal{\PRD}{66}{076013}{2002}

\bibitem{Kuchiev:2002fg}
M.~Y.~Kuchiev and V.~V.~Flambaum, \Journal{\PRL}{89}{283002}{2002}

\bibitem{Milstein:2002mv}
A.~I.~Milstein, O.~P.~Sushkov and I.~S.~Terekhov, preprint hep-ph/0212072

\bibitem{Kuchiev:2003pk}
M.~Y.~Kuchiev and V.~V.~Flambaum, preprint hep-ph/0305053.

\bibitem{Ginges:2001} 
V.~A.~Dzuba, V.~V.~Flambaum and J.~S.~M.~Ginges, 
\Journal{\PRA}{63}{062101}{2001}

\bibitem{Michaels:1999}
R.~Michaels and P.~Souder, spokespersons, proposal JLAB--PR--99--012, \\ 
http://www.jlab.org/exp\_prog/proposals/00/PR99-012.pdf

\bibitem{Marciano:1999ih}
W.~J.~Marciano, \Journal{\PRD}{60}{093006}{1999}

\bibitem{Berman:1958ti}
S.~M.~Berman, \Journal{\PREV}{112}{267}{1958}

\bibitem{Kinoshita:1958ru}
T.~Kinoshita and A.~Sirlin, \Journal{\PREV}{113}{1652}{1959}

\bibitem{Roos:mj}
M.~Roos and A.~Sirlin, \Journal{\NPB}{29}{296}{1971}

\bibitem{vanRitbergen:1998yd}
T.~van Ritbergen and R.~G.~Stuart, \Journal{\PRL}{82}{488}{1999}

\bibitem{Kurylov:2001zx}
A.~Kurylov and M.~J.~Ramsey-Musolf, \Journal{\PRL}{88}{071804}{2002}

\bibitem{Drees:1991zk}
M.~Drees, K.~Hagiwara and A.~Yamada, \Journal{\PRD}{45}{1725}{1992}

\bibitem{Altarelli:1993bh}
G.~Altarelli, R.~Barbieri and F.~Caravaglios, \Journal{\PLB}{314}{357}{1993}

\bibitem{Garcia:1993sb}
D.~Garcia and J.~Sola, \Journal{\MPL}{9}{211}{1994}

\bibitem{Chankowski:tn}
P.~H.~Chankowski \etal, \Journal{\NPPS}{37}{232}{1994} 

\bibitem{Cho:1999km}
G.~C.~Cho and K.~Hagiwara, \Journal{\NPB}{574}{623}{2000}

\bibitem{deBoer:1995fr}
W.~de Boer \etal, \Journal{\ZPC}{71}{415}{1996}

\bibitem{Pierce:1996zz}
D.~M.~Pierce, J.~A.~Bagger, K.~T.~Matchev and R.~Zhang,
\Journal{\NPB}{491}{3}{1997}

\bibitem{Erler:1998ur}
J.~Erler and D.~M.~Pierce, \Journal{\NPB}{526}{53}{1998}

\bibitem{FAST} 
PSI-Experiment R99.06.1, {\em Precision Measurement of the $\mu^+$ Lifetime 
($G_F$) with the FAST Detector}, J.~Kirkby and M.~Pohl, co-spokespersons

\bibitem{mulan} 
PSI-Experiment R99.07.1, {\em A Measurement of the Positive Muon Lifetime
utilizing the $\mu$Lan (MUON Lifetime ANalysis) Detector}, D.~Hertzog, contact 
person

\bibitem{Aysto:2001zs}
J.~Aysto \etal, preprint hep-ph/0109217

\bibitem{Michel:1950} 
L.~Michel, \Journal{\PPS}{63}{514}{1950}

\bibitem{Bouchiat:1957}
C.~Bouchiat and L.~Michel, \Journal{\PREV}{106}{170}{1957}

\bibitem{Derenzo:za}
S.~E.~Derenzo, \Journal{\PREV}{181}{1854}{1969}

\bibitem{Balke:1987vr}
B.~K.~E.~Balke, preprint UMI-87-26134

\bibitem{Beltrami:ne}
I.~Beltrami \etal, \Journal{\PLB}{194}{326}{1987} 

\bibitem{Burkard:kf}
H.~Burkard \etal, \Journal{\PLB}{150}{242}{1985}

\bibitem{Poutissou:kg}
TWIST Collaboration: J.~M.~Poutissou \etal, \Journal{\NPA}{721}{465}{2003}

\bibitem{Commins:ns}
E.~D.~Commins and P.~H.~Bucksbaum, {\em Weak Interactions of Leptons and 
Quarks} (Cambridge University Press, Cambridge, 1983)

\bibitem{Herczeg:cx}
P.~Herczeg, \Journal{\PRD}{34}{3449}{1986}

\bibitem{Marciano:1993sh}
W.~J.~Marciano and A.~Sirlin, \Journal{\PRL}{71}{3629}{1993}

\bibitem{Czapek:kc}
G.~Czapek \etal, \Journal{\PRL}{70}{1993}{17}

\bibitem{Britton:1992pg}
D.~I.~Britton \etal, \Journal{\PRL}{68}{3000}{1992}

\bibitem{Barger:1989rk}
V.~D.~Barger, G.~F.~Giudice and T.~Han, \Journal{\PRD}{40}{2987}{1989}

\bibitem{Sirlin:1977sv}
A.~Sirlin, \Journal{\RMP}{50}{573}{1978}

\bibitem{Cirigliano:2002ng}
V.~Cirigliano, M.~Knecht, H.~Neufeld and H.~Pichl, 
\Journal{\EPJ}{27}{255}{2003}

\bibitem{McFarlane:1985} 
W.~K.~McFarlane \etal, \Journal{\PRD}{32}{547}{1985}

\bibitem{Pocanic:2003jp}
PIBETA Collaboration: D.~Pocanic \etal, preprint hep-ex/0311013, talk presented
at NAPP~2003

\bibitem{Pocanic:2003pf}
D.~Pocanic \etal, preprint hep-ex/0312030

\bibitem{Poblaguev:1990tv}
A.~A.~Poblaguev, \Journal{\PLB}{238}{108}{1990}

\bibitem{Herczeg:ur}
P.~Herczeg, \Journal{\PRD}{49}{247}{1994}

\bibitem{Frlez:2003pe}
E.~Frlez \etal, preprint hep-ex/0312029

\bibitem{Jackson:1957} 
J.~D.~Jackson, S.~B.~Treiman and H.~W.~Wyld, \Journal{\PREV}{106}{1957}{517}

\bibitem{Lising:2000pa}
emiT Collaboration: L.~J.~Lising \etal, \Journal{\PRC}{62}{055501}{2000}

\bibitem{Huffman:2002bb}
P.~R.~Huffman \etal, talk presented at International Workshop On Quark Mixing, 
CKM Unitarity

\bibitem{Abele:2002wc}
H.~Abele \etal, \Journal{\PRL}{88}{211801}{2002}

\bibitem{Saunders:2003jg}
A.~Saunders \etal, preprint nucl-ex/0312021

\bibitem{Gardner:2000nk}
S.~Gardner and C.~Zhang, \Journal{\PRL}{86}{5666}{2001}

\bibitem{Ando:2004rk}
S.~Ando \etal, \Journal{\PLB}{595}{250}{2004}

\bibitem{Towner:1995za}
I.~S.~Towner and J.~C.~Hardy, preprint nucl-th/9504015

\bibitem{Towner:1998qj}
I.~S.~Towner and J.~C.~Hardy, preprint nucl-th/9809087, talk presented at 
WEIN~98

\bibitem{Hardy:ci}
J.~C.~Hardy and I.~S.~Towner, \Journal{\EPA}{15}{223}{2002}

\bibitem{Towner:2002rg}
I.~S.~Towner and J.~C.~Hardy, \Journal{\PRC}{66}{035501}{2002}

\bibitem{Sher:2003}
A.~Sher \etal, \Journal{\PRL}{91}{261802}{2003}

\bibitem{Cirigliano:2001mk}
V.~Cirigliano \etal, \Journal{\EPJ}{23}{121}{2002}

\bibitem{Leutwyler:1984je}
H.~Leutwyler and M.~Roos, \Journal{\ZPC}{25}{91}{1984}

\bibitem{Gasser:1984ux}
J.~Gasser and H.~Leutwyler, \Journal{\NPB}{250}{517}{1985}

\bibitem{Post:2001si}
P.~Post and K.~Schilcher, \Journal{\EPJ}{25}{427}{2002}

\bibitem{Bijnens:2003uy}
J.~Bijnens and P.~Talavera, \Journal{\NPB}{669}{341}{2003}

\bibitem{Moulson:2003zu}
KLOE Collaboration: M.~Moulson \etal, \Journal{\AIP}{698}{74}{2004}

\bibitem{Madigozhin:2003ri}
NA48/2 Collaboration: D.~Madigozhin \etal,
\Journal{\eConf}{C0304052}{WG605}{2003}

\bibitem{Vincenzo:2004} 
V.~Cirigliano, H.~Neufeld, and R.~Kessler, private communication

\bibitem{Kurylov:2003zh}
A.~Kurylov, M.~J.~Ramsey-Musolf and S.~Su, \Journal{\PRD}{68}{035008}{2003}

\bibitem{Kurylov:2003xa}
A.~Kurylov, M.~J.~Ramsey-Musolf and S.~Su, \Journal{\PLB}{582}{222}{2004}

\bibitem{Batley:2002gn}
NA48 Collaboration: J.~R.~Batley \etal, \Journal{\PLB}{544}{97}{2002}

\bibitem{Alavi-Harati:2002ye}
KTeV Collaboration: A.~Alavi-Harati \etal, \Journal{\PRD}{67}{012005}{2003}

\bibitem{Jarlskog:1985ht}
C.~Jarlskog, \Journal{\PRL}{55}{1039}{1985}

\bibitem{Fortson:fi}
N.~Fortson, P.~Sandars and S.~Barr, \Journal{\PT}{56N6}{33}{2003}

\bibitem{Ginges:2003qt}
J.~S.~M.~Ginges and V.~V.~Flambaum, preprint physics/0309054

\bibitem{Khriplovich:ga}
I.~B.~Khriplovich and S.~K.~Lamoreaux, {\em CP Violation Without Strangeness: 
Electric Dipole Moments of Particles, Atoms, and Molecules} (Springer, Berlin,
1997) 

\bibitem{Dmitriev:2003hs}
V.~F.~Dmitriev and I.~B.~Khriplovich, \Journal{\PREP}{391}{243}{2004}

\bibitem{Gell-Mann:vs}
M.~Gell-Mann, P.~Ramond and R.~Slansky, p.~315 in {\it Supergravity}, P.~van 
Nieuwenhuizen and D.~Z.~Freedman (eds.), (North-Holland, Amsterdam, 1979)

\bibitem{Mohapatra:1979ia}
R.~N.~Mohapatra and G.~Senjanovic, \Journal{\PRL}{44}{912}{1980}

\bibitem{Yanagida:xy}
T.~Yanagida, \Journal{\PRO}{64}{1103}{1980}

\bibitem{Regan:ta}
B.~C.~Regan, E.~D.~Commins, C.~J.~Schmidt and D.~DeMille, 
\Journal{\PRL}{88}{071805}{2002}

\bibitem{Sandars:1965} 
P.~G.~H.~Sandars, \Journal{\PL}{14}{194}{1965}

\bibitem{Liu:1992} 
Z.~W.~Liu and H.~P.~Kelly, \Journal{\PRA}{45}{R4210}{1992}

\bibitem{DeMille:2000} 
D.~DeMille \etal, \Journal{\PRA}{61}{052507}{2000}

\bibitem{Kawall:2003ga}
D.~Kawall \etal, preprint hep-ex/0309079

\bibitem{Sandars:1967} 
P.~G.~H.~Sandars, \Journal{\PRL}{19}{1396}{1967}

\bibitem{Sushkov:1978} 
O.~P.~Sushkov and V.~V.~Flambaum, \Journal{\ZETF}{75}{1208}{1978}

\bibitem{Kozlov:1995xz}
M.~G.~Kozlov and L.~N.~Labzowsky, \Journal{\JPB}{28}{1933}{1995}

\bibitem{Kozlov:2002} 
M.~G.~Kozlov and D.~DeMille, \Journal{\PRL}{89}{133001}{2002}

\bibitem{Isaev:2003} 
T.~A.~Isaev \etal, preprint physics/0306071

\bibitem{Liu:2004} 
C.~Y.~Liu and S.~Lamoreaux, private communication

\bibitem{Shapiro:68} 
F.~L.~Shapiro, {\em Usp. Fiz. Nauk.} 95 (1968) 145

\bibitem{Semertzidis:2003iq}
EDM Collaboration: Y.~K.~Semertzidis \etal, \Journal{\AIP}{698}{200}{2004}

\bibitem{Purcell:1950}
E.~M.~Purcell and N.~F.~Ramsey, \Journal{\PREV}{78}{807}{1950}

\bibitem{Smith:ht}
J.~H.~Smith, E.~M.~Purcell and N.~F.~Ramsey, \Journal{\PREV}{108}{120}{1957}

\bibitem{Harris:jx}
P.~G.~Harris \etal, \Journal{\PRL}{82}{904}{1999}

\bibitem{Harris:ap}
P.~G.~Harris \etal, {\em Nucl. Instrum. Meth.} A 440 (2000) 479

\bibitem{Mischke:ac}
EDM Collaboration: R.~E.~Mischke, \Journal{\AIP}{675}{246}{2003}

\bibitem{Aleksandrov:2002} 
E.~Aleksandrov \etal, 
http://ucn.web.psi.ch/papers/annual\_report\_02/R-00-05\_mag\_ru-main.pdf

\bibitem{Romalis:2000mg}
M.~V.~Romalis, W.~C.~Griffith and E.~N.~Fortson, \Journal{\PRL}{86}{2505}{2001}

\bibitem{Romalis:2001} 
M.~V.~Romalis and M.~P.~Ledbetter, \Journal{\PRL}{87}{067601}{2001}

\bibitem{Romalis:2004} 
M.~V.~Romalis, private communication

\bibitem{Holt:2004} 
R.~Holt, private communication

\bibitem{Behr:2003}
J.~Behr, {\em Nucl. Inst. Meth.} B 204 (2003) 526

\bibitem{Shabalin:rs}
E.~P.~Shabalin, \Journal{\SJNP}{28}{75}{1978} and 
{\em Yad. Fiz.} 31 (1980) 1665

\bibitem{Shabalin:sg}
E.~P.~Shabalin, {\em Sov. Phys. Usp.} 26 (1983) 297

\bibitem{Bernreuther:1990jx}
W.~Bernreuther and M.~Suzuki, \Journal{\RMP}{63}{313}{1991}

\bibitem{Gavela:1981sm}
M.~B.~Gavela \etal, \Journal{\PLB}{109}{83}{1982} and 215

\bibitem{Khriplovich:1981ca}
I.~B.~Khriplovich and A.~R.~Zhitnitsky, \Journal{\PLB}{109}{490}{1982}

\bibitem{He:1989xj}
X.~G.~He, B.~H.~J.~McKellar and S.~Pakvasa, \Journal{\INT}{4}{5011}{1989}

\bibitem{Haxton:dq}
W.~C.~Haxton and E.~M.~Henley, \Journal{\PRL}{51}{1937}{1983}

\bibitem{Flambaum:1984fb}
V.~V.~Flambaum, I.~B.~Khriplovich and O.~P.~Sushkov,
\Journal{\JETP}{60}{873}{1984} and \Journal{\NPA}{449}{750}{1986}

\bibitem{Donoghue:dd}
J.~F.~Donoghue, B.~R.~Holstein and M.~J.~Musolf, \Journal{\PLB}{196}{196}{1987}

\bibitem{Schiff:1963} 
L.~I.~Schiff, \Journal{\PRD}{132}{2194}{1963}

\bibitem{Engel:2003rz}
J.~Engel \etal, \Journal{\PRC}{68}{025501}{2003}

\bibitem{Engel:1999np}
J.~Engel, J.~L.~Friar and A.~C.~Hayes, \Journal{\PRC}{61}{035502}{2000}

\bibitem{Dzuba:2002kg}
V.~A.~Dzuba, V.~V.~Flambaum, J.~S.~M.~Ginges and M.~G.~Kozlov,
\Journal{\PRA}{66}{012111}{2002}

\bibitem{Dmitriev:2003sc}
V.~F.~Dmitriev and R.~A.~Sen'kov, \Journal{\PRL}{91}{212303}{2003}

\bibitem{Khriplovich:1999qr}
I.~B.~Khriplovich and R.~A.~Korkin, \Journal{\NPA}{665}{365}{2000}

\bibitem{Crewther:1979pi}
R.~J.~Crewther, P.~Di Vecchia, G.~Veneziano and E.~Witten, 
\Journal{\PLB}{88}{123}{1979}

\bibitem{Kawarabayashi:1980dp}
K.~Kawarabayashi and N.~Ohta, \Journal{\NPB}{175}{477}{1980} and 
{\em Prog.\ Theor.\ Phys.} 66 (1981) 1789 

\bibitem{Falk:1999tm}
T.~Falk, K.~A.~Olive, M.~Pospelov and R.~Roiban, \Journal{\NPB}{560}{3}{1999}

\bibitem{Conti:xn}
R.~S.~Conti and I.~B.~Khriplovich, \Journal{\PRL}{68}{3262}{1992}

\bibitem{Engel:1995vv}
J.~Engel, P.~H.~Frampton and R.~P.~Springer, \Journal{\PRD}{53}{5112}{1996} 

\bibitem{Ramsey-Musolf:1999nk}
M.~J.~Ramsey-Musolf, \Journal{\PRL}{83}{3997}{1999}

\bibitem{Kurylov:2000ub}
A.~Kurylov, G.~C.~McLaughlin and M.~J.~Ramsey-Musolf, 
\Journal{\PRD}{63}{076007}{2001}

\bibitem{Sakharov:dj}
A.~D.~Sakharov, {\em JETP Lett.} 5 (1967) 24 

\bibitem{Barr:1988mc}
S.~M.~Barr and W.~J.~Marciano, {\em Adv. Ser. Direct. High Energy Phys.} 3 
(1989) 455

\bibitem{Brignole:1993wv}
A.~Brignole, J.~R.~Espinosa, M.~Quiros and F.~Zwirner, 
\Journal{\PLB}{324}{181}{1994}

\bibitem{Riotto:1999yt}
A.~Riotto and M.~Trodden, \Journal{\ARNPS}{49}{35}{1999}

\bibitem{Huet:1995sh}
P.~Huet and A.~E.~Nelson, \Journal{\PRD}{53}{4578}{1996}

\bibitem{Riotto:1998zb}
A.~Riotto, \Journal{\PRD}{58}{095009}{1998}

\bibitem{Brooks:1999pu}
MEGA Collaboration: M.~L.~Brooks \etal, \Journal{\PRL}{83}{1521}{1999}

\bibitem{Wintz:rp}
P.~Wintz, talk presented at ICHEP 98

\bibitem{Bellgardt:1987du}
SINDRUM Collaboration: U.~Bellgardt \etal, \Journal{\NPB}{299}{1}{1988}

\bibitem{Kaulard:rb}
SINDRUM II Collaboration: J.~Kaulard \etal, \Journal{\PLB}{422}{334}{1998}

\bibitem{Yashima:2000qz}
J.~Yashima \etal, in {\sl New Initiatives on Lepton Flavor Violation and 
Neutrino Oscillation With High Intense Muon and Neutrino Sources}, 
2--6 Oct 2000, Honolulu, HI, Y.~Kuno \etal\ (eds.)

\bibitem{Molzon:sf}
W.~Molzon, \Journal{\NPPS}{111}{188}{2002}

\bibitem{Riazuddin:hz}
Riazuddin, R.~E.~Marshak and R.~N.~Mohapatra, \Journal{\PRD}{24}{1310}{1981}

\bibitem{Leontaris:1985pq}
G.~K.~Leontaris, K.~Tamvakis and J.~D.~Vergados, \Journal{\PLB}{171}{412}{1986}

\bibitem{Borzumati:1986qx}
F.~Borzumati and A.~Masiero, \Journal{\PRL}{57}{961}{1986}

\bibitem{Barbieri:1994pv}
R.~Barbieri and L.~J.~Hall, \Journal{\PLB}{338}{212}{1994}

\bibitem{Barbieri:1995tw}
R.~Barbieri, L.~J.~Hall and A.~Strumia, \Journal{\NPB}{445}{219}{1995}

\bibitem{Hisano:1995nq}
J.~Hisano \etal, \Journal{\PLB}{357}{579}{1995}

\bibitem{Barenboim:1996vu}
G.~Barenboim and M.~Raidal, \Journal{\NPB}{484}{63}{1997}

\bibitem{Huitu:1997bi}
K.~Huitu, J.~Maalampi, M.~Raidal and A.~Santamaria,
\Journal{\PLB}{430}{355}{1998}

\bibitem{Raidal:1997hq}
M.~Raidal and A.~Santamaria, \Journal{\PLB}{421}{250}{1998}

\bibitem{Cirigliano:2004} 
V.~Cirigliano, A.~Kurylov, M.~Ramsey-Musolf and P.~Vogel, preprint 
hep-ph/0404223

\bibitem{Glashow:gm}
S.~L.~Glashow, J.~Iliopoulos and L.~Maiani, \Journal{\PRD}{2}{1285}{1970}

\bibitem{McKeown:2004yq}
R.~D.~McKeown and P.~Vogel, \Journal{\PREP}{394}{315}{2004}

\bibitem{Spergel:2003cb}
D.~N.~Spergel \etal, {\em Astrophys. J. Suppl.} 148 (2003) 175

\bibitem{Majorana:vz}
E.~Majorana, \Journal{\NC}{14}{171}{1937}

\bibitem{Gelmini:1980re}
G.~B.~Gelmini and M.~Roncadelli, \Journal{\PLB}{99}{411}{1981}

\bibitem{Langacker:1991nt}
P.~Langacker, p.~863 in {\it Testing The Standard Model}, 
M.~Cveti$\check{\rm c}$ and P.~Langacker (eds.), (World Scientific, Singapore, 
1991)

\bibitem{Maki:mu}
Z.~Maki, M.~Nakagawa and S.~Sakata, \Journal{\PRO}{28}{870}{1962}

\bibitem{Barbieri:1979hc}
R.~Barbieri, J.~R.~Ellis and M.~K.~Gaillard, \Journal{\PLB}{90}{249}{1980}

\bibitem{Weinheimer:tn}
C.~Weinheimer \etal, \Journal{\PLB}{460}{219}{1999}

\bibitem{Lobashev:tp}
V.~M.~Lobashev \etal, \Journal{\PLB}{460}{227}{1999}

\bibitem{Vogel:ee}
P.~Vogel and A.~Piepke, p.~380 of Ref.~\cite{Hagiwara:fs}

\bibitem{Osipowicz:2001sq}
KATRIN Collaboration: A.~Osipowicz \etal, preprint hep-ex/0109033

\bibitem{Bilenky:2002aw}
S.~M.~Bilenky, C.~Giunti, J.~A.~Grifols and E.~Masso, 
\Journal{\PREP}{379}{69}{2003}

\bibitem{Pontecorvo:cp}
B.~Pontecorvo, \Journal{\JETP}{6}{429}{1957}

\bibitem{Gribov:1968kq}
V.~N.~Gribov and B.~Pontecorvo, \Journal{\PLB}{28}{493}{1969}

\bibitem{Bilenky:nj}
S.~M.~Bilenky and B.~Pontecorvo, \Journal{\PREP}{41}{225}{1978}

\bibitem{Kayser:1981ye}
B.~Kayser, \Journal{\PRD}{24}{110}{1981}

\bibitem{Vergados:1985pq}
J.~D.~Vergados, \Journal{\PREP}{133}{1}{1986}

\bibitem{Bilenky:ty}
S.~M.~Bilenky and S.~T.~Petcov, \Journal{\RMP}{59}{671}{1987}

\bibitem{Bilenky:1998dt}
S.~M.~Bilenky, C.~Giunti and W.~Grimus, \Journal{\PPNP}{43}{1}{1999}

\bibitem{Bilenky:1976yj}
S.~M.~Bilenky and B.~Pontecorvo, \Journal{\LNC}{17}{569}{1976}

\bibitem{Murayama:eb}
H.~Murayama, p.~401 of Ref.~\cite{Hagiwara:fs}

\bibitem{Nakamura:dz}
K.~Nakamura, p.~408 of Ref.~\cite{Hagiwara:fs}

\bibitem{Goswami:2003bh}
S.~Goswami, preprint hep-ph/0303075

\bibitem{Bahcall:2001cb}
J.~N.~Bahcall, M.~C.~Gonzalez-Garcia and C.~Pe\~na-Garay, 
\Journal{\JHEP}{0204}{007}{2002}

\bibitem{Bahcall:2000nu}
J.~N.~Bahcall, M.~H.~Pinsonneault and S.~Basu, \Journal{\AJ}{555}{990}{2001}

\bibitem{Junghans:2001ee}
A.~R.~Junghans \etal, \Journal{\PRL}{88}{041101}{2002}

\bibitem{Adelberger:1998qm}
E.~G.~Adelberger \etal, \Journal{\RMP}{70}{1265}{1998}

\bibitem{Baby:2002hj}
ISOLDE Collaboration: L.~T.~Baby \etal, \Journal{\PRL}{90}{022501}{2003}

\bibitem{Cleveland:nv}
B.~T.~Cleveland \etal, \Journal{\AJ}{496}{505}{1998}

\bibitem{Abdurashitov:2002nt}
SAGE Collaboration: J.~N.~Abdurashitov \etal, \Journal{\JETP}{95}{181}{2002}

\bibitem{Hampel:1998xg}
GALLEX Collaboration: W.~Hampel \etal, Journal{\PLB}{447}{127}{1999}

\bibitem{Altmann:2000ft}
GNO Collaboration: M.~Altmann \etal, \Journal{\PLB}{490}{16}{2000}

\bibitem{Fukuda:1996sz}
Kamiokande Collaboration: Y.~Fukuda \etal, \Journal{\PRL}{77}{1683}{1996}

\bibitem{Fukuda:2002pe}
Super-Kamiokande Collaboration: S.~Fukuda \etal, \Journal{\PLB}{539}{179}{2002}

\bibitem{Smy:2002rz}
Super-Kamiokande Col.: M.~B.~Smy \etal, preprint hep-ex/0208040, talk given at
Neutrino 2002 

\bibitem{Ahmad:2002jz}
SNO Collaboration: Q.~R.~Ahmad \etal, \Journal{\PRL}{89}{011301}{2002}

\bibitem{Gando:2002ub}
Super-Kamiokande Collaboration: Y.~Gando \etal, 
\Journal{\PRL}{90}{171302}{2003}

\bibitem{Bludman:1993tk}
S.~Bludman, N.~Hata and P.~Langacker, \Journal{\PRD}{49}{3622}{1994}

\bibitem{Fogli:1993ck}
G.~L.~Fogli, E.~Lisi and D.~Montanino, \Journal{\PRD}{49}{3626}{1994}

\bibitem{Akhmedov:1994ix}
E.~K.~Akhmedov, A.~Lanza and S.~T.~Petcov, \Journal{\PLB}{348}{124}{1995}

\bibitem{Berezinsky:1995zt}
V.~Berezinsky, G.~Fiorentini and M.~Lissia, \Journal{\PLB}{365}{185}{1996}

\bibitem{Krastev:1994nx}
P.~I.~Krastev and S.~T.~Petcov, \Journal{\NPB}{449}{605}{1995}

\bibitem{Alimonti:2000xc}
Borexino Collaboration: G.~Alimonti \etal, \Journal{\APP}{16}{205}{2002}

\bibitem{Ahmad:2002ka}
SNO Collaboration: Q.~R.~Ahmad \etal, \Journal{\PRL}{89}{011302}{2002}

\bibitem{Mikheev:gs}
S.~P.~Mikheev and A.~Y.~Smirnov, \Journal{\SJNP}{42}{913}{1985}

\bibitem{Wolfenstein:1977ue}
L.~Wolfenstein, \Journal{\PRD}{17}{2369}{1978}

\bibitem{Barger:2002at}
V.~Barger and D.~Marfatia, \Journal{\PLB}{555}{144}{2003}

\bibitem{Fogli:2002au}
G.~L.~Fogli \etal, \Journal{\PRD}{67}{073002}{2003}

\bibitem{Maltoni:2002aw}
M.~Maltoni, T.~Schwetz and J.~W.~Valle, \Journal{\PRD}{67}{093003}{2003}

\bibitem{Bandyopadhyay:2002en}
A.~Bandyopadhyay \etal, \Journal{\PLB}{559}{121}{2003}

\bibitem{Nunokawa:2002mq}
H.~Nunokawa, W.~J.~Teves and R.~Zukanovich Funchal, 
\Journal{\PLB}{562}{28}{2003}

\bibitem{Aliani:2002na}
P.~Aliani, V.~Antonelli, M.~Picariello and E.~Torrente-Lujan, 
\Journal{\PRD}{69}{013005}{2004} 

\bibitem{deHolanda:2002iv}
P.~C.~de Holanda and A.~Y.~Smirnov, \Journal{\JCAP}{0302}{001}{2003}

\bibitem{Balantekin:2003dc}
A.~B.~Balantekin and H.~Yuksel, \Journal{\JPG}{29}{665}{2003}

\bibitem{Bahcall:2003ce}
J.~N.~Bahcall and C.~Pe\~na-Garay, \Journal{\JHEP}{0311}{004}{2003}

\bibitem{Ahmed:2003kj}
SNO Collaboration: S.~N.~Ahmed \etal, preprint nucl-ex/0309004

\bibitem{Hirata:1992ku}
Kamiokande-II Collaboration: K.~S.~Hirata \etal, \Journal{\PLB}{280}{146}{1992}

\bibitem{Fukuda:1994mc}
Kamiokande Collaboration: Y.~Fukuda \etal, \Journal{\PLB}{335}{237}{1994}

\bibitem{Becker-Szendy:hq}
R.~Becker-Szendy \etal, \Journal{\PRD}{46}{3720}{1992}

\bibitem{Daum:bf}
Frejus Collaboration: K.~Daum \etal, \Journal{\ZPC}{66}{417}{1995}

\bibitem{Fukuda:1998tw}
Super-Kamiokande Collaboration: Y.~Fukuda \etal, \Journal{\PLB}{433}{9}{1998}

\bibitem{Fukuda:1998ub}
Super-Kamiokande Collaboration: Y.~Fukuda \etal, \Journal{\PLB}{436}{33}{1998}

\bibitem{Allison:1999ms}
Soudan-2 Collaboration: W.~W.~Allison \etal, \Journal{\PLB}{449}{137}{1999}

\bibitem{Fukuda:1998mi}
Super-Kamiokande Collaboration: Y.~Fukuda \etal, \Journal{\PRL}{81}{1562}{1998}

\bibitem{Fukuda:1998ah}
Super-Kamiokande Collaboration: Y.~Fukuda \etal, \Journal{\PRL}{82}{2644}{1999}

\bibitem{Fukuda:1999pp}
Super-Kamiokande Collaboration: S.~Fukuda \etal, \Journal{\PLB}{467}{185}{1999}

\bibitem{Fukuda:2000np}
Super-Kamiokande Collaboration: S.~Fukuda \etal, \Journal{\PRL}{85}{3999}{2000}

\bibitem{Ambrosio:2001je}
MACRO Collaboration: M.~Ambrosio \etal, \Journal{\PLB}{517}{59}{2001}

\bibitem{Ambrosio:2003yz}
MACRO Collaboration: M.~Ambrosio \etal, \Journal{\PLB}{566}{35}{2003}

\bibitem{Fogli:2003th}
G.~L.~Fogli, E.~Lisi, A.~Marrone and D.~Montanino, 
\Journal{\PRD}{67}{093006}{2003}

\bibitem{Groom:ec}
D.~E.~Groom, p.~399 of Ref.~\cite{Hagiwara:fs}

\bibitem{Armbruster:2002mp}
KARMEN Collaboration: B.~Armbruster \etal, \Journal{\PRD}{65}{112001}{2002}

\bibitem{Aguilar:2001ty}
LSND Collaboration: A.~Aguilar \etal, \Journal{\PRD}{64}{112007}{2001}

\bibitem{Church:1997jc}
BooNE Collaboration: E.~Church \etal, preprint nucl-ex/9706011

\bibitem{Barenboim:2001ac}
G.~Barenboim, L.~Borissov, J.~Lykken and A.~Y.~Smirnov, 
\Journal{\JHEP}{0210}{001}{2002}

\bibitem{Barger:2003zg}
V.~Barger \etal, \Journal{\PLB}{566}{8}{2003}

\bibitem{Eskut:2000de}
CHORUS Collaboration: E.~Eskut \etal, \Journal{\PLB}{497}{8}{2001}

\bibitem{Astier:2001yj}
NOMAD Collaboration: P.~Astier \etal,\Journal{\NPB}{611}{3}{2001}

\bibitem{Dydak:1983zq}
CERN-PS-169 Collaboration: F.~Dydak \etal, \Journal{\PLB}{134}{281}{1984}

\bibitem{Ahn:2002up}
K2K Collaboration: M.~H.~Ahn \etal, \Journal{\PRL}{90}{041801}{2003}

\bibitem{Saakian:sd}
MINOS Collaboration: R.~Saakian \etal, \Journal{\NPPS}{111}{169}{2002}

\bibitem{Duchesneau:2002yq}
OPERA and ICARUS Collns.: D.~Duchesneau \etal, 
{\em Nucl. Phys. Proc. Suppl.} 123 (2003) 279

\bibitem{Declais:1994su}
Y.~Declais \etal, \Journal{\NPB}{434}{503}{1995}

\bibitem{Apollonio:2002gd}
M.~Apollonio \etal, \Journal{\EPJ}{27}{331}{2003}

\bibitem{Boehm:2001ik}
F.~Boehm \etal, \Journal{\PRD}{64}{112001}{2001}

\bibitem{Eguchi:2002dm}
KamLAND Collaboration: K.~Eguchi \etal, \Journal{\PRL}{90}{021802}{2003}

\bibitem{Elliott:2002xe}
S.~R.~Elliott and P.~Vogel, \Journal{\ARNPS}{52}{115}{2002}

\bibitem{Baudis:1999xd}
L.~Baudis \etal, \Journal{\PRL}{83}{41}{1999}

\bibitem{Klapdor-Kleingrothaus:2000sn}
H.~V.~Klapdor-Kleingrothaus \etal,  \Journal{\EPA}{12}{147}{2001}

\bibitem{Aalseth:ji}
IGEX Collaboration: C.~E.~Aalseth \etal, \Journal{\PRC}{59}{2108}{1999}

\bibitem{Aalseth:ud}
C.~E.~Aalseth \etal, {\em Phys. Atom. Nucl.} 63 (2000) 1225

\bibitem{Luscher:sd}
R.~Luscher \etal, \Journal{\PLB}{434}{407}{1998}

\bibitem{Alessandrello:kt}
A.~Alessandrello \etal, \Journal{\PLB}{486}{2000}{13}

\bibitem{Ejiri:2001fx}
H.~Ejiri \etal, \Journal{\PRC}{63}{065501}{2001}

\bibitem{Klapdor-Kleingrothaus:2000ue}
H.~V.~Klapdor-Kleingrothaus, preprint hep-ph/0103074, talk presented at 
NOON~2000

\bibitem{Aalseth:2002sy}
Majorana Collaboration: C.~E.~Aalseth \etal, preprint hep-ex/0201021, talk
given at NANP~01

\bibitem{Danevich:2000cf}
F.~A.~Danevich \etal, \Journal{\PRC}{62}{045501}{2000}

\bibitem{Arnaboldi:hb}
C.~Arnaboldi \etal, \Journal{\AIP}{605}{469}{2002}

\bibitem{Ejiri:1999rk}
H.~Ejiri \etal, \Journal{\PRL}{85}{2917}{2000}

\bibitem{Klapdor-Kleingrothaus:2001ke}
H.~V.~Klapdor-Kleingrothaus \etal, \Journal{\MPL}{16}{2409}{2001}

\bibitem{Aalseth:2002dt}
C.~E.~Aalseth \etal, \Journal{\MPL}{17}{1475}{2002}

\bibitem{Klapdor-Kleingrothaus:2004na}
H.~V.~Klapdor-Kleingrothaus \etal, preprint hep-ph/0404062

\bibitem{Faessler:1996ph}
A.~Faessler, S.~Kovalenko, F.~Simkovic and J.~Schwieger, 
\Journal{\PRL}{78}{183}{1997}

\bibitem{Paes:2000vn}
H.~Paes, M.~Hirsch, H.~V.~Klapdor-Kleingrothaus, S.~G.~Kovalenko,
\Journal{\PLB}{498}{35}{2001}

\bibitem{Prezeau:2003xn}
G.~Pr\'ezeau, M.~Ramsey-Musolf and P.~Vogel, \Journal{\PRD}{68}{034016}{2003}

\bibitem{Kinoshita:1990}
T. Kinoshita and W.~J. Marciano, p.~419 of {\it Quantum Electrodynamics}, 
T.~Kinoshita (ed.), Advanced series on directions in high energy physics: 7 
(World Scientific, Singapore, 1990)

\bibitem{Bennett:2004pv}
Muon g-2 Collaboration: G.~W.~Bennett \etal, preprint hep-ex/0401008

\bibitem{Czarnecki:2001pv}
A.~Czarnecki and W.~J.~Marciano, \Journal{\PRD}{64}{013014}{2001}

\bibitem{Lopez:1993vi}
J.~L.~Lopez, D.~V.~Nanopoulos, and X.~Wang, \Journal{\PRD}{49}{366}{1994}

\bibitem{Erler:2000nx}
J.~Erler and M.~Luo, \Journal{\PRL}{87}{071804}{2001}

\bibitem{Gourdin:dm}
M.~Gourdin and E.~De Rafael, \Journal{\NPB}{10}{667}{1969}

\bibitem{Davier:2003pw}
M.~Davier, S.~Eidelman, A.~H\"ocker and Z.~Zhang, \Journal{\EPJ}{31}{503}{2003}

\bibitem{Akhmetshin:2001ig}
CMD-2 Collaboration: R.~R.~Akhmetshin \etal, \Journal{\PLB}{527}{161}{2002}

\bibitem{Hagiwara:2003da}
K.~Hagiwara, A.~D.~Martin, D.~Nomura and T.~Teubner, preprint hep-ph/0312250

\bibitem{Ghozzi:2003yn}
S.~Ghozzi and F.~Jegerlehner, \Journal{\PLB}{583}{222}{2004}

\bibitem{Matorras:2003}
F.~Matorras, talk presented at EPS 2003

\bibitem{Binner:1999bt}
S.~Binner, J.~H.~K\"uhn and K.~Melnikov, \Journal{\PLB}{459}{279}{1999}

\bibitem{Cirigliano:2002pv}
V.~Cirigliano, G.~Ecker and H.~Neufeld, \Journal{\JHEP}{0208}{002}{2002}

\bibitem{Marciano:vm}
W.~J.~Marciano and A.~Sirlin, \Journal{\PRL}{61}{1815}{1988}

\bibitem{Erler:2002mv}
J.~Erler, preprint hep-ph/0211345

\bibitem{Knecht:2001qf}
M.~Knecht and A.~Nyffeler, \Journal{\PRD}{65}{073034}{2002}

\bibitem{Hayakawa:2001bb}
M.~Hayakawa and T.~Kinoshita, preprint hep-ph/0112102

\bibitem{Bijnens:2001cq}
J.~Bijnens, E.~Pallante and J.~Prades, \Journal{\NPB}{626}{410}{2002}

\bibitem{Ramsey-Musolf:2002cy}
M.~Ramsey-Musolf and M.~B.~Wise, \Journal{\PRL}{89}{041601}{2002}

\bibitem{Krause:1996rf}
B.~Krause, \Journal{\PLB}{390}{392}{1997}

\bibitem{Kinoshita:2004wi}
T.~Kinoshita and M.~Nio, preprint hep-ph/0402206

\bibitem{Melnikov:2003xd}
K.~Melnikov and A.~Vainshtein, preprint hep-ph/0312226

\bibitem{Donoghue:dd1}
J.~F.~Donoghue \etal, {\em Cambridge Monogr. Part. Phys. Nucl. Phys. Cosmol.} 
2 (1992) 1

\bibitem{Borasoy:1998ku}
B.~Borasoy and B.~R.~Holstein, \Journal{\EPJ}{6}{1999}{85}

\bibitem{Potter:1974gu}
J.~M.~Potter \etal, \Journal{\PRL}{33}{1307}{1974}

\bibitem{Nagle:1978vn}
D.~E.~Nagle \etal, \Journal{\AIP}{51}{224}{1978}

\bibitem{Balzer:au}
R.~Balzer \etal, \Journal{\PRC}{30}{1409}{1984}

\bibitem{Kistryn:1987tq}
S.~Kistryn \etal, \Journal{\PRL}{58}{1616}{1987}

\bibitem{Eversheim:tg}
P.~D.~Eversheim \etal, \Journal{\PLB}{256}{11}{1991}

\bibitem{Berdoz:2001nu}
TRIUMF-E497-Collaboration: A.~R.~Berdoz \etal, \Journal{\PRL}{87}{272301}{2001}

\bibitem{Berdoz:2002sn}
TRIUMF-E497-Collaboration: A.~R.~Berdoz \etal, \Journal{\PRC}{68}{034004}{2003}

\bibitem{Lang:jv}
J.~Lang \etal, \Journal{\PRL}{54}{170}{1985}

\bibitem{Lang:nw}
J.~Lang \etal, \Journal{\PRC}{34}{1545}{1986}

\bibitem{Alfimenkov:1983}
V.~P.~Alfimenkov \etal, \Journal{\NPA}{398}{93}{1983}

\bibitem{Masuda:1989}
Y.~Masuda \etal, \Journal{\NPA}{504}{269}{1989}

\bibitem{Yuan:1991}
V.~W.~Yuan \etal, \Journal{\PRC}{44}{2187}{1991}

\bibitem{Adelberger:1983}
E.~G.~Adelberger \etal, \Journal{\PRC}{27}{2833}{1983}

\bibitem{Elsener:vp}
K.~Elsener \etal, \Journal{\PRL}{52}{1476}{1984}

\bibitem{Elsener:sx}
K.~Elsener \etal, \Journal{\NPA}{461}{579}{1987}

\bibitem{Snover:1978}
K~A.~Snover \etal, \Journal{\PRL}{41}{145}{1978}

\bibitem{Earle:ji}
E.~D.~Earle \etal, \Journal{\NPA}{396}{221C}{1983}

\bibitem{Barnes:sq}
C.~A.~Barnes \etal, \Journal{\PRL}{40}{840}{1978}

\bibitem{Bini:1985}
M.~Bini \etal, \Journal{\PRL}{55}{795}{1985}

\bibitem{Ahrens:1982}
G.~Ahrens \etal, \Journal{\NPA}{390}{496}{1982}

\bibitem{Page:ak}
S.~A.~Page \etal, \Journal{\PRC}{35}{1119}{1987}

\bibitem{Haxton:2001mi}
W.~C.~Haxton, C.~P.~Liu and M.~J.~Ramsey-Musolf, \Journal{\PRL}{86}{5247}{2001}

\bibitem{Wilburn:1998xq}
W.~S.~Wilburn and J.~D.~Bowman, \Journal{\PRC}{57}{3425}{1998}

\bibitem{Mitchell:2004fn}
NPDGamma Collaboration: G.~S.~Mitchell \etal, {\em Nucl. Inst. Meth.} A 521 
(2004) 468

\bibitem{Desplanques:1979hn}
B.~Desplanques, J.~F.~Donoghue and B.~R.~Holstein, 
\Journal{\ANNP}{124}{449}{1980}

\bibitem{Dubovik:pj}
V.~M.~Dubovik and S.~V.~Zenkin, \Journal{\ANNP}{172}{100}{1986}

\bibitem{Feldman:tj}
G.~B.~Feldman, G.~A.~Crawford, J.~Dubach and B.~R.~Holstein,
\Journal{\PRC}{43}{863}{1991}

\bibitem{Holstein:2004}
B.~R.~Holstein, \Journal{\AIP}{698}{176}{2004}

\bibitem{Holstein:2003} 
B.~R.~Holstein, talk presented at FB~17

\bibitem{Holstein:2004a} 
B.~R.~Holstein \etal, to be published

\bibitem{Kaplan:1988ku}
D.~B.~Kaplan and A.~Manohar, \Journal{\NPB}{310}{527}{1988}

\bibitem{Ramsey-Musolf:2003dd}
M.~J.~Ramsey-Musolf, preprint nucl-th/0302049, talk presented at PAVI~2002

\bibitem{Beck:2001dz}
D.~H.~Beck and B.~R.~Holstein, \Journal{INTE}{10}{1}{2001}

\bibitem{Beck:2001yx}
D.~H.~Beck and R.~D.~McKeown, \Journal{\ARNPS}{51}{189}{2001}

\bibitem{McKeown:2002by}
R.~D.~McKeown and M.~J.~Ramsey-Musolf, \Journal{\MPL}{18}{75}{2003}

\bibitem{Spayde:2003nr}
SAMPLE Collaboration: D.~T.~Spayde \etal, \Journal{\PLB}{583}{79}{2004}

\bibitem{Ito:2003mr}
SAMPLE Collaboration]: T.~M.~Ito \etal, \Journal{\PRL}{92}{102003}{2004}

\bibitem{Aniol:2004hp}
HAPPEX Collaboration: K.~A.~Aniol \etal, preprint nucl-ex/0402004

\bibitem{Cates:1999} 
HAPPEX Collaboration, G.~Cates, K.~Kumar and D.~Lhuillier, spokespersons, \\ 
proposal JLAB--E--99--115, 
http://www.jlab.org/exp\_prog/proposals/99/PR99-115.pdf

\bibitem{Kumar:2004} K.~Kumar, private communication

\bibitem{Armstrong:2000} 
HAPPEX Collaboration, D.~Armstrong and R.~Michaels, spokespersons, \\
proposal JLAB--E--00--114, 
http://www.jlab.org/exp\_prog/proposals/00/PR00-114.pdf

\bibitem{Maas:2004ta}
A4 Collaboration: F.~E.~Maas \etal, preprint nucl-ex/0401019.

\bibitem{Beck:2000} 
G\O\ Collaboration: D.~Beck, spokesperson, proposal JLAB--E--01--116, \\ 
http://www.jlab.org/exp\_prog/proposals/01/PR01-116.pdf

\bibitem{Pitt:2004} M. Pitt, private communication

\bibitem{Garvey:1992cg}
G.~T.~Garvey, W.~C.~Louis and D.~H.~White, \Journal{\PRC}{48}{761}{1993}

\bibitem{Bugel:2004yk}
FINeSSE Collaboration: L.~Bugel \etal, preprint hep-ex/0402007, 
proposal at Fermilab

\bibitem{Alberico:1998qw}
W.~M.~Alberico \etal, \Journal{\NPA}{651}{277}{1999}

\bibitem{Pate:2003rk}
S.~F.~Pate, \Journal{\PRL}{92}{082002}{2004}

\end{thebibliography}
\end{document}